%% file: thesis.tex
\documentclass[11pt,a4paper,palatino,oneside]{book}
\usepackage{anuthesismod}
\usepackage{fancyhdr}
\usepackage{graphicx}
\usepackage{amsthm}
\usepackage{amsmath}
\usepackage{amsfonts}
\usepackage{subfigure}
\usepackage{aas_macros}
\usepackage{multirow}
\usepackage[germanb,english]{babel}

\begin{document}

\newcommand{\pd}[2]{\frac{\partial #1}{\partial #2}}
\newcommand{\pdsq}[2]{\frac{\partial^2 #1}{\partial #2^2}}
\newcommand{\link}{\prec\!\!*\,}
\newcommand{\mink}{\mathbb{M}^4}
\newcommand{\Lob}{\mathbb{H}^3}

\newtheorem{thm}{Theorem}[section]
\newtheorem{lem}[thm]{Lemma}

%define psfrag replacements for swerve model picture
%\psfrag{en}{\scriptsize{$e_n$}}
%\psfrag{en-1}{\scriptsize{$e_{n-1}$}}
%\psfrag{en+1}{\scriptsize{$e_{n+1}$}}
%\psfrag{pn}{\scriptsize{$p_n$}}
%\psfrag{pn+1}{\scriptsize{$p_{n+1}$}}
%\psfrag{tauf}{\scriptsize{$\tau_{\text{f}}$}}

	%\title{\textbf{Causal Set Phenomenology}\\[3cm]}
	%\author{\textbf{Lydia Philpott}\\
	%Other text here\\}
	%\maketitle
	
\include{Title}
\setcounter{page}{2}
\include{Dedication}
\include{Declaration}
\include{Abstract}
\include{Acknowledgements}

\tableofcontents
\listoffigures
\include{ListofSymbols}

%INTRODUCTION
\include{Introduction}
\include{CausalSetTheory}
%don't include\input{StochasticEvolution}
%SWERVES
\include{Swerves}
\include{Models}
\include{ProperTimeEquation}
\include{MassiveCosmicTimeEquation}
\include{FiniteDiffConstant}
\include{BoundsonDiffConstant}
\include{MassiveSims}
\include{SwervesConclusion}
%MASSLESS PARTICLES
\include{MasslessParticles}
\include{MasslessAffineTimeEquation}
\include{MasslessCosmicTimeEquation}
\include{CMBBounds}
\include{SpectralLineShift}
\include{MasslessConclusion}
%POLARIZATION
\include{Polarization}
\include{PolarizationStateSpace}
\include{PolarizationAffineTimeEquation}
\include{PolarizationCosmicTimeEquation}
\include{ObservationalConstraints}
\include{PolarizationConclusion}
%CONCLUSION
\include{Conclusion}
\appendix
\include{PolarizationLI}
%\nocite{*}
%\bibliographystyle{h-physrev}
\bibliographystyle{hunsrt}
\bibliography{../BibliographyMasterCopy/refs} 

\end{document}

%% file: Title.tex
\begin{titlepage}
	\begin{center}
	%\title{
	\hrule height0.425pt \vskip 14pt {\Huge \scshape {\fontfamily{ptm}\selectfont Causal Set Phenomenology}}
	  \par \vskip 14pt\hrule height0.425pt
	\vspace{3cm}
	%}
	%\author{
	\textbf{Lydia Philpott}\\
	Imperial College London\\
	\vfill
	A thesis submitted for the degree of\\ Doctor of Philosophy
	 of the University of London\\
	  and the Diploma of Imperial College London\\
	2010
	%}
	\date{}
	%\maketitle
	\end{center}
\end{titlepage}

%% file: Dedication.tex
\begin{minipage}[h]{0.8\textwidth}
\vspace{5cm}
\begin{center}
\textit{To my grandfather,\\ Alan William Conway\\22 January 1924 -- 9 May 2008.}
\end{center}
\end{minipage}

\thispagestyle{plain}{\fancyhead{}\fancyfoot[C]{\thepage}}

%% file: Declaration.tex
\chapter*{Declaration}
\addcontentsline{toc}{chapter}{Declaration}

\vspace{3cm}
I declare that all work in this thesis is my own except where specifically mentioned otherwise.
\vspace{5cm}
\begin{flushright}
Lydia Philpott\\
\today
\end{flushright}

%% file: Abstract.tex
\chapter*{Abstract}
\addcontentsline{toc}{chapter}{Abstract}

Central to the development of any new theory is the investigation of the observable consequences of the theory. In the search for quantum gravity, research in phenomenology has been dominated by models violating Lorentz invariance -- despite there being, at present, no evidence that Lorentz invariance is violated. 
Causal set theory is a Lorentz invariant candidate theory of quantum gravity that seeks not to quantise gravity as such, but rather to develop a new understanding of the universe from which both general relativity and quantum mechanics could arise separately. The key hypothesis is that spacetime is a discrete partial order: a set of spacetime events where the partial ordering is the physical causal ordering between the events.  This thesis investigates Lorentz invariant quantum gravity phenomenology motivated by the causal set approach. 

Massive particles propagating in a discrete spacetime will experience diffusion in both position and momentum in proper time. This thesis considers this idea in more depth, providing a rigorous derivation of the diffusion equation in terms of observable cosmic time. The diffusion behaviour does not depend on any particular underlying particle model. Simulations of three different models are conducted, revealing behaviour that matches the diffusion equation despite limitations on the size of causal set simulated.

The effect of spacetime discreteness on the behaviour of massless particles is also investigated. Diffusion equations in both affine time and cosmic time are derived, and it is found that massless particles undergo diffusion and drift in energy. Constraints are placed on the magnitudes of the drift and diffusion parameters by considering the blackbody nature of the cosmic microwave background. Spacetime discreteness also has a potentially observable effect on photon polarisation. For linearly polarised photons, underlying discreteness is found to cause a rotation in polarisation angle and a suppression in overall polarisation.

\clearpage
\thispagestyle{empty}

%% file: Acknowledgements.tex
\chapter*{Acknowledgements}
\addcontentsline{toc}{chapter}{Acknowledgements}

Special thanks to my supervisor, Fay Dowker, without whose support this work would not have been possible.
Thanks are also due to Rafael Sorkin, whose ideas provided much of the inspiration for this work.
I would like to thank Steven Johnston, Joe Henson, and Sumati Surya for helpful discussions throughout the course of this research. Thanks are also due to Steven Johnston for proofreading parts of this thesis. Thanks to David Rideout for making the Cactus CausalSets arrangement available and for taking the time to assist me with its use.

This research was funded by a Tertiary Education Commission of New Zealand Top Achiever Doctoral Scholarship and an FfWG Foundation Main Grant.

Finally, many thanks to my parents for their unfailing support, and to Richard for proofreading large parts of this work and for all his patience, support, and encouragement.
\clearpage
\thispagestyle{empty}

%% file: ListofSymbols.tex
\chapter*{Notation}
\addcontentsline{toc}{chapter}{Notation}

\begin{tabular}{cp{0.7\textwidth}}
$\prec$ & The `precedes' relation on a causal set.\\
$\link$ & The `link' relation on a causal set.\\
$d(x,y)$ & The length of the longest chain between elements $x$ and $y$ on a causal set.\\
$\dot{x}$ & The derivative of $x$ with respect to time.\\
$A^{*},\,A^{\dagger},\,*A$ & The complex conjugate, adjoint, and Hodge dual of $A$, \mbox{respectively}.\\
$\mink$ & 4-dimensional Minkowski spacetime with metric $\eta_{\mu\nu}$.\\
$\Lob$ & The mass shell, momentum state space for massive particles.\\
$\Lob_0$ & The massless particle momentum state space.\\
$\mathcal{B}$ & The Bloch sphere polarisation state space.\\
\multicolumn{2}{l}{A metric signature $(-+++)$ is used throughout.}\\
\multicolumn{2}{l}{Planck units with $c=h=G=1$ and Boltzmann constant $k_B=1$ are used.}\\
\end{tabular}

\clearpage
\thispagestyle{empty}

%% file: Introduction.tex
\chapter{Introduction}
\label{c:introduction}

\begin{quote}
\ldots I shall admit a system as empirical or scientific only if it is capable of being tested by experience.
\end{quote}
\begin{flushright}
Karl Popper~\cite{Popper:1992}
\end{flushright}

Is there really a singularity inside a black hole? Did the universe have a beginning? These are just two of the many questions that can't be answered until a theory of quantum gravity is developed. Quantum gravity isn't demanded by any unexplained experimental results. The predictions of general relativity and quantum theory agree with experiments wherever they have been tested. The main motivation behind the long standing search for quantum gravity is a philosophical desire for a single unified approach to physics. General relativity and quantum theory provide entirely different world views: is the universe a dynamic four dimensional Lorentzian manifold, or is there a fixed background spacetime on which fields exist? Or is reality stranger than either of these options? Not only are our main theories of physics seemingly incompatible, they contain within themselves the signs of their own breakdown. The singularity inside a black hole is not so much a prediction of general relativity as an indication that we are attempting to use the theory in a realm where it doesn't apply. Renormalisation and the associated problems in quantum theory suggest to me, at least, that a key part of our understanding is lacking. 

The search for quantum gravity has become not merely a search for a mathematical method of tying the theories together, but a search for a completely new understanding of the fundamental nature of spacetime. A number of approaches to quantum gravity exist. Most, including the prominent approaches of string theory and loop quantum gravity, give some preference to quantum theory, attempting to literally `quantise gravity' to fit it into a unified framework. Others, such as the causal set approach that will be discussed in this thesis, begin with a completely new fundamental structure and hope that general relativity and quantum theory will arise in the appropriate limits of the theory. No approach to quantum gravity, as yet, can be considered complete. 

This thesis will focus on the possibility that spacetime may be fundamentally discrete and, more importantly, whether such a hypothesis is testable. 
If the explanation of experimental results has not yet required any theory of quantum gravity, how are we to test the many theories that are being developed? If no effort is made to test developing theories they risk becoming only interesting mathematical constructs and losing any connection with a description of reality. As research areas expand and it becomes more and more difficult for an individual to comprehend an entire subject, it is crucial to remind ourselves that if we claim to be physicists we should be attempting to check our theories against the real world at every step. 

Since theories of quantum gravity must reproduce the results of general relativity and quantum theory in all areas where they have been shown to hold, and introduce new phenomena only on very small spacetime scales (or large energy scales), directly testing them is, at least in the near future, impossible. To test quantum gravity we need to seek ways in which such small effects could become amplified and result in deviations from the standard predictions of general relativity and quantum theory. 

It may seem strange to seek a signature for fundamental spacetime discreteness in large scale phenomena. Consider, however, the discreteness of matter. When Einstein provided an explanation for Brownian motion, he provided the last piece of evidence necessary to convince any doubters of the atomic nature of matter. That matter was made of atoms and molecules was, of course, not a new idea at that point. Einstein did not resolve the question of the discreteness of matter by directly observing atoms or molecules, but by recognising that the already observable phenomenon of Brownian motion demonstrated their existence. Likewise, to determine if spacetime is discrete we should not look to magical future technology that may let us see spacetime `atoms' directly, but rather seek some currently observable phenomena that reveals the answer.

Investigations into quantum gravity phenomenology have thus far focused overwhelmingly on producing and observing small violations of Lorentz invariance, often by introducing modified dispersion relations. As yet there is no evidence that Lorentz invariance is violated at all. This is not to say that Lorentz violating quantum gravity phenomenology should not be explored: Lorentz violations can be investigated by current experiments and thus offer a very useful way of testing quantum gravity theories. It is often unclear, however, whether Lorentz violations are a necessary consequence of any particular quantum gravity theory, or simply a possible outcome that happens to be the only one that can currently be tested. Observations have forced very tight constraints on Lorentz violating models, for a recent review see~\cite{Liberati:2009pf}. Violations of Lorentz invariance can never be ruled out, as experiments can only constrain the effects to be smaller. While it is important to explore these constraints to experimental limits, it is clear that Lorentz \textit{invariant} quantum gravity phenomenology deserves more attention than it currently receives. This thesis attempts to begin to remedy this problem.
 
Causal set theory, a Lorentz invariant, discrete theory of quantum gravity, provides the primary motivation for the phenomenology discussed in this thesis. Although the work in this thesis is discussed within the framework of causal set theory, it should be noted that many of the conclusions are more generally applicable. 

To investigate Lorentz invariant phenomenology, I focus on the effect spacetime discreteness would have on the propagation of particles. Particles travelling through a spacetime with an underlying discreteness are expected to deviate from the continuum geodesics predicted by general relativity. For massive particles this was first considered by Dowker et al.~\cite{Dowker:2003hb}, who found the propagation of massive particles could be described by a diffusion equation in the proper time of the particle. This thesis extends the work of Dowker et al.~to derive a diffusion equation in observable laboratory time (`cosmic time') and also numerically investigates causal set models for massive particle propagation. A phenomenological model for massless particles travelling in discrete spacetime is also developed and astrophysical and cosmological data allow the free parameters in the model to be constrained. The models discussed here do not, unfortunately, make falsifiable predictions. Stronger and stronger constraints could be placed on the free parameters without ever ruling out the effects, unless, of course, there is an indisputable observation of Lorentz invariance violation. It is hoped that future developments in causal set theory will allow `natural' values for the phenomenological parameters to be derived. Observations that conclusively rule out the models would then be possible. In the meantime the models help us determine where to look for the signature of discreteness. The questions raised by the investigations into phenomenology will hopefully also contribute to progress in the underlying theory.
\vspace{1cm}

Chapter~\ref{s:causalsettheory} introduces causal set theory and provides the basic definitions that will be required later in the thesis. Chapter~\ref{c:swerves} discusses the effect of spacetime discreteness on the propagation of massive particles. This chapter begins by reviewing the swerves diffusion model proposed by Dowker et al.~in~\cite{Dowker:2003hb}, while the remainder of the chapter consists of original work. In Chapter~\ref{c:massless} massless particles are considered and found to also experience diffusion.  Chapter~\ref{c:polarisation} develops the work on massless particles further by considering the effect of discreteness on photon polarisation. 

Portions of Chapter~\ref{c:swerves}, together with a large part of the content of Chapter~\ref{c:massless} appear in
\vspace{-5mm}
\begin{quote}
F.~Dowker, L.~Philpott, and R.~Sorkin. Energy-momentum diffusion from spacetime discreteness. \textit{Phys.~Rev.}, D79:124047, 2009, 0810.5591.\\
\end{quote}
\vspace{-10mm}
Section~\ref{s:swervesnumeric} appears in\\
\vspace{-10mm}
\begin{quote}
L.~Philpott. Particle simulations in causal set theory. \textit{Class.~Quantum Grav.}, 27:042001, 2010, 0911.5595.\\
\end{quote}
\vspace{-10mm}
The majority of the work in Chapter~\ref{c:polarisation} appears in\\
\vspace{-10mm}
\begin{quote}
C.~Contaldi, F.~Dowker, and L.~Philpott. Polarisation Diffusion from Spacetime Uncertainty, 1001.4545.\\
\end{quote}

Throughout this thesis Planck units defined by $c=h=G=1$ will be used. Unless otherwise specified Boltzmann's constant is also set to one: $k_B=1$.

%% file: CausalSetTheory.tex
\chapter{A brief introduction to causal set theory}
\label{s:causalsettheory}

\begin{quote}
\ldots in a discrete manifold the principle of metric relations is already contained in the concept of the manifold, but in a continuous one it must come from something else. Therefore, either the reality underlying space must form a discrete manifold, or the basis for the metric relations must be sought outside it, in binding forces acting upon it.
\end{quote}
\begin{flushright}
Riemann (translated in~\cite{Spivak:1979})
\end{flushright}

\begin{quote}
The continuity of space apparently rests upon sheer assumption unsupported by any \textit{a priori} or experimental grounds.
\end{quote}
\begin{flushright}
Whitehead~\cite{Whitehead:1948}
\end{flushright}

%\section{Spacetime discreteness}

Given that both general relativity and quantum theory are based on continuum spacetimes, it may seem slightly strange that spacetime discreteness is invoked in attempts to reconcile the two. Why throw away the one thing these theories have in common? In brief, the singularities and associated infinite curvatures of general relativity could indicate that the continuum is not a good description of spacetime on very small scales; in quantum field theory discreteness would give a natural cutoff and allow issues with renormalisation to be avoided. 

Discrete spacetime is by no means a new idea. Riemann (as quoted above) points out the elegance of a discrete manifold: a discrete manifold actually contains more information than a continuum. Einstein, in view of the discrete nature of matter, suggested that ``the continuum of the present theory contains too great a manifold of possibilities''.~\footnote{Einstein to Walter D{\"a}llenbach, November 1916, excerpts published in~\cite{Stachel:1986}.} Schr{\"o}dinger also discusses the subject, stating that the continuum ``may very well turn out to be out of place for physical space and physical time''~\cite{Schrodinger:1954}. A comprehensive review of the history of discrete spacetime would fill a thesis in and of itself; for an interesting discussion of early ideas see~\cite{Abramenko:1958}. One further historical aspect should be mentioned briefly: the first Lorentz invariant `quantised spacetime' was described by Snyder in 1946~\cite{Snyder:1946qz}. It is, however, a rather different approach to the one discussed here. 

A considerable difficulty in formulating a discrete theory lies in the fact that many of the mathematical tools we are accustomed to using in the description of physics rely on the existence of the continuum. What is the analogue of a differential equation of motion in a discrete spacetime? Developing a discrete theory of spacetime involves not only new physics, but a new language to describe that physics.

Causal set theory, first proposed by Bombelli et al.~in 1987~\cite{Bombelli:1987aa}, is a discrete, Lorentz invariant approach to quantum gravity. Although there are a number of approaches to quantum gravity that are in some sense discrete (Regge calculus, causal dynamical triangulations, loop quantum gravity, to name a few), the exact nature of the discreteness varies greatly. Causal set theory draws its primary motivation from theorems proved by Malament~\cite{Malament:1977} and Levichev~\cite{Levichev:1987}, and based on the results of Hawking, King, and McCarthy~\cite{Hawking:1976fe}: for a past and future distinguishing spacetime, the causal ordering together with a four-dimensional volume element provides sufficient information to determine all metric, topological, and differentiable structure of a spacetime. If spacetime is discrete, this result translates to (in a phrase coined by Sorkin):
\begin{center}
\textbf{order + number = geometry.}
\end{center}
Number here is just the equivalent of volume -- in a discrete spacetime the volume of a region is determined by counting the elements within it.
In other words, a collection of elements endowed with a causal ordering should be all that is needed to describe all the complexities of spacetime to the discreteness scale.
Early proposals for a theory based on discrete spacetime, endowed only with a causal ordering of events, were made independently by both Myrheim~\cite{Myrheim:1978} and 't~Hooft~\cite{tHooft:1979} in 1978. These theories remained apparently undeveloped until causal set theory was put forward in 1987.
For comprehensive reviews of the field see, for example,~\cite{Henson:2006kf, Dowker:2006sb, Sorkin:1990bh, Sorkin:1990bj, Sorkin:2003bx}.

\section{What is a causal set?}

Specifically, a causal set is a set $C$ endowed with a binary relation $\prec$ satisfying:
\begin{enumerate}
        \item transitivity: if $x\prec y$ and $y\prec z$ then $x\prec z$, $\forall x,y,z\in C$;
        \item reflexivity: $x\prec x$, $\forall x \in C$;
        \item acyclicity: if $x\prec y$ and $y \prec x$ then $x=y$, $\forall x, y \in C$;
        \item local finiteness: $\forall x, z\in C$ the set $\left\{y\mid x\prec y\prec z\right\}$ of elements is finite.
\end{enumerate}
The relation $\prec$ is commonly called `precedes'.
This definition seems a little more intuitive if one considers, for a moment, a standard continuum Lorentzian manifold, $\mathcal{M}$. The usual causal future relation, $J^+$, satisfies the first three of the above points. The causal future $J^+(x),\,x\in\mathcal{M}$, is the union of $x$ with the set of all $y\in\mathcal{M}$ that can be reached from $x$ by a future-directed, non-spacelike curve in $\mathcal{M}$. This relation is clearly transitive, reflexive ($x\in J^+(x)$) and acyclic (provided the spacetime has no closed causal curves). The condition that the causal future relation does not satisfy is local finiteness. Local finiteness provides the discreteness of the theory -- in any causal interval of the causal set there are only finitely many elements.

\begin{figure}[t]
\begin{center}
\includegraphics[width=0.6\textwidth]{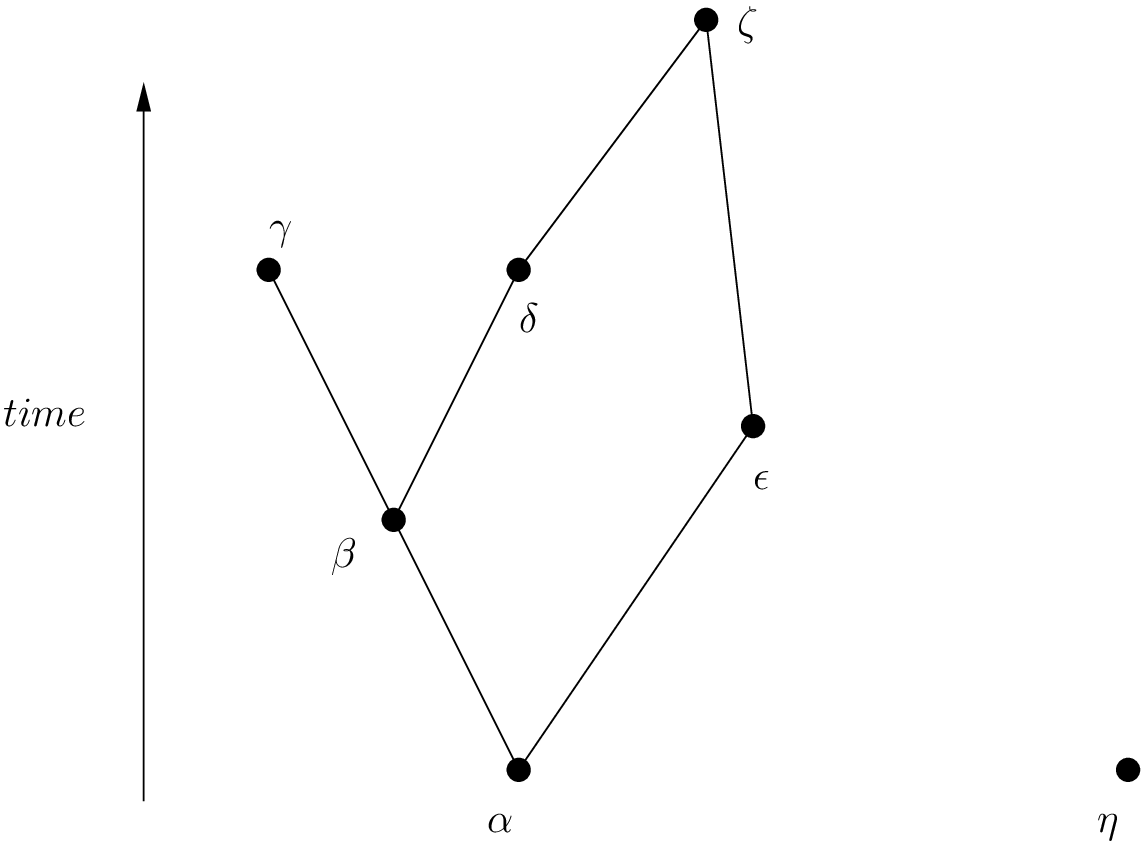}
\caption[Hasse diagram]{An example Hasse diagram of a 7 element causal set.}\label{f:hasse}
\end{center}
\end{figure}

Causal sets can be visualised using Hasse diagrams (see Figure~\ref{f:hasse}). Here time runs vertically, elements are drawn as points and relations as lines. Relations implied by transitivity are omitted to avoid over-complicating the diagram. Even so, visualising a causal set using a Hasse diagram is only feasible for very small causal sets. Although Hasse diagrams will not be used in the remainder of this thesis, Figure~\ref{f:hasse} provides useful examples for the following important definitions.

Let $C$ be a causal set.
\begin{enumerate}
        \item A \textit{chain} is a totally ordered subset of $C$, e.g.~$\alpha\prec\beta\prec\zeta$ in Figure~\ref{f:hasse}.
        \item A \textit{longest} chain between two elements $x,\;y\in C$ is a chain whose length is longest amongst chains between those endpoints, e.g.~$\alpha\prec\beta\prec\delta\prec\zeta$ is a longest chain between $\alpha$ and $\zeta$. There may be more than one longest chain between two elements. The length (i.e.~number of steps) of the longest chain between elements $x,\;y\in C$ will be denoted $d(x,y)$. For example $d(\alpha,\zeta)=3$.
        \item A \textit{link} is an irreducible relation: elements $x$ and $y$ are linked if and only if $d(x,y)=1$, e.g.~$\alpha\prec\beta$ is a link. If two elements $x,\;y\in C$ are linked, it will be denoted $x\link y$.
        \item A \textit{path} is a chain consisting of links, e.g.~$\alpha\link\beta\link\gamma$.  
\end{enumerate}

\section{The causal set hypothesis}

Above I provided a mathematical definition of a causal set. Causal set theory is based on the hypothesis that spacetime \textit{is} a causal set (or, indeed, a quantum sum-over-causal sets). The observed continuum Lorentzian manifold, it is assumed, arises as an approximation to an underlying causal set. The partial order gives rise to the causal ordering of events in the approximating continuum spacetime, and the number of elements comprising a spacetime region gives the volume of that region in fundamental units. The fundamental volume unit is expected to be of the order of the Planck volume (where necessary in the later chapters of this thesis, it will be assumed that the fundamental volume is equal to the Planck volume).
Note that any continuum manifold that approximates a causal set is necessarily Lorentzian, as only a Lorentzian metric can give rise to a partial order on spacetime points. 

The central, and as yet unproved, conjecture of causal set theory is that if two continuum manifolds are in some sense a `good approximation' to a given causal set, then those two continuum manifolds should themselves be approximately the same. The idea here is simply that although multiple manifolds could approximate a given  causal set, the manifolds should capture all the significant information from the causal set and only vary between themselves on very small scales. It is far from simple to mathematically define what is meant by `approximate' in this situation. Some intuitive understanding can be gained by working backwards: constructing causal sets from manifolds via a process called `sprinkling'.

\section{Sprinklings}
\label{s:sprinkling}

\begin{figure}[t]
\begin{center}
\includegraphics[width=0.8\textwidth]{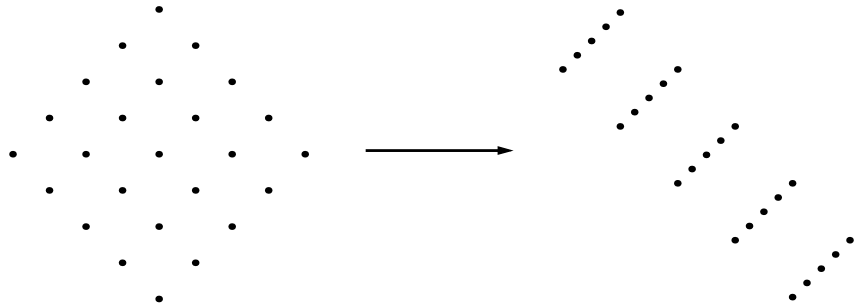}
\caption[A diamond lattice]{A diamond lattice no longer appears uniform when viewed in a boosted frame.}\label{f:lattice}
\end{center}
\end{figure}

A causal set can be constructed by selecting points from a Lorentzian manifold. The causal relations between the points on the manifold induce the order relations between the elements in the causal set. 
The most obvious way to select points from a manifold is to construct a regular lattice, but this does not provide the features needed in causal set theory. Consider choosing a frame and selecting points from two dimensional Minkowski spacetime in a diamond lattice. If a Lorentz boost is applied the points are no longer a good sampling of the manifold, as is shown in Figure~\ref{f:lattice}. In the boosted frame the number of elements in any region is clearly not a good approximation of the volume of that region. 

\begin{figure}[t]
\begin{center}
\includegraphics[width=\textwidth]{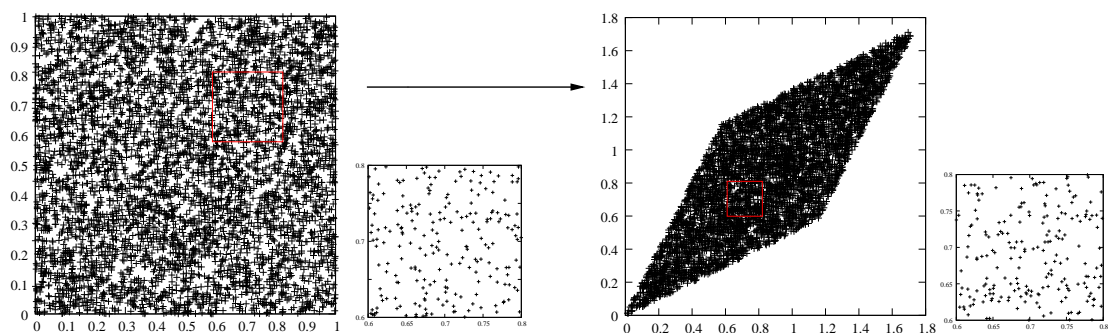}
\caption[An example sprinkling]{A causal set produced by sprinkling into a region of two dimensional Minkowski spacetime remains uniform density under a Lorentz boost. The empty regions in the boosted case are edge effects due to sprinkling into a finite region.}\label{f:sprinkling}
\end{center}
\end{figure}

Consider, instead, selecting points at random from a manifold via a Poisson process in which the probability measure is equal to the spacetime volume measure in some fundamental units -- this process is called `sprinkling'. The number of points chosen from any region of the manifold will be approximately equal (up to Poisson  fluctuations) to the volume of that region in fundamental units. If a Lorentz boost is applied to points selected this way, the distribution remains of uniform density (see Figure~\ref{f:sprinkling}). 
For a proof that the sprinkling process is Lorentz invariant, see~\cite{Bombelli:2006nm}. Sprinklings prove very useful in analytic and numeric studies of causal sets -- Chapter~\ref{c:swerves} will make use of them in both ways.

Earlier in this section, the longest chain between two elements $x,\,y$ in a causal set was defined. For a causal set generated by sprinkling into Minkowski spacetime, the length of the longest chain, $d(x,y)$, is found to be proportional to the proper time between $x$ and $y$ in the limit of large distances~\cite{Brightwell:1991}. The constant of proportionality is dependent on the spacetime dimension. A longest chain in a causal set is thus a close approximation to a timelike geodesic. For a numerical investigation of this correspondence in conformally flat spacetimes see~\cite{Ilie:2005qg}.

\section{Causal set phenomenology}

As discussed in Chapter~\ref{c:introduction} the majority of current research in quantum gravity phenomenology focuses on violations of Lorentz invariance. Causal set theory, on the other hand, provides a way of investigating Lorentz invariant quantum gravity phenomenology. The difficulty of course, is that causal set theory is incomplete, and obtaining testable predictions from an incomplete theory is far from simple. Although much will have to wait until the development of a quantum dynamics for causal sets, the concreteness and simplicity of the causal set hypothesis allow considerable progress to be made even at our early state of research.

Before moving on in Chapter~\ref{c:swerves} to the particle phenomenology that forms the basis of this thesis, another significant result in causal set phenomenology should be mentioned. As early as 1997 Sorkin~\cite{Sorkin:1997gi} suggested that causal set theory could give rise to a cosmological constant of the order $10^{-120}$ -- a number that matches current observations (see also~\cite{Ahmed:2002mj, Sorkin:2007bd}). This prediction for, in fact, a non-constant cosmological constant relies on the simple idea that a discrete spacetime leads to fluctuations in volume. As discussed above, the sprinkling procedure places $N$ points in a continuum volume $V$ with fluctuations of the order $\sqrt{N}$. Although there are questions that remain to be addressed~\cite{Barrow:2006vy, Zuntz:2008zza} this model illustrates the powerful results that can be obtained from causal set theory.

%% file: Swerves.tex
\chapter{Swerves}
\label{c:swerves}
\begin{center}
\begin{quote}
quare etiam atque etiam paulum inclinare necessest\\
corpora; nec plus quam minimum, ne fingere motus\\
obliquos videamur et id res vera refutet.
\end{quote}
\end{center}
\begin{flushright}
Lucretius, De Natura Rerum, ll. 243-245
\end{flushright}
\begin{quote}
Wherefore it is necessary that bodies swerve a little perpetually,\\
Not more than a little, lest we seem to imagine sideway motions, and the facts confute our conjecture.\footnote{Translation courtesy of Dr.~R.~M.~Pollard, who notes that \textit{inclinare} would be better translated as 'move back and forth', which is somewhat less succinct yet even more fitting for our context. }
\end{quote}

General relativity predicts that a free, massive particle will travel along a timelike geodesic -- a prediction well tested by observations. If the underlying structure of spacetime is discrete rather than continuous, particles will no longer be able to travel along perfect geodesics. Discreteness, one expects, will introduce some random fluctuations into a particle's trajectory. Of course, given how well tested general relativity is, if such fluctuations exist they must be so small as to have escaped notice. Spacetime discreteness is likely to occur on the Planck scale, so it is certainly reasonable to suppose that an effect on particle trajectories exists that has not yet been observed.

In~\cite{Dowker:2003hb} Dowker et al.~consider this problem in the context of causal set theory. Motivated by a simple model of particle motion on a causal set (see Section~\ref{s:models}) they suggest that particles travelling through a discrete spacetime will be subject to \textit{swerves}: random fluctuations in momentum. From a simple random walk on $\mathbb{R}$ we can derive the standard diffusion equation in the continuum limit: similarly, a basic stochastic process that gives random fluctuations in momentum is found in the continuum limit to lead to a diffusion equation. 

The swerves diffusion equation given in~\cite{Dowker:2003hb} does not depend on any specific underlying particle model -- nor, in fact, does it explicitly depend on the hypothesis that spacetime is a causal set. The power of the swerves diffusion equation lies in it being the \textit{unique} Markovian Poincar\'e invariant diffusion equation on the state space. Any underlying process that gives small random fluctuations in momentum will be described by this equation in the continuum limit. %Note that although the derivation of the diffusion equation doesn't utilise any properties specific to causal set theory, it has been suggested [HAS IT?] that any theory of quantum gravity that is both discrete and Lorentz invariant must in some sense \textit{be} causal set theory. 

A full derivation of the swerves diffusion equation, first stated in~\cite{Dowker:2003hb}, is given in Section~\ref{s:massivepropertime}. The diffusion equation, as derived by Dowker et al.~in terms of the proper time of the particle, is not the most useful form. To make contact with observations an equation in terms of an observable cosmic time is needed. In~\cite{Dowker:2003hb} Dowker et al.~suggest the form of such an equation in the special case of a homogeneous distribution of particles. Here this work is extended and a full derivation of the inhomogeneous case is given in Section~\ref{s:massivecosmictime}. 

I begin, however, by investigating models for particle propagation on causal sets more fully. Three models, including the model first described in~\cite{Dowker:2003hb}, are discussed in Section~\ref{s:models} below. 
The connection between fundamental models and the diffusion equation is investigated in Section~\ref{s:finitediffconstant}, where it is shown that the swerves model on the causal set gives rise to a finite nonzero diffusion constant. The work in Sections~\ref{s:models},\ref{s:massivepropertime},\ref{s:massivecosmictime} appears in~\cite{Philpott:2008vd} and together with Section~\ref{s:finitediffconstant} is work done in collaboration with Fay Dowker and Rafael Sorkin. Section~\ref{s:massivebounds} discusses the observable consequences of the swerves diffusion equation and reviews the bounds placed on the diffusion constant in~\cite{Dowker:2003hb,Kaloper:2006pj}. Lastly, the models introduced in Section~\ref{s:models} are investigated numerically in Section~\ref{s:swervesnumeric}, and the behaviour is compared with that predicted by the swerves equation. The work in Section~\ref{s:massivebounds} appears in~\cite{Philpott:2009fz}.

%% file: Models.tex
\section{Models for massive particle propagation on a causal set}
\label{s:models}
What is a particle in causal set theory? It must be admitted that this is not a question that can currently be answered. In general relativity matter determines spacetime curvature. In the context of a causal set this curvature must be encoded in the structure of the causal links -- perhaps a particle is some pattern or knot of causal links? Of course, if the mass of the particle is a manifestation of a tangle of links between causal set elements, what about the other particle properties, the charge or spin?  Approaching from the direction of quantum field theory we could assign amplitudes for a particle to be `on' a particular causal set element and to move from one element to another. For work on developing quantum field theory on causal sets, see~\cite{Johnston:2008za,Johnston:2009fr}. 

An understanding of particles in causal set theory will really require a theory of quantum causal set dynamics (QCSD). If QCSD takes the form of a `sum-over-causal sets', will a particle trajectory be some sum over all trajectories in all possible causal sets?  Even without a theory of QCSD an intuition into the effect of discreteness can be gained by considering simple models.
Massive particles will be considered as classical point particles located `on' a causal set element, a particle trajectory will consist of jumps from one element to another. The models discussed in this section are clearly not physically realistic, but they do capture the important aspects of the causal set approach and therefore provide one very useful way of investigating observable consequences of discrete spacetime.
The models introduced in this section will be investigated numerically in Section~\ref{s:swervesnumeric}.

\subsection{The swerves model}
This model was introduced in~\cite{Dowker:2003hb}. 
\subsubsection{Model 1}
Construct a causal set $C$ by sprinkling into Minkowski spacetime. A massive particle trajectory is taken to be a chain of elements $e_n$ in the causal set, i.e.~a linearly ordered subset of $C$. It is assumed that the trajectory's past determines its future, but that only a certain proper time $\tau_f$ into the past is relevant. When this `forgetting time' $\tau_f$ is small the process should be approximately Markovian. $\tau_f$ is also assumed to be much larger than the discreteness scale. The particle trajectory is constructed iteratively. Suppose the particle is currently located `on' an element $e_n$, with a four-momentum $p_n$. The next element, $e_{n+1}$ is chosen such that
\begin{itemize}
\item $e_{n+1}$ is in the causal future of $e_n$ and within a proper time $\tau_f$ of $e_n$,
\item the momentum change $|p_{n+1}-p_n|$ is minimised.
\end{itemize}

\begin{figure}[t]
\begin{center}
\includegraphics[width = 0.55\textwidth]{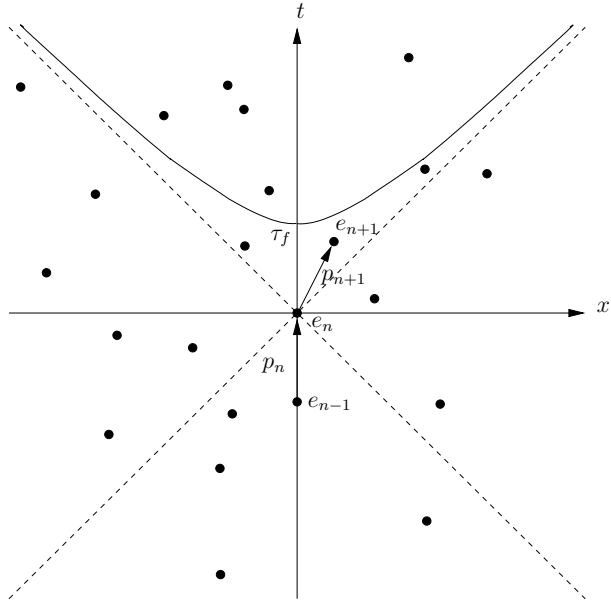}
\caption[The swerves model for particle propagation]{The swerves model for particle propagation on a causal set. This illustration is not to scale: $\tau_f$ should be much greater than the discreteness scale.}\label{f:mmodel}
\end{center}
\end{figure}

\noindent Here the momentum $p_{n+1}$ is defined to be proportional to the vector between $e_n$ and $e_{n+1}$, and thus this model relies on knowledge of the embedding of the causal set into Minkowski spacetime. The constant of proportionality is chosen such that the mass of the particle remains fixed, i.e~$p^2=-m^2$.
This method for determining the trajectory is illustrated in Figure~\ref{f:mmodel}, where the frame has been chosen to be the rest-frame of the particle at $e_n$. Heuristically, the requirement that the momentum change be minimised creates a trajectory that stays as straight as possible at each step. If we work in the rest-frame of the particle, this method chooses the next element to be as close as possible to the $t$ axis within the forgetting time. At each step there is a small random fluctuation in the momentum of the particle. There is no direct dependence on the mass of the particle in this model. Any nonzero mass will result in the same trajectory, although we could imagine that the forgetting time $\tau_f$ is mass dependent.

This model is clearly not a realistic fundamental law of motion for a particle on a causal set. It requires information not contained in the causal set -- the embedding of the elements into the approximating Minkowski spacetime. This issue is overcome in the `intrinsic' models described below. All these models treat particles as classical and zero size. 

\subsection{Intrinsic models}

For a model of particle propagation to be intrinsic to a causal set, it cannot refer to a continuum forgetting time $\tau_f$. Instead, a `forgetting number' (an integer $n_f$) must be defined. The forgetting number can be interpreted as the number of discrete steps into the past of the trajectory that are relevant in determining the future trajectory (although in Model 2 below this number is in fact $2n_f$). If the embedding into a continuum spacetime was known, $n_f$ could be roughly written $n_f = \tau_f/d_{pl}$, where $d_{pl}$ is the discreteness scale. It is important to note, however, that the models are defined on a general causal set and it is not assumed here that the underlying causal set can be faithfully embedded into a `reasonable' spacetime. 

In defining these intrinsic models there is no reference to the mass of the particle. As above, the forgetting number could be taken to be mass dependent. The models discussed cannot, however, be considered as models of particle propagation for \textit{massless} particles -- these models give trajectories whose long time behaviour approximates a timelike geodesic, rather than the null geodesic appropriate to massless particles.

Two slightly different intrinsic models will be described in this section, to give an idea of the wealth of possibilities available. First recall some causal set definitions from Chapter~\ref{s:causalsettheory}: 
\begin{itemize}
\item $d(a,b)$ denotes the length of the longest chain between two elements $a$ and $b$,
\item a path is a chain consisting of links $a\link b$, i.e.~$d(a,b)=1$. 
\end{itemize}

\subsubsection{Model 2}
\begin{figure}[th]
\begin{center}
\subfigure[Model 2.]{\label{f:model2}
\includegraphics[height=4.7cm]{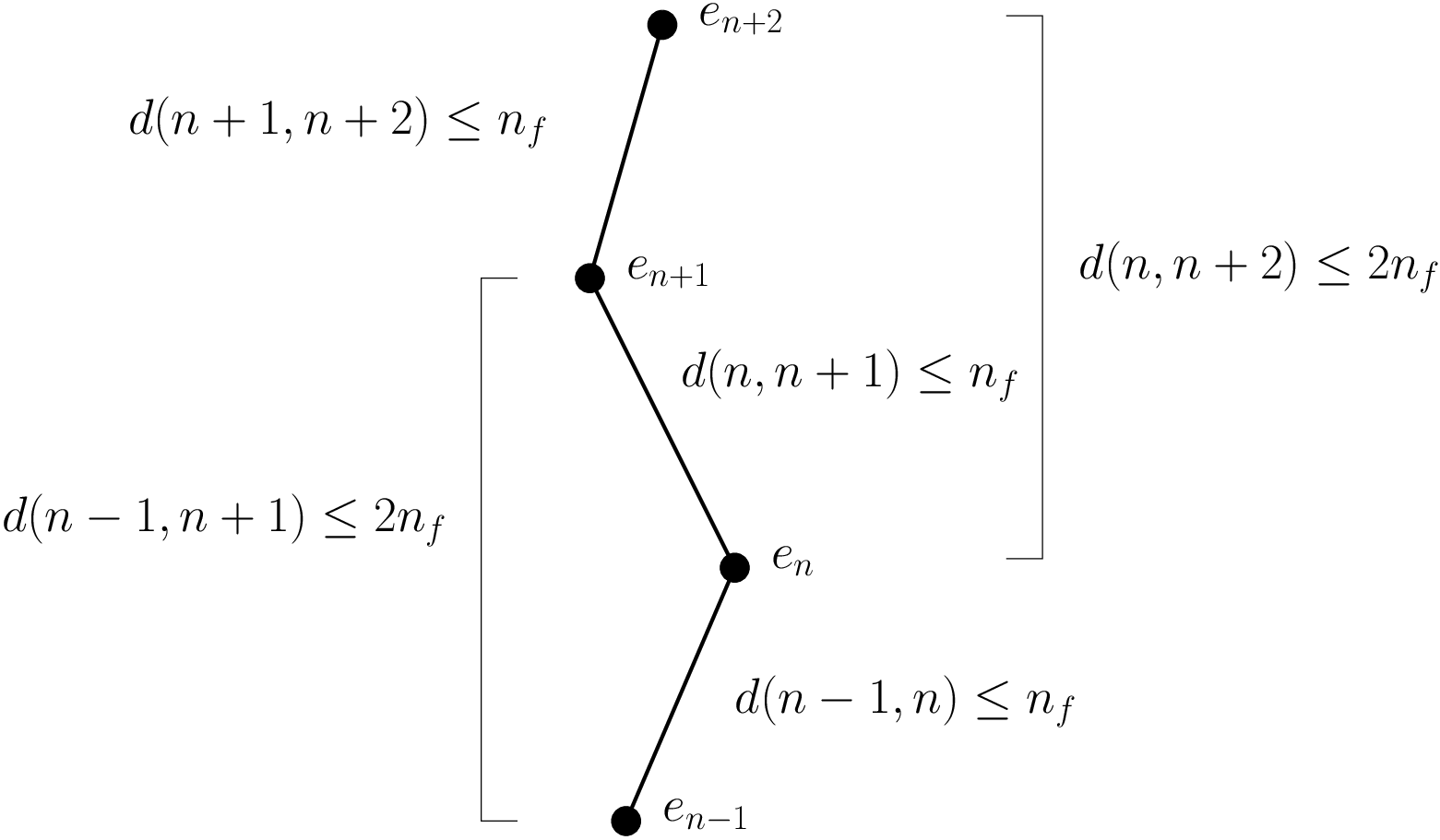}}
\hspace{5mm}
\subfigure[Model 3.]{\label{f:model3}
\includegraphics[height=4.7cm]{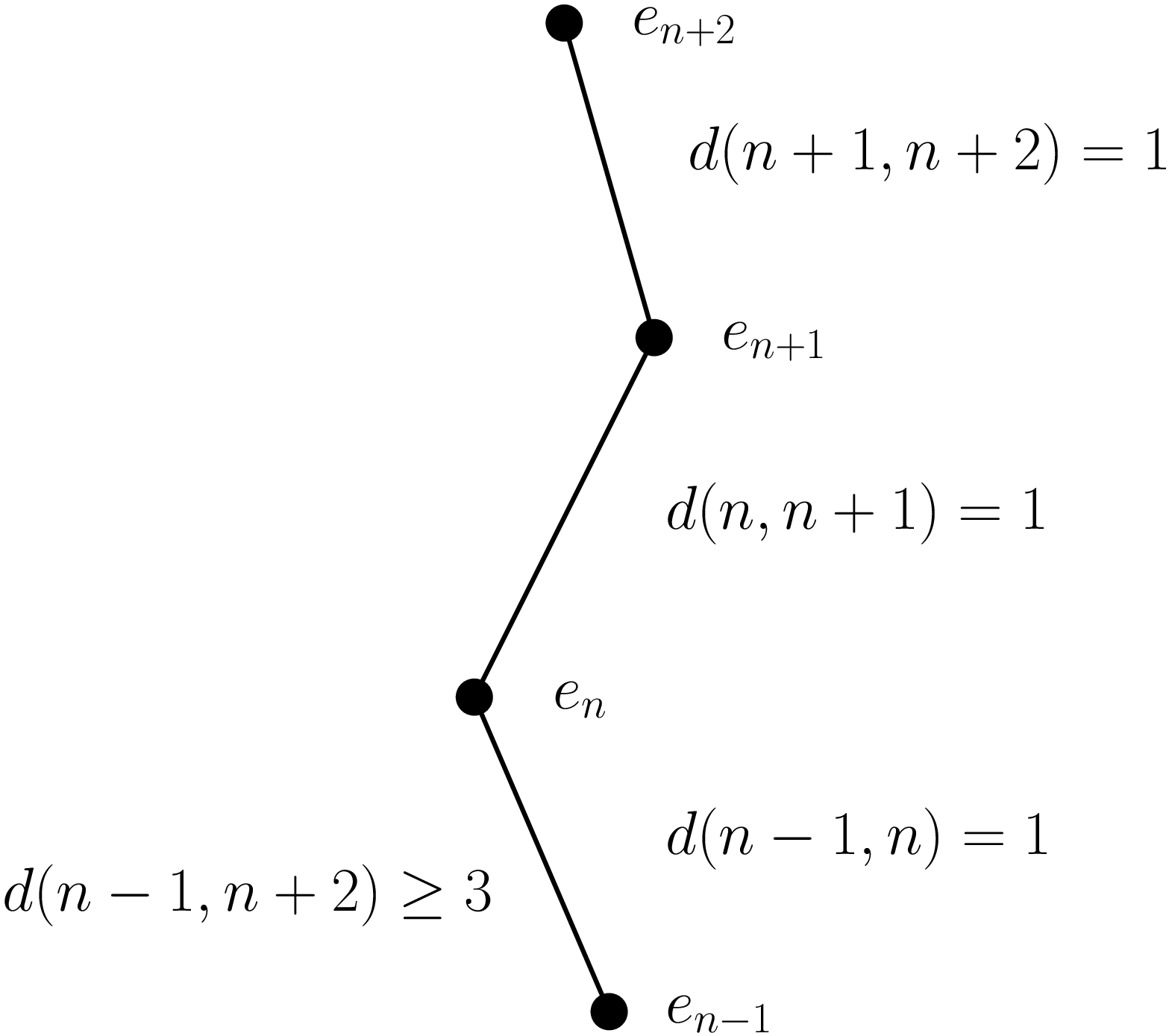}}
\caption[Intrinsic models for particle propagation]{Trajectories constructed using intrinsic models.}
\end{center}
\end{figure}

A massive particle trajectory is taken to be a chain of elements $\ldots e_{n-2}\prec e_{n-1}\prec e_n\ldots $. Given a partial particle trajectory $\ldots e_{n-1},e_n$ the next element $e_{n+1}$ is chosen such that
\begin{itemize}
\item $d(e_{n-1},e_{n+1})\leq 2n_f$,
\item $d(e_n,e_{n+1})$ is maximised subject to $d(e_n,e_{n+1})\leq n_f$,
\end{itemize}
(see Figure~\ref{f:model2}).
These requirements do not guarantee the existence of a unique $e_{n+1}$. There will, however, almost surely be finitely many eligible elements and the trajectory can be constructed by choosing an element uniformly at random from these. Note that this model is slightly different from the first intrinsic model presented in~\cite{Philpott:2008vd} (where equalities in the above conditions were given). Model 1 of~\cite{Philpott:2008vd} does not guarantee the existence of an $e_{n+1}$ under reasonable conditions. 

Under this model the particle trajectory should swerve a little, but remain approximately straight so long as $n_f$ is large. It is easiest to see this if we consider the `ideal' case where there exist $e_{n-1},e_n,e_{n+1}$ such that $d(e_{n-1},e_{n})=n_f$ and $d(e_n,e_{n+1})=n_f$. The elements have been chosen such that $d(e_{n-1},e_{n+1})\leq 2n_f$; since $d(e_{n-1},e_{n})=n_f$ and $d(e_n,e_{n+1})=n_f$ we know there exists some chain between $e_{n-1}$ and $e_{n+1}$ of length $2n_f$ and hence the equality $d(e_{n-1},e_{n+1})=2n_f$ must hold. In other words, the chain we have chosen between $e_{n-1}$ and $e_{n+1}$ is a longest chain, and thus the trajectory is approximately geodesic over $\{e_{n-1},e_{n+1}\}$ segments. In practice it will not always be possible to choose an $e_{n+1}$ such that $d(e_n,e_{n+1})= n_f$ and $d(e_{n-1},e_{n+1})= 2n_f$, necessitating the inequalities.

In this model the trajectory can be considered as composed of just the elements $\ldots e_{n-1},e_n,e_{n+1}\ldots$ or of the `filled in chain' consisting of a (randomly chosen) longest chain (of length $\leq n_f$) between $e_{n-1}$ and $e_n$, another between $e_n$ and $e_{n+1}$, and so on.  Variations on this model can also be constructed. One possibility is to choose the forgetting number at random at each step from a distribution with a mean $n_f$ and some fixed variance. 
%\begin{figure}
%\begin{center}
%\includegraphics[width=0.55\textwidth]{model2thesis.ps}
%\caption[An intrinsic model for particle propagation]{A trajectory constructed using the intrinsic model for particle propgation, Model 2.}\label{f:model2}
%\end{center}
%\end{figure}

\subsubsection{Model 3}
The trajectory is constructed as a path in this model, i.e.~$d(e_n,e_{n+1})=1$ for any $e_n,\,e_{n+1}$. 
Given a partial particle trajectory $\ldots e_{n-n_f}, \ldots, e_{n-1},e_n$ the next element $e_{n+1}$ is chosen such that
\begin{itemize}
\item $d(e_n,e_{n+1})=1$,
\item $d(e_{n-n_f},e_{n+1}) + \ldots + d(e_{n-1},e_{n+1}) + d(e_n, e_{n+1})$ is minimised,
\end{itemize}
(see Figure~\ref{f:model3}).
Note that this minimisation does not necessarily yield a unique $e_{n+1}$, in which case we construct the trajectory by choosing an element uniformly at random from those eligible. Also, if the trajectory has length less than $n_f$ the minimisation is done over all elements available. 

In this model each element is linked to the previous, i.e. $d(e_{n-1}, e_{n}) = 1$, and thus we know there exists a chain (our trajectory) of length $n_f$ between $e_{n-n_f}$ and $e_n$. The longest chain length, $d(e_{n-n_f},e_n)$, must therefore be greater than or equal to $n_f$. If we choose $e_n$ to minimise $d(e_{n-n_f},e_n)$ we ask that the trajectory be as close as possible to geodesic between $e_{n-n_f}$ and $e_n$ while fulfilling $d(e_{n-1}, e_{n}) = 1$. Minimizing the sum of the partial lengths distributes the geodesic property along the path.

%\begin{figure}
%\begin{center}
%\includegraphics[width=0.4\textwidth]{model3thesis.ps}
%\caption{A trajectory constructed using model 3.}\label{f:model3}
%\end{center}
%\end{figure}

%% file: ProperTimeEquation.tex
\section{The swerves diffusion equation}
\label{s:massivepropertime}

Although the intrinsic models described in Section~\ref{s:models} are defined on any causal set, the process is really only of interest when the causal set is a good approximation to Minkowski spacetime (or another physically relevant spacetime, such as Friedmann-Robertson-Walker). The universe we live in is approximately Minkowskian, and thus even though the behaviour of particles on nonembeddable causal sets may be very interesting, it is of little relevance when investigating the \textit{observable} phenomenology of discrete spacetime.
If it is assumed that the causal set can be produced by sprinkling into Minkowski spacetime, the particle models can be thought of as defining piecewise linear curves in Minkowski spacetime. The particle's momentum -- which cannot (currently) be defined in a manner intrinsic to the causal set -- is then defined everywhere on the curve except at the vertices. It is clear here that in the continuum limit the intrinsic models lead to random fluctuations of momentum (defined, as usual, in Minkowski spacetime).

Rather than choosing a specific particle model on a causal set and investigating its continuum behaviour, an entirely general situation will be considered. Motivated by the models for particle trajectories on a causal set, it is assumed that there exists some underlying Markovian, Poincar\'e invariant process that causes random fluctuations in momentum (and consequently position).
The process takes place in the proper time of the particle. 
Note that the models described above are approximately Markovian when the forgetting time (number) is small. If the discreteness scale is of the order of the Planck scale, the forgetting time can be many orders of magnitude greater than the discreteness scale and yet still small compared to the trajectory length.

\input{StochasticEvolution}

\subsection{Diffusion in proper time}
The state space, $\mathcal{M}$, of a swerving particle of mass $m$ is $\mathcal{M} = \mink\times\Lob$, where $\Lob$ is the mass shell. A point on $\mathcal{M}$ thus represents a position in $\mink$ and a momentum in $\Lob$.
The coordinates on $\mink$ are the usual Cartesians $\{x^\mu \}$, $\mu = 0,1,2,3$ and indices are raised and lowered with $\eta_{\mu\nu}$, the Minkowski metric.
The spatial coordinates on $\mink$ will be written as $\{x^i\}$. Cartesian coordinates in momentum
space are $p_\mu$, where $p_\mu$ is subject to the constraint that it lies on the mass shell, i.e.~the hyperboloid $\Lob$ in momentum space defined by $p_\mu p^\mu + m^2 = 0$.
$p^0 = E$ is the energy (taken to be positive) and $p = \sqrt{p_1^2 + p_2^2 + p_3^2}$ is the norm of the three momentum.
 The three coordinates on $\mathbb{H}^3$ will be written abstractly as
$p^a$.
Coordinates on $\mathcal{M}$ are denoted collectively as $X^A = \{ x^\mu, p^a\}$ and in what follows capital letters $A, B$ will be used to indicate general indices on $\mathcal{M}$; $\mu, \nu=0,1,2,3$ are indices on
$\mink$; $i,j=1,2,3$ are spatial indices on $\mink$; $a, b=1,2,3$ are indices on $\Lob$.

The metric on $\mathcal{M}$ is the product of the Minkowski metric $\eta_{\mu\nu}$ on $\mink$ and the Lobachevski metric
$g_{ab}$ on $\Lob$. This is the unique Poincar\'e invariant metric on $\mathcal{M}$ (up to an overall constant).
The `density of states', $n$,
plays a role in the formalism of \cite{Sorkin:1986}, and by symmetry, it
must be proportional
to the volume measure on $\mathcal{M}$, so $n \propto \sqrt{g}$ where
$g = {\textrm{det}}(g_{ab})$. 

As described above, a process that undergoes stochastic evolution on a manifold of states,
$\mathcal{M}$, in time parameter $T$, can be described by a current, $J^A$ and a continuity equation~\cite{Sorkin:1986}:
\begin{eqnarray}
J^A &=& -\partial_B\left(K^{AB}\rho\right)+v^A\rho\,,\\
\frac{\partial\rho}{\partial T} &=& -\partial_A J^A\,,
\end{eqnarray}
or alternatively by the equation
\begin{equation}
\pd{\rho}{T} =
 \partial_A\left(K^{AB}n\,\partial_B\left(\frac{\rho}{n}\right)-u^A\rho\right)\,,
 \label{e:diffequA}
\end{equation}
where
\begin{equation}
u^A = v^A - \partial_B K^{AB} - K^{AB}\partial_B s\,.\label{e:uA}
\end{equation}
To find the diffusion equation for the swerves particle process, it is necessary to determine $K^{AB}$ and
$u^A$. The requirement that  the equation be Poincar\'e invariant is a very stringent condition and allows $K^{AB}$ and $u^A$ to be determined up to the choice of one constant parameter. 

Consider the process in terms of $\tau$, the proper time along the worldline of the particle. 
From Equation~\ref{e:KABlim} in Section~\ref{s:stochevol}
\begin{equation}
K^{AB} = \lim_{\Delta\tau\rightarrow 0+}\left<\frac{\Delta X^A\Delta X^B}{2\Delta\tau}\right>\,.
\end{equation}
Recall $K^{AB}$ is a symmetric, positive semidefinite matrix, and here is also required to be Poincar\'e invariant. Looking first at the spacetime component of this matrix,
\begin{equation}
K^{\mu\nu} = \lim_{ \Delta \tau \to 0+} \left< \frac{\Delta x^\mu \Delta x^\nu}{2\Delta \tau}
\right>\, .
\end{equation}
$\Delta x^\mu = \frac{1}{m} p^\mu \Delta  \tau$ at every step of the process and
so $K^{\mu\nu} = \frac{1}{2}\lim_{\Delta \tau \to 0} p^\mu p^\nu \Delta \tau = 0$.
Given $K^{\mu\nu}=0$, $K^{\mu a}=0$ is required by the condition that $K^{AB}$ be positive semidefinite. To see this, first recall that a matrix is positive semi-definite if all the principal minors of the matrix are nonnegative. Consider a symmetric 2x2 matrix of the form
\begin{equation}
\begin{pmatrix}
0& a\\
a& b
\end{pmatrix}\,.
\end{equation}

To be positive semi-definite the matrix must have $-a^2\geq0$, i.e.~$a=0$. Similarly, beginning with
\begin{equation}
K^{AB}=
\begin{pmatrix}
0 & K^{\mu a}\\
K^{a\mu}& K^{ab}
\end{pmatrix}\,,
\end{equation}
we can consider all the $2\times2$ principal minors and determine $K^{\mu a} = K^{a\mu} = 0$. 

The only Lorentz invariant tensor on $\Lob$ is proportional to the metric, $g^{ab}$, and the coefficient is independent of
$x^\mu$ by translation invariance, thus $K^{ab}=kg^{ab}$, where $k>0$ is a constant. This gives
\begin{equation}
K^{AB}= \left(
\begin{array}{cc}
0 & 0\\ 0 & kg^{ab}
\end{array}
\right)\,.
\end{equation}

To determine $u^A$ first recall Equation~\ref{e:vAlim}:
\begin{equation}
v^A = \lim_{\Delta\tau\rightarrow 0+}\left<\frac{\Delta X^A}{\Delta\tau}\right>\,.
\end{equation}
The spacetime component
\begin{equation}
v^\mu = \lim_{ \Delta \tau \to 0+} \left< \frac{\Delta x^\mu}{\Delta \tau} \right>\, ,
\end{equation}
is simply $v^\mu = p^\mu/m$. The components of the true vector $u^\mu$ are equal to $v^\mu$ since $K^{\mu A} = 0$.
There is no Lorentz invariant vector on $\Lob$ and so $u^a = 0$, giving
\begin{equation}
u^A=\left(p^\mu/m,0\right)\,.
\end{equation}
The proper time diffusion equation can now be written down from Equation~\ref{e:diffequA}:
\begin{equation}
\label{e:mtau}
\pd{\rho_{\tau}}{\tau}=
k \; \partial_a \left( g^{ab} \sqrt{g} \partial_b \left(\frac{\rho_\tau}{\sqrt{g}}
\right)\right) - \frac{1}{m} p^\mu \partial_\mu \rho_\tau\,.
\end{equation}

This can be seen to be equivalent to Equation~1 of~\cite{Dowker:2003hb} if a scalar $\bar{\rho} = \rho_\tau/\sqrt{g}$ is defined:
\begin{equation}
\pd{\bar{\rho}}{\tau} = k\;\nabla^2_H \bar{\rho}
-  \frac{1}{m} p^\mu \partial_\mu \bar{\rho}\,,
\end{equation}
where $\nabla^2_H$ is the Laplacian on $\Lob$.

%% file: StochasticEvolution.tex
\subsection{Stochastic evolution on a manifold of states}
\label{s:stochevol}
The swerves equation is obtained using the general formalism of~\cite{Sorkin:1986} for stochastic evolution on a manifold of states.
Chapters~\ref{c:massless} and~\ref{c:polarisation} also draw on this framework, and thus for completeness the main points in this work will be discussed here. 

Consider a `mesoscopic' process occurring on a manifold of states $\mathcal{M}$. The current state of the system is described by a probability density $\rho$.
It is assumed that the future of the physical system depends only on its present state and not its past, i.e.~the evolution is a Markov process. It is also assumed that the system traces a continuous path through $\mathcal{M}$.
Such a process can be described by a linear, first order in time equation
\begin{equation}
\pd{\rho}{T} = K^{AB}\partial_A\partial_B\rho + B^{A}\partial_A\rho-C\rho\,,
\label{e:stochevol}
\end{equation}
where $\rho=\rho(x,T)$, the probability density for the system to be in a state $x\in\mathcal{M}$ at time $T$, is a scalar density in $\mathcal{M}$. The indices $A,B=1,\ldots,\textrm{dim}(\mathcal{M})$. $K^{AB} = K^{BA}$, $B^A, C$ are functions of $x$ (and possibly $T$). For extensive justification of this assumption see~\cite{Sorkin:1986}. Note that if Equation~\ref{e:stochevol} were to contain higher order terms it would allow unphysical negative probability densities, $\rho<0$, proof of this is given in Appendix A of~\cite{Sorkin:1986}.

The requirement that the probability density $\rho$ be nonnegative leads to a constraint on $K^{AB}$ in the above, general equation. 
Suppose the initial distribution, in some neighbourhood of the origin, is given by
\begin{equation}
\rho(x,0) = \frac{1}{2}Z_{AB}x^Ax^B\,,
\end{equation}
where $Z$ must be positive by the requirement $\rho\geq 0$. Clearly $\rho(0,0) = 0$ and in order for $\rho(0,T)\geq0$ we must have $\dot{\rho}(0,0)\geq 0$. From Equation~\ref{e:stochevol} we can conclude $Z_{AB}K^{AB}\geq 0$. Suppose $Z_{AB} = z_A z_B$. This implies $z_A z_B K^{AB}\geq 0$, i.e.~$K^{AB}$ is a (symmetric) positive semidefinite matrix (a matrix $A$ is positive semidefinite if $x^{T}Ax\geq0\: \forall x\in\mathbb{R}^n$).  

We can also impose the conservation of probability
\begin{equation}
\int\dot{\rho}\,dx = 0\,.
\end{equation} 
With $\rho$ given by Equation~\ref{e:stochevol}, integration by parts implies $C = \partial_A\partial_BK^{AB}-\partial_AB^A$. Substituting this expression back into Equation~\ref{e:stochevol} gives
\begin{eqnarray}
\pd{\rho}{T} &=& K^{AB}\partial_A\partial_B\rho + B^{A}\partial_A\rho-\left(\partial_A\partial_BK^{AB}-\partial_AB^A\right)\rho\nonumber\\
&=& \partial_A\partial_B(K^{AB}\rho) + \partial_A(B^A\rho) - 2\partial_A\partial_BK^{AB}\rho - 2\partial_BK^{AB}\partial_A\rho\nonumber\\
&=&\partial_A\partial_B(K^{AB}\rho) + \partial_A(B^A\rho) - 2\partial_A(\partial_BK^{AB}\rho).
\end{eqnarray}

Equation~\ref{e:stochevol} can therefore be written in terms of a current, $J^A$, and a continuity equation:
\begin{eqnarray}
J^A&=&-\partial_B(K^{AB}\rho) + v^A\rho\,,\label{e:current}\\
\pd{\rho}{T}&=&-\partial_A J^A\label{e:continuity},
\end{eqnarray}
where $v^A = 2\partial_B K^{AB}-B^A$. Note that $J^A$ is a vector density (see pg.~124 of~\cite{Sorkin:1986}) and the vector density transformation implies that while $K^{AB}$ is a tensor, $v^A$ is not. The above equations can be expressed in terms of a true vector $u^A$ if the density of states, $n$, is introduced. The vector $u^A$ is defined by
\begin{equation}
u^A = v^A - \partial_B K^{AB} - K^{AB}\partial_B s\,,
\end{equation}
where $s$ is the entropy scalar
\begin{equation}
s = k_B \log n\,.
\end{equation}
Here $k_B$ is the Boltzmann constant (note that in the sections that follow $k_B=1$). Equations~\ref{e:current} and~\ref{e:continuity} can then be written in the form
\begin{equation}
\pd{\rho}{T} = \partial_A\left(K^{AB}D_B\rho - u^A\rho\right)\,,
\end{equation}
where
\begin{equation}
D_A\rho = \partial_A\rho -\rho\partial_A s\,.
\end{equation}
In the sections that follow, it will often be useful to use the following relations for $K^{AB}$ and $v^A$:
\begin{eqnarray}
K^{AB} &=& \lim_{\Delta T\rightarrow 0}\left<\frac{\Delta x^A\Delta x^B}{2\Delta T}\right>\label{e:KABlim}\,,\\
v^A &=&\lim_{\Delta T\rightarrow 0}\left<\frac{\Delta x^A}{\Delta T}\right>\,.\label{e:vAlim}
\end{eqnarray}
Here $<>$ denotes the expectation value. These equations give a way of relating the abstract objects $K^{AB}$ and $v^A$ to the basic properties of the physical stochastic process -- the `spatial' step $\Delta x$ and the time step $\Delta T$. Equations~\ref{e:KABlim}  and~\ref{e:vAlim} are derived in Appendix B of~\cite{Sorkin:1986}.

%% file: MassiveCosmicTimeEquation.tex
\section{Diffusion in cosmic time}
\label{s:massivecosmictime}

Given an initial distribution of particles, for instance from an astronomical source, the equation derived above is not very useful for predicting the results of observations. Even if particles all leave the source at the same time with the same momentum, the momentum variation induced by the swerves will result in particles arriving after different proper times and at different observatory times. The proper time that elapses along the particles' worldlines from source to detector is not observable.
 To compare the swerves model with experiment and observation it is necessary to describe the evolution of the distribution in time in the rest frame of our detector, which time will be referred to as {\it cosmic time}. One may ask why the equation was derived in terms of proper time if it is more useful to obtain a cosmic time equation. The reason becomes clear if one considers that when working with proper time, the explicit Lorentz invariance makes it simple to determine $K^{AB}$ and $u^A$. The result can then be easily transformed to a specific frame, as shown here.

A first step in this direction was to look at the nonrelativistic limit of the proper time diffusion equation, when proper time and cosmic time are comparable. The nonrelativistic limit in fact proves sufficient to place very strong bounds on the value of the diffusion constant and severely limit any observable effects (see~\cite{Dowker:2003hb} and~\cite{Kaloper:2006pj} and Section~\ref{s:massivebounds}).

In the fully relativistic case, Dowker et al.~wrote down the diffusion equation in terms of cosmic time for the special case of an initially spatially homogeneous distribution~\cite{Dowker:2003hb}. The derivation of the cosmic time evolution equation for
the general case of a spatially inhomogeneous distribution will be given here.

\begin{figure}[t]
\begin{center}
\includegraphics[width = 0.7\textwidth]{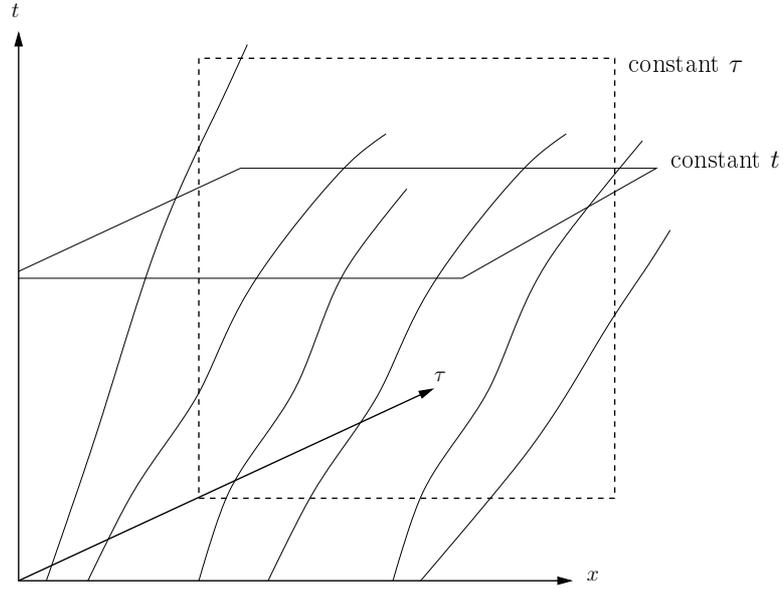}
\caption[Flowlines in
$\mathcal{M}^{\prime} = \mathbb{M}^4\times\mathbb{H}^3\times\mathbb{R}$]{Particle trajectories as flowlines in
$\mathcal{M}^{\prime} = \mathbb{M}^4\times\mathbb{H}^3\times\mathbb{R}$ (where the momentum and two spatial dimensions have been suppressed).}
\label{f:ttau}
\end{center}
\end{figure}

The conversion between proper time and cosmic time is possible because both are good time parameters along
all possible particle worldlines, which are causal. If the diffusion process is visualised as a collection of worldlines through spacetime and momentum space, both cosmic time, $t = x^0$, in our chosen frame and proper time $\tau$
increase monotonically along each trajectory. Assume that the particle starts at parameter $\tau = 0$ and cosmic time $t=0$. Proper time can be added to the state space and the process is represented by flowlines in
$\mathcal{M}^{\prime} = \mathbb{M}^4\times\mathbb{H}^3\times\mathbb{R}$ (see
Figure~\ref{f:ttau}). Along each flowline, both $\tau$ and $t$ are good time parameters. The proper time diffusion
equation, Equation~\ref{e:mtau}, describes the evolution of the distribution on constant $\tau$ hypersurfaces in $\mathcal{M}^{\prime}$.
What is needed is a diffusion equation for evolution of the distribution on constant $t$ hypersurfaces integrated over all proper times.

First, working in the larger space $\mathcal{M}^{\prime}$, define a new current component
%%%%%%%%%%%%%%%%%%%%%%%%%%%%%%%%%%%%%%%%%%%%%%%%%%%%%%%%%%%%%%%%%%%%%%%%%%%%%%%%%%%%%%%%%%%%%%%%%%%%%%%%%%%%%%%%
\begin{equation}
J^{\tau}(t,x^i,p^a,\tau) = \rho_{\tau}\;.
\end{equation}
%%%%%%%%%%%%%%%%%%%%%%%%%%%%%%%%%%%%%%%%%%%%%%%%%%%%%%%%%%%%%%%%%%%%%%%%%%%%%%%%%%%%%%%%%%%%%%%%%%%%%%%%%%%%%%%%
If coordinates on this extended space
$\mathcal{M}^{\prime}= \mathcal{M} \times\mathbb{R}$, are denoted by $X^\alpha = \{X^A, \tau\}$
then the continuity equation (\ref{e:continuity}) can be written
%%%%%%%%%%%%%%%%%%%%%%%%%%%%%%%%%%%%%%%%%%%%%%%%%%%%%%%%%%%%%%%%%%%%%%%%%%%%%%%%%%%%%%%%%%%%%%%%%%%%%%%%%%%%%%%%
\begin{equation}
\partial_\alpha J^\alpha=0\;.
\end{equation}
%%%%%%%%%%%%%%%%%%%%%%%%%%%%%%%%%%%%%%%%%%%%%%%%%%%%%%%%%%%%%%%%%%%%%%%%%%%%%%%%%%%%%%%%%%%%%%%%%%%%%%%%%%%%%%%%
Using Equation (\ref{e:current}) (and still treating $\tau$ as the time parameter) the $t$ component of the current can be expressed in terms of $J^{\tau}$ (equivalently $\rho_{\tau}$).
%%%%%%%%%%%%%%%%%%%%%%%%%%%%%%%%%%%%%%%%%%%%%%%%%%%%%%%%%%%%%%%%%%%%%%%%%%%%%%%%%%%%%%%%%%%%%%%%%%%%%%%%%%%%%%%%
\begin{eqnarray}
J^{t}(t,x^i,p^a,\tau)
&=& -\partial_B\left(K^{tB}J^{\tau}\right)+v^tJ^{\tau}\nonumber\\
&=& v^t J^{\tau}\nonumber\\
&=& \gamma J^{\tau}\,,
\end{eqnarray}
%%%%%%%%%%%%%%%%%%%%%%%%%%%%%%%%%%%%%%%%%%%%%%%%%%%%%%%%%%%%%%%%%%%%%%%%%%%%%%%%%%%%%%%%%%%%%%%%%%%%%%%%%%%%%%%%
where $\gamma = \partial t/ \partial \tau$ is the usual relativistic gamma factor.

The remaining components of the current can now be written in terms of $J^t$. The spatial components are:
%%%%%%%%%%%%%%%%%%%%%%%%%%%%%%%%%%%%%%%%%%%%%%%%%%%%%%%%%%%%%%%%%%%%%%%%%%%%%%%%%%%%%%%%%%%%%%%%%%%%%%%%%%%%%%%%
\begin{eqnarray}
J^{i}(t,x^i,p^a,\tau)
&=&-\partial_B\left(K^{iB}J^{\tau}\right)+v^iJ^{\tau}\nonumber\\
&=& v^i J^{\tau}\nonumber\\
&=& \frac{p^i}{m} \frac{J^t}{\gamma}.
\end{eqnarray}
%%%%%%%%%%%%%%%%%%%%%%%%%%%%%%%%%%%%%%%%%%%%%%%%%%%%%%%%%%%%%%%%%%%%%%%%%%%%%%%%%%%%%%%%%%%%%%%%%%%%%%%%%%%%%%%%
In the case of the $p$ components the algebra is simpler if Equation~\ref{e:current} is expressed in the form
(see Equation~\ref{e:diffequA})
%%%%%%%%%%%%%%%%%%%%%%%%%%%%%%%%%%%%%%%%%%%%%%%%%%%%%%%%%%%%%%%%%%%%%%%%%%%%%%%%%%%%%%%%%%%%%%%%%%%%%%%%%%%%%%%%
\begin{equation}
J^A = -K^{AB}n\,\partial_B\!\left(\frac{\rho}{n}\right) +\rho u^A,
\label{e:altform}
\end{equation}
%%%%%%%%%%%%%%%%%%%%%%%%%%%%%%%%%%%%%%%%%%%%%%%%%%%%%%%%%%%%%%%%%%%%%%%%%%%%%%%%%%%%%%%%%%%%%%%%%%%%%%%%%%%%%%%%
and so
%%%%%%%%%%%%%%%%%%%%%%%%%%%%%%%%%%%%%%%%%%%%%%%%%%%%%%%%%%%%%%%%%%%%%%%%%%%%%%%%%%%%%%%%%%%%%%%%%%%%%%%%%%%%%%%%
\begin{eqnarray}
J^{a}(t,x^i,p^a,\tau)
&=& -kg^{ab}\, n \,\partial_b\!\left(\frac{J^t}{\gamma\,n}\right)\nonumber\\
&=& -kg^{ab}\sqrt{g}\,\partial_b\!\left(\frac{J^t}{\gamma
\sqrt{g}}\right)\;.
\end{eqnarray}
%%%%%%%%%%%%%%%%%%%%%%%%%%%%%%%%%%%%%%%%%%%%%%%%%%%%%%%%%%%%%%%%%%%%%%%%%%%%%%%%%%%%%%%%%%%%%%%%%%%%%%%%%%%%%%%%
The metric $g^{ab}$ that appears here is the Lobachevski metric on $\mathbb{H}^3$.

Since $\tau$ is unobservable we need to integrate $J$ over $\tau$: the current $J$ describes the flow through a region $\Delta t\Delta x\Delta p$ at any point $(t,x^i,p^a,\tau)$; what is needed is the cumulative flow through $\Delta t\Delta x\Delta p$ at any given $(t,x^i,p^a)$ over all proper times. The integrated current will be denoted $\bar{J}$.
Integrating the $t$ component of the current over proper time from zero to infinity gives the probability density on a hypersurface of constant t:
%%%%%%%%%%%%%%%%%%%%%%%%%%%%%%%%%%%%%%%%%%%%%%%%%%%%%%%%%%%%%%%%%%%%%%%%%%%%%%%%%%%%%%%%%%%%%%%%%%%%%%%%%%%%%%%%
\begin{eqnarray}
\rho_{t}&=&\bar{J}^t(x^i,p^a,t)\nonumber\\
 &\equiv&\int{J^t d\tau}.
\end{eqnarray}
%%%%%%%%%%%%%%%%%%%%%%%%%%%%%%%%%%%%%%%%%%%%%%%%%%%%%%%%%%%%%%%%%%%%%%%%%%%%%%%%%%%%%%%%%%%%%%%%%%%%%%%%%%%%%%%%
The components of the new current can be written:
%%%%%%%%%%%%%%%%%%%%%%%%%%%%%%%%%%%%%%%%%%%%%%%%%%%%%%%%%%%%%%%%%%%%%%%%%%%%%%%%%%%%%%%%%%%%%%%%%%%%%%%%%%%%%%%%
\begin{eqnarray}
\bar{J}^{i}(x^i,p^a,t)&\equiv& \int{J^{i} d\tau}\nonumber\\
&=& \int{\frac{p^i J^t}{m \gamma} d\tau}\nonumber\\
&=& \frac{p^i}{m} \frac{\bar{J}^t}{\gamma}\nonumber\\
&=& \frac{p^i}{m} \frac{\rho_t}{\gamma}\,,\label{e:barJi}\\
\bar{J}^{a}(x^i,p^a,t)&\equiv&\int{J^{a} d\tau}\nonumber\\
&=&-kg^{ab}n\partial_b\left(\frac{\bar{J}^t}{\gamma n}\right)\nonumber\\
&=& -kg^{ab}n\partial_b\left(\frac{\rho_t}{\gamma n}\right)\label{e:barJa}
\;.
\end{eqnarray}
%%%%%%%%%%%%%%%%%%%%%%%%%%%%%%%%%%%%%%%%%%%%%%%%%%%%%%%%%%%%%%%%%%%%%%%%%%%%%%%%%%%%%%%%%%%%%%%%%%%%%%%%%%%%%%%%
Integrating the continuity equation over $\tau$ gives
%%%%%%%%%%%%%%%%%%%%%%%%%%%%%%%%%%%%%%%%%%%%%%%%%%%%%%%%%%%%%%%%%%%%%%%%%%%%%%%%%%%%%%%%%%%%%%%%%%%%%%%%%%%%%%%%
\begin{equation}
\left[J^\tau\right]_0^\infty + \partial_t \bar{J}^t + \partial_i \bar{J}^i +
\partial_a \bar{J}^a = 0\;.
\end{equation}
%%%%%%%%%%%%%%%%%%%%%%%%%%%%%%%%%%%%%%%%%%%%%%%%%%%%%%%%%%%%%%%%%%%%%%%%%%%%%%%%%%%%%%%%%%%%%%%%%%%%%%%%%%%%%%%%
$J^\tau|_{\tau =0}$ is zero for all $t>0$ and $J^\tau$ tends to zero as $\tau$ goes to infinity for finite $t$. So for all $t>0$
%%%%%%%%%%%%%%%%%%%%%%%%%%%%%%%%%%%%%%%%%%%%%%%%%%%%%%%%%%%%%%%%%%%%%%%%%%%%%%%%%%%%%%%%%%%%%%%%%%%%%%%%%%%%%%%%
\begin{equation}
\partial_t \bar{J}^t + \partial_i \bar{J}^i +
\partial_a \bar{J}^a = 0\,.\label{e:barcont}
\end{equation}
%%%%%%%%%%%%%%%%%%%%%%%%%%%%%%%%%%%%%%%%%%%%%%%%%%%%%%%%%%%%%%%%%%%%%%%%%%%%%%%%%%%%%%%%%%%%%%%%%%%%%%%%%%%%%%%%
Finally, substituting Equations~\ref{e:barJi} and~\ref{e:barJa} into Equation~\ref{e:barcont} and recalling $\bar{J}^t=\rho_t$, gives the cosmic-time diffusion equation
%%%%%%%%%%%%%%%%%%%%%%%%%%%%%%%%%%%%%%%%%%%%%%%%%%%%%%%%%%%%%%%%%%%%%%%%%%%%%%%%%%%%%%%%%%%%%%%%%%%%%%%%%%%%%%%%
\begin{equation}
\label{e:swervescosmic}
  \pd{\rho_{t}}{t} =
   -\frac{p^i}{m \gamma}\partial_i\rho_{t} +
   k\; \partial_a
   \left(g^{ab}\sqrt{g}\partial_b\left(\frac{\rho_{t}}{\gamma \sqrt{g}}\right)\right).
\end{equation}
%%%%%%%%%%%%%%%%%%%%%%%%%%%%%%%%%%%%%%%%%%%%%%%%%%%%%%%%%%%%%%%%%%%%%%%%%%%%%%%%%%%%%%%%%%%%%%%%%%%%%%%%%%%%%%%%
This is a powerful phenomenological model because it depends on only
one parameter, the diffusion constant $k$. Data can therefore strongly
constrain $k$.

%% file: FiniteDiffConstant.tex
\section{Finiteness of the diffusion constant}
\label{s:finitediffconstant}
The swerves diffusion equation derived in Section~\ref{s:massivepropertime} is independent of the details of the underlying particle model. To make clearer how the underlying model on the causal set gives rise to a diffusion equation on $\mink\times\Lob$ the swerves model (Model 1) will be investigated more fully. Specifically, in this section it will be shown that Model 1 gives rise to a finite nonzero diffusion constant in the macroscopic limit. 

Recall that a process undergoing stochastic evolution on a manifold of states can be described by
%%%%%%%%%%%%%%%%%%%%%%%%%%%%%%%%%%%%%%%%%%%%%%%%%%%%%%%%%%%%%%%%%%%%%%%%%%%%%%%%%%%%%%%%%%%%%%%%%%%%%%%%%%%%%%%%
\begin{eqnarray}
J^{A}&=&-\partial_B\left(K^{AB}\rho\right) + v^A\rho\,,\\
\frac{\partial\rho}{\partial T}&=&-\partial_A J^{A}\,,
\end{eqnarray}
%%%%%%%%%%%%%%%%%%%%%%%%%%%%%%%%%%%%%%%%%%%%%%%%%%%%%%%%%%%%%%%%%%%%%%%%%%%%%%%%%%%%%%%%%%%%%%%%%%%%%%%%%%%%%%%%
where $K^{AB}$ is a symmetric, positive semi-definite matrix given by
%%%%%%%%%%%%%%%%%%%%%%%%%%%%%%%%%%%%%%%%%%%%%%%%%%%%%%%%%%%%%%%%%%%%%%%%%%%%%%%%%%%%%%%%%%%%%%%%%%%%%%%%%%%%%%%%
\begin{equation}
\label{e:kablim}
K^{AB} = \lim_{\Delta T\rightarrow0+}\left\langle \frac{\Delta X^{A}\Delta X^{B}}{2\Delta T}\right\rangle.
\end{equation}
%%%%%%%%%%%%%%%%%%%%%%%%%%%%%%%%%%%%%%%%%%%%%%%%%%%%%%%%%%%%%%%%%%%%%%%%%%%%%%%%%%%%%%%%%%%%%%%%%%%%%%%%%%%%%%%%
In Section~\ref{s:massivepropertime} it was shown that due to the constraint of Lorentz invariance $K^{AB}$ has the form: 
%%%%%%%%%%%%%%%%%%%%%%%%%%%%%%%%%%%%%%%%%%%%%%%%%%%%%%%%%%%%%%%%%%%%%%%%%%%%%%%%%%%%%%%%%%%%%%%%%%%%%%%%%%%%%%%%
\begin{equation}\label{e:kab}
K^{AB}= \left(
\begin{array}{cc}
0 & 0\\ 0 & kg^{ab}
\end{array}
\right),
\end{equation}
%%%%%%%%%%%%%%%%%%%%%%%%%%%%%%%%%%%%%%%%%%%%%%%%%%%%%%%%%%%%%%%%%%%%%%%%%%%%%%%%%%%%%%%%%%%%%%%%%%%%%%%%%%%%%%%%
where $k$ is some constant, irrespective of the underlying model. It is not immediately clear, however, that the constant of proportionality, $k$, need be finite and nonzero for any particular model. For the specific case of the swerves model, the component $K^{ab}$ is indeed nonzero and finite (and thus $K^{AB}$ is given by Equation~\ref{e:kab}). Demonstrating this is the task of this section.

In the swerves model there are two length scales: the discreteness scale $d_{pl}$, and the forgetting time $\tau_{f}$, where $\tau_f >> d_{pl}$. The trajectory length (or total proper time of the particle) must also be much greater than $\tau_f$. It is only in the continuum limit $d_{pl}\rightarrow 0$ that the model can be described by a diffusion equation.

Suppose, for the swerves model, that $K^{ab} = k g^{ab}$. From Equation~\ref{e:kablim} we also have
%%%%%%%%%%%%%%%%%%%%%%%%%%%%%%%%%%%%%%%%%%%%%%%%%%%%%%%%%%%%%%%%%%%%%%%%%%%%%%%%%%%%%%%%%%%%%%%%%%%%%%%%%%%%%%%%
\begin{equation}
K^{ab} = \lim_{\Delta \tau\rightarrow0+}\left\langle \frac{\Delta p^{a}\Delta p^{b}}{2\Delta \tau}\right\rangle.
\end{equation}
%%%%%%%%%%%%%%%%%%%%%%%%%%%%%%%%%%%%%%%%%%%%%%%%%%%%%%%%%%%%%%%%%%%%%%%%%%%%%%%%%%%%%%%%%%%%%%%%%%%%%%%%%%%%%%%%
Here $\Delta p^{a}$ is the change in the momentum component $p^a$ in a single swerves step and $\Delta \tau$ is the change in proper time for that step. Note that in the swerves model $\Delta\tau$ is not constant for each step: it is a random variable in the range $0<\Delta\tau<\tau_{f}$. We wish to determine $K^{ab}$ as $d_{pl}\rightarrow 0$. First note the following two lemmas for the swerves model:
%%%%%%%%%%%%%%%%%%%%%%%%%%%%%%%%%%%%%%%%%%%%%%%%%%%%%%%%%%%%%%%%%%%%%%%%%%%%%%%%%%%%%%%%%%%%%%%%%%%%%%%%%%%%%%%%
\begin{lem}\label{lem:1}
If $d_{pl}\rightarrow 0$ and $\tau_{f}$ is fixed then $\left\langle \frac{\Delta p^{a}\Delta p^{b}}{2\Delta \tau}\right\rangle\rightarrow 0$.
\end{lem}
%%%%%%%%%%%%%%%%%%%%%%%%%%%%%%%%%%%%%%%%%%%%%%%%%%%%%%%%%%%%%%%%%%%%%%%%%%%%%%%%%%%%%%%%%%%%%%%%%%%%%%%%%%%%%%%%
\begin{lem}\label{lem:2}
If $d_{pl}$ is fixed and $\tau_{f}\rightarrow 0$ then $\left\langle \frac{\Delta p^{a}\Delta p^{b}}{2\Delta \tau}\right\rangle\rightarrow\infty$.
\end{lem}
%%%%%%%%%%%%%%%%%%%%%%%%%%%%%%%%%%%%%%%%%%%%%%%%%%%%%%%%%%%%%%%%%%%%%%%%%%%%%%%%%%%%%%%%%%%%%%%%%%%%%%%%%%%%%%%%
Section~\ref{s:lemmaproof} below contains the proofs of these lemmas.
From these two lemmas it is clear that as we take the continuum limit $d_{pl}\rightarrow0$ we can also take $\tau_{f}\rightarrow0$ (and thus $\Delta\tau\rightarrow 0$) in such a way that $\left\langle \frac{\Delta p^{a}\Delta p^{b}}{2\Delta \tau}\right\rangle$ remains fixed at some finite nonzero value. Thus, if $K^{ab}=kg^{ab}$ the diffusion constant $k$ is finite and nonzero, i.e.~the swerves model results in the diffusion equation. 

\subsection{Proofs of Lemmas~\ref{lem:1} and~\ref{lem:2}}
\label{s:lemmaproof}
\subsubsection{Hyperbolic coordinates}
\begin{figure}[th]
\begin{center}
\includegraphics[width=0.8\textwidth]{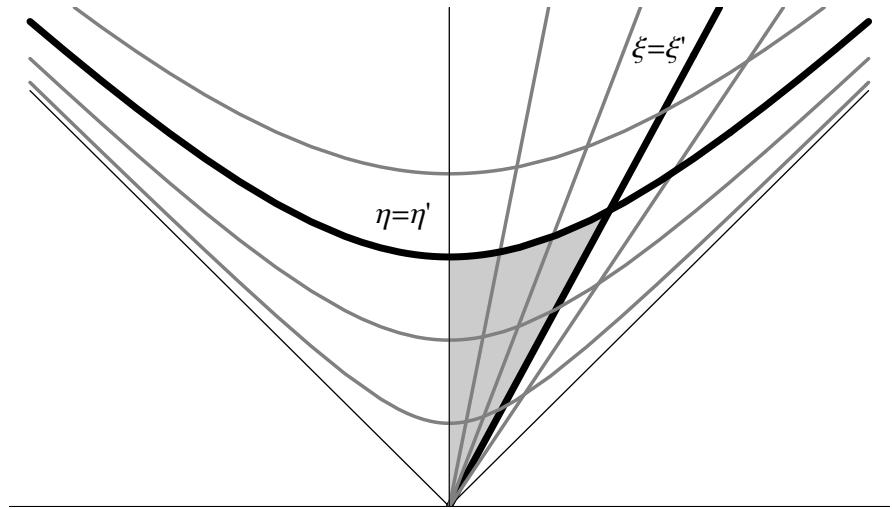}
\caption[Hyperbolic coordinates]{An illustration of hyperbolic coordinates $\eta$ and $\xi$, with the angular dimensions suppressed.}\label{f:hypcoord}
\end{center}
\end{figure}

For ease of calculation, hyperbolic coordinates $\left(\eta,\xi,\theta,\phi\right)$ in four dimensional Minkowski spacetime can be defined:
%%%%%%%%%%%%%%%%%%%%%%%%%%%%%%%%%%%%%%%%%%%%%%%%%%%%%%%%%%%%%%%%%%%%%%%%%%%%%%%%%%%%%%%%%%%%%%%%%%%%%%%%%%%%%%%%
\begin{eqnarray}
\eta^2 &=& t^2-r^2\,,\\
\xi &=& \tanh^{-1}\left(\frac{r}{t}\right)\,,\\
\theta &=& \arccos\left(\frac{z}{r}\right)\,,\\
\phi &=& \arctan\left(\frac{y}{x}\right)\,,
\end{eqnarray}
%%%%%%%%%%%%%%%%%%%%%%%%%%%%%%%%%%%%%%%%%%%%%%%%%%%%%%%%%%%%%%%%%%%%%%%%%%%%%%%%%%%%%%%%%%%%%%%%%%%%%%%%%%%%%%%%
where $r = \sqrt{x^2+y^2+z^2}$. These coordinates are illustrated in Figure~\ref{f:hypcoord}: $\eta$ defines hyperbolae and $\xi$ defines radial directions, varying from $\xi = 0$ on the $t$-axis to $\xi=\infty$ at $t=r$.

\subsubsection{Calculating volumes in hyperbolic coordinates}
Before addressing the Lemmas~\ref{lem:1} and~\ref{lem:2} it is useful to calculate the volume of the shaded region shown in Figure~\ref{f:hypcoord}: the volume of 4-dimensional Minkowski spacetime such that $0<\eta<\eta^{\prime}$ and $0<\xi<\xi^{\prime}$. The first step is to determine the volume element in coordinates $\left(\eta,\xi,\theta,\phi\right)$. It is easiest to begin with spherical coordinates $\left(t,r,\theta,\phi\right)$ defined by
%%%%%%%%%%%%%%%%%%%%%%%%%%%%%%%%%%%%%%%%%%%%%%%%%%%%%%%%%%%%%%%%%%%%%%%%%%%%%%%%%%%%%%%%%%%%%%%%%%%%%%%%%%%%%%%%
\begin{eqnarray}
t&=&t\,,\\
x&=&r\sin\theta\cos\phi\,,\\
y&=&r\sin\theta\sin\phi\,,\\
z&=&r\cos\theta.
\end{eqnarray}
%%%%%%%%%%%%%%%%%%%%%%%%%%%%%%%%%%%%%%%%%%%%%%%%%%%%%%%%%%%%%%%%%%%%%%%%%%%%%%%%%%%%%%%%%%%%%%%%%%%%%%%%%%%%%%%%
The spherical coordinates can then be expressed in terms of hyperbolic coordinates
\begin{eqnarray}
t&=&\eta\cosh\xi\,,\\
r&=&\eta\sinh\xi\,,\\
\theta&=&\theta\,,\\
\phi&=&\phi.
\end{eqnarray}
%%%%%%%%%%%%%%%%%%%%%%%%%%%%%%%%%%%%%%%%%%%%%%%%%%%%%%%%%%%%%%%%%%%%%%%%%%%%%%%%%%%%%%%%%%%%%%%%%%%%%%%%%%%%%%%%
The Jacobian determinant for this change of variables is
%%%%%%%%%%%%%%%%%%%%%%%%%%%%%%%%%%%%%%%%%%%%%%%%%%%%%%%%%%%%%%%%%%%%%%%%%%%%%%%%%%%%%%%%%%%%%%%%%%%%%%%%%%%%%%%%
\begin{eqnarray}
\det J &=& \left|\begin{array}{cccc}
\frac{\partial t}{\partial\eta} & \frac{\partial t}{\partial\xi} & \frac{\partial t}{\partial\theta} & \frac{\partial t}{\partial\phi}\\
\frac{\partial r}{\partial\eta} & \frac{\partial r}{\partial\xi} & \frac{\partial r}{\partial\theta} & \frac{\partial r}{\partial\phi}\\
\frac{\partial\theta}{\partial\eta} & \frac{\partial\theta}{\partial\xi} & \frac{\partial\theta}{\partial\theta} & \frac{\partial\theta}{\partial\phi}\\
\frac{\partial\phi}{\partial\eta} & \frac{\partial\phi}{\partial\xi} & \frac{\partial\phi}{\partial\theta} & \frac{\partial\phi}{\partial\phi}
\end{array}\right|\nonumber\\
&=&\left|\begin{array}{cccc}
\cosh\xi & \eta\sinh\xi & 0 & 0\\
\sinh\xi & \eta\cosh\xi & 0 & 0\\
0 & 0 & 1 & 0\\
0 & 0 & 0 & 1
\end{array}\right|\nonumber\\
&=&\eta\cosh^2\xi-\eta\sinh^2\xi\nonumber\\
&=&\eta\,.
\end{eqnarray}
%%%%%%%%%%%%%%%%%%%%%%%%%%%%%%%%%%%%%%%%%%%%%%%%%%%%%%%%%%%%%%%%%%%%%%%%%%%%%%%%%%%%%%%%%%%%%%%%%%%%%%%%%%%%%%%%
The volume element is thus given by
%%%%%%%%%%%%%%%%%%%%%%%%%%%%%%%%%%%%%%%%%%%%%%%%%%%%%%%%%%%%%%%%%%%%%%%%%%%%%%%%%%%%%%%%%%%%%%%%%%%%%%%%%%%%%%%%
\begin{eqnarray}
dt\,dx\,dy\,dz &=& r^2\sin\theta\,dt\,dr\,d\theta\,d\phi\nonumber\\
&=& \eta^3\sinh^2\xi\sin\theta\,d\eta\,d\xi\,d\theta\,d\phi\,.
\end{eqnarray}
%%%%%%%%%%%%%%%%%%%%%%%%%%%%%%%%%%%%%%%%%%%%%%%%%%%%%%%%%%%%%%%%%%%%%%%%%%%%%%%%%%%%%%%%%%%%%%%%%%%%%%%%%%%%%%%%
The volume of the shaded region is
%%%%%%%%%%%%%%%%%%%%%%%%%%%%%%%%%%%%%%%%%%%%%%%%%%%%%%%%%%%%%%%%%%%%%%%%%%%%%%%%%%%%%%%%%%%%%%%%%%%%%%%%%%%%%%%%
\begin{eqnarray}
V &=& \int^{2\pi}_{0}\int^{\pi}_{0}\int^{\xi^{\prime}}_{0}\int^{\eta^{\prime}}_{0}\eta^3\sinh^2\xi\sin\theta \,d\eta\,d\xi\,d\theta\,d\phi\nonumber\\
&=& 4\pi\int^{\xi^{\prime}}_{0}\int^{\eta^{\prime}}_{0}\eta^3\sinh^2\xi\,d\eta\,d\xi\nonumber\\
&=&\pi{\eta^{\prime}}^4\int^{\xi^{\prime}}_{0}\sinh^2\xi\,d\xi\nonumber\\
\label{e:volint}
&=& \pi{\eta^{\prime}}^4\left(-\frac{\xi^{\prime}}{2} + \frac{\sinh 2\xi^{\prime}}{4}\right)\,.
\end{eqnarray}
%%%%%%%%%%%%%%%%%%%%%%%%%%%%%%%%%%%%%%%%%%%%%%%%%%%%%%%%%%%%%%%%%%%%%%%%%%%%%%%%%%%%%%%%%%%%%%%%%%%%%%%%%%%%%%%%
This result will be used repeatedly in the calculations below.
To prove Lemmas~\ref{lem:1} and~\ref{lem:2} the following property of sprinklings is required: if points are sprinkled with a density $\rho$ into a region, the probability that $n$ points are sprinkled into a volume $V$ is given by the Poisson distribution
%%%%%%%%%%%%%%%%%%%%%%%%%%%%%%%%%%%%%%%%%%%%%%%%%%%%%%%%%%%%%%%%%%%%%%%%%%%%%%%%%%%%%%%%%%%%%%%%%%%%%%%%%%%%%%%%
\begin{equation}
\label{e:poisson}
\text{Pr}_{\rho} = \frac{\left(\rho V\right)^n e^{-\rho V}}{n!}\,.
\end{equation}
%%%%%%%%%%%%%%%%%%%%%%%%%%%%%%%%%%%%%%%%%%%%%%%%%%%%%%%%%%%%%%%%%%%%%%%%%%%%%%%%%%%%%%%%%%%%%%%%%%%%%%%%%%%%%%%%
For example, the probability that \emph{no} points are sprinkled into the shaded region in Figure~\ref{f:hypcoord} is simply
%%%%%%%%%%%%%%%%%%%%%%%%%%%%%%%%%%%%%%%%%%%%%%%%%%%%%%%%%%%%%%%%%%%%%%%%%%%%%%%%%%%%%%%%%%%%%%%%%%%%%%%%%%%%%%%%
\begin{eqnarray}
\text{Pr}_{\rho}&=&e^{-\rho V}\nonumber\\
&=& \exp{\left(-\rho \pi{\eta^{\prime}}^4
\left(-\frac{\xi^{\prime}}{2} + \frac{\sinh 2\xi^{\prime}}{4}\right)\right)}\,.
\label{e:volempty}
\end{eqnarray}
%%%%%%%%%%%%%%%%%%%%%%%%%%%%%%%%%%%%%%%%%%%%%%%%%%%%%%%%%%%%%%%%%%%%%%%%%%%%%%%%%%%%%%%%%%%%%%%%%%%%%%%%%%%%%%%%

To prove Lemmas~\ref{lem:1} and~\ref{lem:2} it is necessary to calculate the expectation value $\left\langle \frac{\Delta p^{a}\Delta p^{b}}{2\Delta \tau}\right\rangle$. In fact, since it is known that $K^{ab}\propto g^{ab}$ it suffices to calculate a single component, $K^{pp}$.

Working in terms of polar coordinates $\{p,\theta,\phi\}$ on $\Lob$, the metric $g^{ab}$ is
\begin{equation}
ds^2 = \frac{m^2}{m^2+p^2}dp^2 + p^2(d\theta^2+\sin^2\theta\,d\phi^2)\,.
\end{equation}
Consider a swerves trajectory in the rest frame of the particle after $n$ steps (see Figure~\ref{f:mmodel}). In this frame the metric component $g^{pp}=\frac{m^2+p^2}{m^2}=1$ and thus $K^{pp}=k$.  
Since the three momentum $p_n$ is zero, the change in momentum in the next step is simply $p_{n+1}$. Working in Cartesian coordinates for a moment, the momentum at step $e_{n+1}$ is proportional to the coordinates $X^{\mu}$ of the point $e_{n+1}$ in this frame: $p_{n+1}^{\mu} = \frac{m}{\eta}X^{\mu}$, where $\eta$ is the proper time between $e_n$ and $e_{n+1}$. 
Thus the magnitude of the three momentum is given by
%%%%%%%%%%%%%%%%%%%%%%%%%%%%%%%%%%%%%%%%%%%%%%%%%%%%%%%%%%%%%%%%%%%%%%%%%%%%%%%%%%%%%%%%%%%%%%%%%%%%%%%%%%%%%%%%
\begin{eqnarray}
p_{n+1} &=& \sqrt{p_1^2+p_2^2+p_3^2}\nonumber\\
&=& \frac{m}{\eta}\sqrt{X_1^2+X_2^2+X_3^2}\nonumber\\
&=&\frac{m}{\eta}r\nonumber\\
&=&m\sinh\xi.
\end{eqnarray}
%%%%%%%%%%%%%%%%%%%%%%%%%%%%%%%%%%%%%%%%%%%%%%%%%%%%%%%%%%%%%%%%%%%%%%%%%%%%%%%%%%%%%%%%%%%%%%%%%%%%%%%%%%%%%%%%
The expectation value 
%%%%%%%%%%%%%%%%%%%%%%%%%%%%%%%%%%%%%%%%%%%%%%%%%%%%%%%%%%%%%%%%%%%%%%%%%%%%%%%%%%%%%%%%%%%%%%%%%%%%%%%%%%%%%%%%
\begin{equation}
\left\langle \frac{\Delta p\Delta p}{2\Delta \tau}\right\rangle = \int_0^{\tau_f}\int_0^{\infty} \frac{\Delta p\Delta p}{2\eta}\mathbb{P}\left(\frac{\Delta p\Delta p}{2\eta}\right)d\eta\,d\xi,
\end{equation}
%%%%%%%%%%%%%%%%%%%%%%%%%%%%%%%%%%%%%%%%%%%%%%%%%%%%%%%%%%%%%%%%%%%%%%%%%%%%%%%%%%%%%%%%%%%%%%%%%%%%%%%%%%%%%%%%
where $P = \mathbb{P}\left(\frac{\Delta p\Delta p}{2\eta}\right)$ is the probability of the particular value occurring. The probability $P$ is the probability that the element $e_{n+1}$ has (hyperbolic) coordinates\\ \mbox{$\eta^{\prime}<\eta<\eta^{\prime}+d\eta$}, $\,\xi^{\prime}<\xi<\xi^{\prime}+d\xi$. Recall that in the swerves model $e_{n+1}$ is chosen such that $\eta<\tau_f$ and the momentum change is minimised. $P$ is thus the probability that there are no points in the region $\{\eta<\tau_f,\;\xi<\xi^{\prime}\}$ \textit{and} there is a point in the region $\{\eta^{\prime}<\eta<\eta^{\prime}+d\eta,\;\xi^{\prime}<\xi<\xi^{\prime}+d\xi\}$.
The probability that there are no points in the region $\{\eta<\tau_f,\;\xi<\xi^{\prime}\}$ is simply
%%%%%%%%%%%%%%%%%%%%%%%%%%%%%%%%%%%%%%%%%%%%%%%%%%%%%%%%%%%%%%%%%%%%%%%%%%%%%%%%%%%%%%%%%%%%%%%%%%%%%%%%%%%%%%%%
\begin{equation}
\exp{\left(-\rho \pi\tau_f^4\left(-\frac{\xi^{\prime}}{2} + \frac{\sinh 2\xi^{\prime}}{4}\right)\right)}\,,
\end{equation}
%%%%%%%%%%%%%%%%%%%%%%%%%%%%%%%%%%%%%%%%%%%%%%%%%%%%%%%%%%%%%%%%%%%%%%%%%%%%%%%%%%%%%%%%%%%%%%%%%%%%%%%%%%%%%%%%
from Equation~\ref{e:volempty} above. The probability that there is a point in the region $\{\eta^{\prime}<\eta<\eta^{\prime}+d\eta,\;\xi^{\prime}<\xi<\xi^{\prime}+d\xi\}$ is 
%%%%%%%%%%%%%%%%%%%%%%%%%%%%%%%%%%%%%%%%%%%%%%%%%%%%%%%%%%%%%%%%%%%%%%%%%%%%%%%%%%%%%%%%%%%%%%%%%%%%%%%%%%%%%%%%
\begin{equation}
4\pi\rho{\eta^{\prime}}^3\sinh^2\xi^{\prime} d\eta\,d\xi\,,
\end{equation}
%%%%%%%%%%%%%%%%%%%%%%%%%%%%%%%%%%%%%%%%%%%%%%%%%%%%%%%%%%%%%%%%%%%%%%%%%%%%%%%%%%%%%%%%%%%%%%%%%%%%%%%%%%%%%%%%
from the integral in Equation~\ref{e:volint} and Equation~\ref{e:poisson} (the exponential term can be neglected as the volume is very small).
The expectation value is therefore (dropping the primes on $\xi$ and $\eta$)
%%%%%%%%%%%%%%%%%%%%%%%%%%%%%%%%%%%%%%%%%%%%%%%%%%%%%%%%%%%%%%%%%%%%%%%%%%%%%%%%%%%%%%%%%%%%%%%%%%%%%%%%%%%%%%%%
\begin{eqnarray}
\left\langle \frac{\Delta p\Delta p}{2\Delta \tau}\right\rangle &=& \int_0^{\tau_f}\int_0^{\infty}2\pi\rho m^2\eta^2 \sinh^4\xi
\exp\left(-\rho \pi\tau_f^4\left(-\frac{\xi}{2} + \frac{\sinh 2\xi}{4}\right)\right)d\eta\, d\xi\nonumber\\
&=&\frac{2}{3}\pi\rho m^2 \tau_f^3\int_0^{\infty}\sinh^4\xi\exp\left(-\rho \pi\tau_f^4\left(-\frac{\xi}{2} + \frac{\sinh 2\xi}{4}\right)\right)d\xi\,.
\end{eqnarray}
%%%%%%%%%%%%%%%%%%%%%%%%%%%%%%%%%%%%%%%%%%%%%%%%%%%%%%%%%%%%%%%%%%%%%%%%%%%%%%%%%%%%%%%%%%%%%%%%%%%%%%%%%%%%%%%%
Integrating by parts, this reduces to
%%%%%%%%%%%%%%%%%%%%%%%%%%%%%%%%%%%%%%%%%%%%%%%%%%%%%%%%%%%%%%%%%%%%%%%%%%%%%%%%%%%%%%%%%%%%%%%%%%%%%%%%%%%%%%%%
\begin{equation}
\left\langle \frac{\Delta p\Delta p}{2\Delta \tau}\right\rangle =
\frac{4m^2}{3\tau_f}\int_0^{\infty}\cosh\xi\sinh\xi\exp\left(-\rho \pi\tau_f^4\left(-\frac{\xi}{2} + \frac{\sinh 2\xi}{4}\right)\right)d\xi\,.
\label{e:expvalue}
\end{equation}
%%%%%%%%%%%%%%%%%%%%%%%%%%%%%%%%%%%%%%%%%%%%%%%%%%%%%%%%%%%%%%%%%%%%%%%%%%%%%%%%%%%%%%%%%%%%%%%%%%%%%%%%%%%%%%%%
The proofs of the Lemmas now follow without difficulty.

\subsubsection{Proof of Lemma~\ref{lem:1}}

\textit{If we take the discreteness scale $d_{pl}$ to zero, while keeping the forgetting time, $\tau_{f}$, fixed, then $\left\langle \frac{\Delta p^{a}\Delta p^{b}}{2\Delta \tau}\right\rangle\rightarrow0$.} 
\vspace{5mm}

Taking $d_{pl}\rightarrow0$ is equivalent to taking the sprinkling density $\rho$ to infinity. To determine the effect of $\rho\rightarrow\infty$ on Equation~\ref{e:expvalue}, the integral can be divided into two parts. There exists some $\bar{\xi}$ such that, for all $\xi>\bar{\xi}$, $\cosh\xi<2\sinh\xi$. 
Let 
%%%%%%%%%%%%%%%%%%%%%%%%%%%%%%%%%%%%%%%%%%%%%%%%%%%%%%%%%%%%%%%%%%%%%%%%%%%%%%%%%%%%%%%%%%%%%%%%%%%%%%%%%%%%%%%%
\begin{equation}
\left\langle \frac{\Delta p\Delta p}{2\Delta \tau}\right\rangle = I_1+I_2\,,
\end{equation}
%%%%%%%%%%%%%%%%%%%%%%%%%%%%%%%%%%%%%%%%%%%%%%%%%%%%%%%%%%%%%%%%%%%%%%%%%%%%%%%%%%%%%%%%%%%%%%%%%%%%%%%%%%%%%%%%
where
%%%%%%%%%%%%%%%%%%%%%%%%%%%%%%%%%%%%%%%%%%%%%%%%%%%%%%%%%%%%%%%%%%%%%%%%%%%%%%%%%%%%%%%%%%%%%%%%%%%%%%%%%%%%%%%%
\begin{eqnarray}
I_1 &=& \frac{4m^2}{3\tau_f}\int_{\bar{\xi}}^{\infty}\cosh\xi\sinh\xi\exp\left(-\rho \pi\tau_f^4\left(-\frac{\xi}{2} + \frac{\sinh 2\xi}{4}\right)\right)d\xi\nonumber\,,\\
I_2&=&\frac{4m^2}{3\tau_f}
\int_{0}^{\bar{\xi}}\cosh\xi\sinh\xi\exp\left(-\rho \pi\tau_f^4\left(-\frac{\xi}{2} + \frac{\sinh 2\xi}{4}\right)\right)d\xi\,.
\end{eqnarray}
%%%%%%%%%%%%%%%%%%%%%%%%%%%%%%%%%%%%%%%%%%%%%%%%%%%%%%%%%%%%%%%%%%%%%%%%%%%%%%%%%%%%%%%%%%%%%%%%%%%%%%%%%%%%%%%%
$I_1$ can be written
%%%%%%%%%%%%%%%%%%%%%%%%%%%%%%%%%%%%%%%%%%%%%%%%%%%%%%%%%%%%%%%%%%%%%%%%%%%%%%%%%%%%%%%%%%%%%%%%%%%%%%%%%%%%%%%%
\begin{eqnarray}
I_1 &\leq& \frac{4m^2}{3\tau_f}\int_{\bar{\xi}}^{\infty}2\sinh^2\xi\exp\left(-\rho \pi\tau_f^4\left(-\frac{\xi}{2} + \frac{\sinh 2\xi}{4}\right)\right)d\xi\nonumber\\
&\leq& \frac{4m^2}{3\tau_f}\frac{-2}{\rho\pi\tau_f^4}\left[\exp\left(-\rho \pi\tau_f^4\left(-\frac{\xi}{2} + \frac{\sinh 2\xi}{4}\right)\right)\right]^{\infty}_{\bar{\xi}}\nonumber\\
&\leq& \frac{8m^2}{3\rho\pi\tau_f^5}\exp\left(-\rho \pi\tau_f^4\left(-\frac{\bar{\xi}}{2} + \frac{\sinh 2\bar{\xi}}{4}\right)\right)\,.
\end{eqnarray}
Thus
\begin{equation}
\lim_{\rho\rightarrow\infty} I_1 = 0\,.
\end{equation}
%%%%%%%%%%%%%%%%%%%%%%%%%%%%%%%%%%%%%%%%%%%%%%%%%%%%%%%%%%%%%%%%%%%%%%%%%%%%%%%%%%%%%%%%%%%%%%%%%%%%%%%%%%%%%%%%
Now consider the integral $I_2$ and the standard result
%%%%%%%%%%%%%%%%%%%%%%%%%%%%%%%%%%%%%%%%%%%%%%%%%%%%%%%%%%%%%%%%%%%%%%%%%%%%%%%%%%%%%%%%%%%%%%%%%%%%%%%%%%%%%%%%
\begin{equation}
\int^a_b|f(x)|dx \leq \max_{x\in[a,b]}|f(x)|(b-a).
\end{equation}
%%%%%%%%%%%%%%%%%%%%%%%%%%%%%%%%%%%%%%%%%%%%%%%%%%%%%%%%%%%%%%%%%%%%%%%%%%%%%%%%%%%%%%%%%%%%%%%%%%%%%%%%%%%%%%%%
Let $I_2=\int_0^{\bar{\xi}}f(\xi)d\xi$, noting that $f(\xi)>0$. For any $\epsilon>0$ there exists some $\rho$ large enough that $f(\xi)<\epsilon/\bar{\xi}$ for all $0\leq\xi\leq\bar{\xi}$. Thus 
%%%%%%%%%%%%%%%%%%%%%%%%%%%%%%%%%%%%%%%%%%%%%%%%%%%%%%%%%%%%%%%%%%%%%%%%%%%%%%%%%%%%%%%%%%%%%%%%%%%%%%%%%%%%%%%%
\begin{eqnarray}
I_2&=& \int_0^{\bar{\xi}}f(\xi)d\xi\nonumber\\
&\leq& \frac{\epsilon}{\bar{\xi}}\bar{\xi}\nonumber\\
&<&\epsilon.
\end{eqnarray}
%%%%%%%%%%%%%%%%%%%%%%%%%%%%%%%%%%%%%%%%%%%%%%%%%%%%%%%%%%%%%%%%%%%%%%%%%%%%%%%%%%%%%%%%%%%%%%%%%%%%%%%%%%%%%%%%
$I_2$ therefore also tends to zero in the limit $\rho\rightarrow\infty$.
Thus in the limit $\rho\rightarrow\infty$, the expectation value $\left\langle \frac{\Delta p\Delta p}{2\Delta \tau}\right\rangle\rightarrow 0$, proving Lemma~\ref{lem:1}.

\pagebreak
\subsubsection{Proof of Lemma~\ref{lem:2}}

\textit{If we take the forgetting time $\tau_{f}\rightarrow 0$ and keep the discreteness scale $d_{pl}$ fixed then $\left\langle \frac{\Delta p^{a}\Delta p^{b}}{2\Delta \tau}\right\rangle\rightarrow\infty$. }
\vspace{5mm}

First note that for all $\xi$, $\cosh\xi\geq\sinh\xi$. The expectation value, Equation~\ref{e:expvalue}, can therefore be written
%%%%%%%%%%%%%%%%%%%%%%%%%%%%%%%%%%%%%%%%%%%%%%%%%%%%%%%%%%%%%%%%%%%%%%%%%%%%%%%%%%%%%%%%%%%%%%%%%%%%%%%%%%%%%%%%
\begin{eqnarray}
\left\langle \frac{\Delta p\Delta p}{2\Delta \tau}\right\rangle
&\geq& \frac{4m^2}{3\tau_f}\int_{0}^{\infty}\sinh^2\xi\exp\left(-\rho \pi\tau_f^4\left(-\frac{\xi}{2} + \frac{\sinh 2\xi}{4}\right)\right)d\xi\nonumber\\
&\geq& \frac{4m^2}{3\tau_f}\left(\frac{-1}{\rho\pi\tau_f^4}\right)\left[\exp\left(-\rho \pi\tau_f^4\left(-\frac{\xi}{2} + \frac{\sinh 2\xi}{4}\right)\right)\right]^{\infty}_0\nonumber\\
&\geq&\frac{4m^2}{3\rho\pi\tau_f^5}\,.
\end{eqnarray}
%%%%%%%%%%%%%%%%%%%%%%%%%%%%%%%%%%%%%%%%%%%%%%%%%%%%%%%%%%%%%%%%%%%%%%%%%%%%%%%%%%%%%%%%%%%%%%%%%%%%%%%%%%%%%%%%
Thus in the limit $\tau_f\rightarrow0$, the expectation value $\left\langle \frac{\Delta p\Delta p}{2\Delta \tau}\right\rangle\rightarrow \infty$ proving Lemma~\ref{lem:2}.

%% file: BoundsonDiffConstant.tex
\section{Consequences and bounds}
\label{s:massivebounds}
It is important to reiterate that the swerves diffusion equation is very general: any Lorentz invariant process for massive particles that results in fluctuations in energy-momentum will be governed by this equation. Deriving such an equation is of little use, however, unless we investigate the observable consequences. Ideally, a new phenomenological model should agree with all current observational data and, in addition, make predictions for new, feasible, experiments. In reality, models will always contain free parameters that limit their predictive powers. The swerves diffusion equation contains only one free parameter, the diffusion constant $k$, and is thus a strong phenomenological model. The usefulness of this model is unfortunately still limited. It will never be possible to prove that swerves do not occur -- null observations only allow the parameter $k$ to be constrained to smaller values. It is hoped, of course, that experiments will actually observe the swerves effect; or that swerves will provide an explanation for some already observed, unexplained phenomena. The difficulty is (as discussed in the Introduction) that quantum gravitational effects are inherently very small and to make positive observations we must think of ways in which these tiny effects could have been amplified.
For swerves the first step is thus to consider how the magnitude of the diffusion constant can be constrained by current data.
Strong constraints have been placed on $k$ by both Dowker et al.~\cite{Dowker:2003hb} and Kaloper and Mattingly~\cite{Kaloper:2006pj}. These constraints are reviewed in this section. Lacking the inspiration of an astrophysical or cosmological problem swerves may solve, it seems unnecessary to seek tighter bounds at this stage.

For simplicity and consistency with~\cite{Dowker:2003hb,Kaloper:2006pj} I will work in terms of a scalar distribution $\bar{\rho}=\rho_t/\sqrt{g}$ rather than the scalar density $\rho_t$. In terms of the scalar $\bar{\rho}$ the homogeneous cosmic time diffusion equation becomes:
\begin{equation}
\pd{\bar{\rho}}{t} = \frac{k}{\sqrt{g}}\partial_a\left(g^{ab}\sqrt{g}\partial_b\left(\frac{\bar{\rho}}{\gamma}\right)\right)\,.
\label{e:ctscalar}
\end{equation}

The swerves model predicts the spontaneous heating of a nonrelativistic gas. To see this, consider the nonrelativistic limit of Equation~\ref{e:ctscalar}.
Working in polar coordinates on $\mathbb{H}^3$
\begin{equation}
g_{ab} = 
\begin{pmatrix}
\frac{m^2}{m^2+p^2}&0&0\\
0&p^2&0\\
0&0&p^2\sin^2\theta
\end{pmatrix}\,,
\end{equation}
and thus
\begin{eqnarray}
\pd{\bar{\rho}}{t}&=&=\frac{k}{\sqrt{g}}\partial_a\left(g^{ab}\sqrt{g}\partial_b\left(\frac{\bar{\rho}}{\gamma}\right)\right)\nonumber\\
&=& \frac{k\sqrt{m^2+p^2}}{mp^2\sin\theta}\left[\pd{}{p}\left(\frac{p^2\sin\theta\sqrt{m^2+p^2}}{m}\pd{}{p}\left(\frac{m\bar{\rho}}{\sqrt{m^2+p^2}}\right)\right) \right.\nonumber\\
&&\left.+ \pd{}{\theta}\left(\frac{m\sin\theta}{\sqrt{m^2+p^2}}\pd{}{\theta}\left(\frac{m\bar{\rho}}{\sqrt{m^2+p^2}}\right)\right)\right.\nonumber\\
&&\left. + \pd{}{\phi}\left(\frac{m}{\sin\theta\sqrt{m^2+p^2}}\pd{}{\phi}\left(\frac{m\bar{\rho}}{\sqrt{m^2+p^2}}\right)\right)\right]\,.
\end{eqnarray}
In the nonrelativistic limit $p<<m$ this becomes
\begin{eqnarray}
\pd{\bar{\rho}}{t}&=&\frac{k}{p^2\sin\theta}\left[\pd{}{p}\left(p^2\sin\theta\pd{\bar{\rho}}{p}\right) \right.\nonumber\\
&&\left.+ \pd{}{\theta}\left(\sin\theta\pd{\bar{\rho}}{\theta}\right) + \pd{}{\phi}\left(\frac{1}{\sin\theta}\pd{\bar{\rho}}{\phi}\right)\right]\nonumber\\
&=&k\left[\frac{1}{p^2}\pd{}{p}\left(p^2\pd{\bar{\rho}}{p}\right) 
+ \frac{1}{p^2\sin\theta}\pd{}{\theta}\left(\sin\theta\pd{\bar{\rho}}{\theta}\right)\right.\nonumber\\
&&\left. + \frac{1}{p^2\sin^2\theta}\pdsq{\bar{\rho}}{\phi}\right]\nonumber\\
&=&k\nabla^2\bar{\rho}\,,\label{e:stddiff}
\end{eqnarray}
where $\nabla^2$ is the standard Laplacian on $\mathbb{R}^3$, i.e.~in the nonrelativistic limit the swerves diffusion equation reduces to the standard three dimensional diffusion equation. The general causal solution to Equation~\ref{e:stddiff} can be derived using Greens's function techniques. The result is
\begin{equation}
\bar{\rho}(\textbf{p},t)=\int^{\infty}_{-\infty}\frac{1}{\left(4\pi kt\right)^{3/2}}\exp{\left(-\frac{\left(\textbf{p}-\textbf{p}^\prime\right)^2}{4 kt}\right)f(\textbf{p}^\prime)d^3\textbf{p}^\prime}\,,
\end{equation}
where $f(\textbf{p})$ is the initial distribution $f(\textbf{p})=\bar{\rho}(\textbf{p},0)$.
Consider an initial distribution of particles at rest in our frame, i.e.~$f(\textbf{p})=\delta^3(\textbf{p})$. In this case, the solution of the diffusion equation simplifies to 
\begin{eqnarray}
\bar{\rho}(\textbf{p},t) &=& 
\int^{\infty}_{-\infty}\frac{1}{\left(4\pi kt\right)^{3/2}}\exp{\left(-\frac{\left(\textbf{p}-\textbf{p}^\prime\right)^2}{4 kt}\right)\delta^3(\textbf{p}^\prime)d^3\textbf{p}^\prime}\nonumber\\
&=& \left(4\pi kt\right)^{-3/2}\exp{\left(-\frac{\textbf{p}^2}{4 kt}\right)}\,.
\end{eqnarray} 
This is just the Maxwell-Boltzmann distribution with a temperature $T=\frac{2kt}{mk_B}$, where $k_B$ is the Boltzmann constant. The temperature is dependent on the length of time the diffusion process has been occurring -- in other words, the nonrelativistic limit of the swerves diffusion equation describes a spontaneous heating of a thermal nonrelativistic gas. 
It is also clear that an initially thermal distribution will remain thermal. Suppose the initial distribution is a thermal distribution with a temperature $T_0$:
\begin{eqnarray}
f(\textbf{p}) &=& \left(2\pi m k_B T_0\right)^{-3/2}\exp{\left(-\frac{\textbf{p}^2}{2mk_BT_0}\right)}\,.
\end{eqnarray}
\noindent
The solution is then
\begin{eqnarray}
\bar{\rho}(\textbf{p},t) 
&=& \int^{\infty}_{-\infty}\frac{1}{\left(4\pi kt\right)^{3/2}}\exp{\left(-\frac{\left(\textbf{p}-\textbf{p}^\prime\right)^2}{4 kt}\right)f(\textbf{p}^\prime)d^3\textbf{p}^\prime}\nonumber\\
&=& \int^{\infty}_{-\infty} \left(8\pi^2ktmk_BT_0\right)^{-3/2} \exp{\left(-\frac{\left(\textbf{p}-\textbf{p}^\prime\right)^2}{4kt} - \frac{{\textbf{p}^\prime}^2}{2mk_BT_0}\right)} d^3\textbf{p}^\prime\nonumber\\
&=& \left(8\pi^2ktmk_BT_0\right)^{-3/2} \exp{\left(-\frac{\textbf{p}^2}{2mk_BT}\right)}\nonumber\\
&&\times \int^{\infty}_{-\infty}\exp{\left(\frac{-T}{4ktT_0}\left(\textbf{p}^\prime - \frac{T_0}{T}\textbf{p}\right)^2\right)}d^3\textbf{p}^\prime\nonumber\\
&=&\left(2\pi mk_BT\right)^{-3/2}\exp{\left(\frac{-\textbf{p}^2}{2mk_BT}\right)}\,,
\end{eqnarray}
where $T \equiv T_0 + 2kt/mk_B$, i.e.~a thermal distribution remains thermal, with the temperature changing at a rate of $dT/dt = 2k/mk_B$.

Dowker et al.~\cite{Dowker:2003hb} placed a bound on the diffusion constant $k$ by considering the spontaneous heating (or rather, the \textit{lack} of spontaneous heating) of hydrogen gas in the laboratory. They assumed that a heating rate of $10^{-6}$ degrees a second would have already been detected if it existed. This resulted in the bound
\begin{eqnarray}
k&=& \frac{mk_B}{2}\frac{dT}{dt}\nonumber\\
&\leq& 10^{-56}kg^2m^2s^{-3}\,,
\label{e:kboundHyd}
\end{eqnarray}
or equivalently
\begin{equation}
k \leq 10^{-44}GeV^3\,.
\end{equation}

The average energy of a thermal distribution at temperature $T$ is given by
\begin{eqnarray}
\left\langle E\right\rangle &=& \frac{3k_B T}{2}\,.
\end{eqnarray} 
\noindent
Constraints on $k$ can thus also be expressed in terms of the maximum energy gain allowed
\begin{eqnarray}
k &\leq& \frac{m\Delta\!\left\langle E\right\rangle}{3\Delta t}\,.
\end{eqnarray}
To more tightly constrain $k$ it is therefore necessary to consider systems that have been around for a long time without much gain in energy.
Kaloper and Mattingly~\cite{Kaloper:2006pj} looked at astrophysical clouds that have existed for roughly the age of the Milky Way without significant energy gain. They found, however, that energy loss due to radiation cannot be neglected and the result is a bound on $k$ roughly comparable to that given above. They obtained a much stronger bound of $k<10^{-61}GeV^3$ by considering cosmic neutrinos. When heated by swerves, the cosmic neutrino background would appear as hot dark matter. The diffusion constant is thus bounded by observations that suggest dark matter is mainly cold.

In attempting to bound the diffusion parameter, $k$, it is assumed that $k$ is the same for all particles. Other than for simplicity, there is no reason to believe this is the case. A strong bound on $k$ for neutrinos does not, therefore, rule out observable swerves for other particles. Without an underlying model for particles on a causal set (a model for the particles themselves, not just their propagation) it is unlikely any progress will be made determining the dependencies of $k$.

It is also interesting to note that, although Lorentz invariant, swerves could give rise to an apparent signal of Lorentz violation~\cite{Mattingly:2007be}. This is because many investigations into Lorentz violation rely not only on the details of the Lorentz violating model, but also on the assumption of energy-momentum conservation. A violation of energy-momentum conservation could be incorrectly interpreted as a signal of Lorentz violation. Consider, for example, a time-of-flight experiment where a particle is measured to arrive with energy $E$. If the particle is subject to swerves, the arrival energy, $E$, is not the energy of the particle throughout its propagation, and the propagation time will therefore be miscalculated if swerves are neglected. Mattingly~\cite{Mattingly:2007be} concludes, however, that any false signal of Lorentz violation that appears due to the energy-momentum violation of swerves would be considerably smaller than current limits of experimental sensitivity. 

%% file: MassiveSims.tex
\section{Simulations of particle models}
\label{s:swervesnumeric}

The diffusion equations derived in this chapter do not rely on any particular model for particle propagation on a causal set. The diffusion parameter $k$ is, however, likely to depend on the underlying propagation mechanism. Simulating the particle models discussed in Section~\ref{s:models} shows explicitly that the models lead to diffusion like behaviour. It also allows an initial investigation into how the continuum diffusion parameter might depend on characteristics of the underlying physical process. 

To simulate the particle models causal sets are constructed via sprinkling into Minkowski spacetime (as discussed in Section~\ref{s:sprinkling}), using the Cactus numerical relativity code~\cite{Cactus} and the CausalSets arrangement written by David Rideout. In this environment, code was developed to implement the three particle trajectory algorithms discussed in Section~\ref{s:models}. The causal sets generated in this manner can of course be considered as partially ordered sets independent of Minkowski spacetime. Here the information about the coordinates of the points in Minkowski spacetime will be retained to allow the causal sets to be easily visualised. The particular frame  to which the coordinates of the sprinkled points in Minkowski spacetime refer will be called the embedding frame.

\begin{figure}[th]
\begin{center}
\includegraphics[height=9cm]{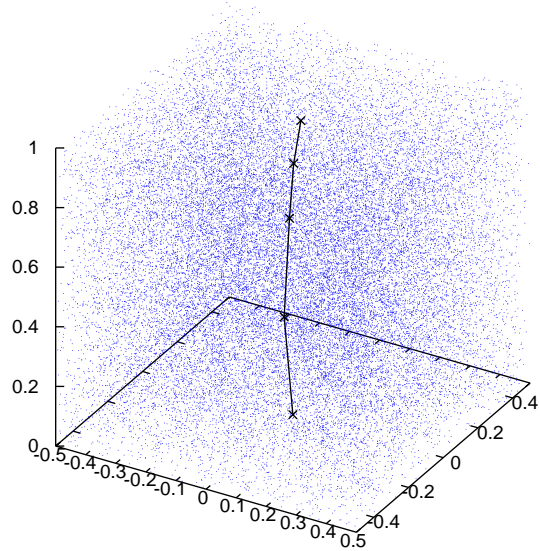}
\caption[An example of a Model 1 particle trajectory in 3+1 dimensions]{An example trajectory using Model 1 in 3+1 dimensions (one spatial dimension is suppressed). The causal set here has size $N=32768$. The trajectory is constructed using $\tau_f = 0.37$ in embedding coordinates.}\label{f:3Dexamplemom}
\end{center}
\end{figure}

\begin{figure}[p]
\begin{center}
\subfigure[Model 2, $n_f=5$.]{
\includegraphics[height = 9cm]{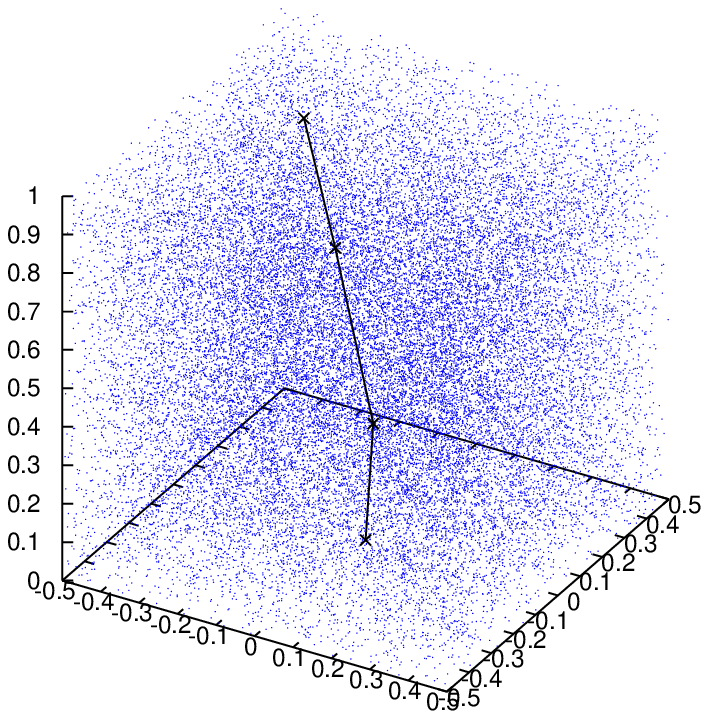}}
\subfigure[Model 3, $n_f = 5$.]{
\includegraphics[height=9cm]{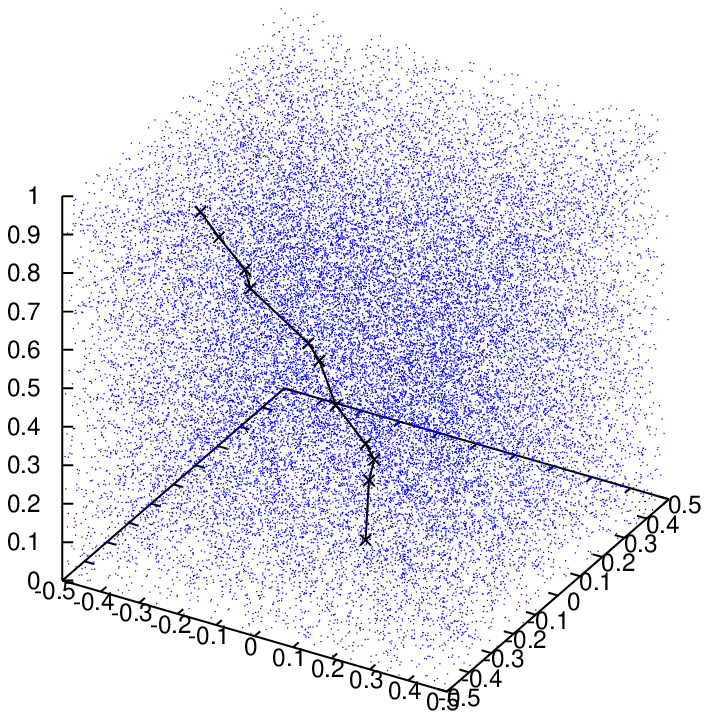}}
\caption[Examples of intrinsic particle trajectories in 3+1 dimensions]{Example intrinsic trajectories in 3+1 dimensions (one spatial dimension is suppressed). The causal set here has size $N=32768$.}\label{f:3Dexampleintrinsic}
\end{center}
\end{figure}

The results discussed here are, for the most part, in 1+1 dimensions. The code is, however, completely general and can also be run in 3+1 dimensions (example trajectories in 3+1 for each particle model are shown in Figures~\ref{f:3Dexamplemom} and~\ref{f:3Dexampleintrinsic}). While simulations in 3+1 are more physically relevant, they require much larger causal sets to obtain the same discreteness scale and the necessary time and computer resources were unfortunately not available during the writing of this thesis. Simulations in 1+1 dimensions in fact prove to be more than adequate to investigate the link between the continuum and discrete processes.

Several important aspects of the simulations need to be noted. Firstly, an element is added to the sprinkling at the origin to allow a fixed beginning point for the trajectories. Also, the models all assume a partial particle trajectory already exists and is being extended -- in these simulations the trajectories start with a single element and thus the development of the first part of the trajectory needs to be carefully considered. For Model 1, the swerves model, the matter is very simple. An initial momentum, $p_0$ is defined and no further information is needed to construct the future trajectory. In these simulations $p_0$ is taken as $p_0=0$, i.e.~the particle is initially at rest in the {embedding} frame. 

Model 2 relies on knowledge of $e_{n-1}$ as well as $e_n$ to determine $e_{n+1}$. Unfortunately it is not possible to just define an $e_{-1}$ to begin the trajectory. In this case, when constructing the first step, the condition $d(e_{n-1},e_{n+1})\leq 2n_f$ is simply ignored. The element $e_1$ is chosen such that it maximises $d(e_0,e_1)\leq n_f$. Points satisfying this condition will be close to the hyperbola defined by a proper time $\tau = \sqrt{t^2-x^2}=n_f d_{pl}$, where $d_{pl}$ is the discreteness scale of the causal set. Again, the particle should be initially at rest (or as close as possible) in the embedding frame. To impose this, instead of choosing $e_1$ at random from those eligible, the element with the smallest time coordinate in the embedding is chosen.  The rest of the trajectory is constructed as per the algorithm.  

Model 3 presents a slightly more complicated situation. Here the future trajectory from $e_0$ depends on elements $e_{-n_f}, \ldots, e_{-1}$.  In simulations of Model 3, I first construct a longest chain between the element at $t=0,\,x=0$ and an element at $t=0.25,\,x=0$ (also artificially added to the sprinkling). This longest chain forms the beginning of the trajectory: essentially, it imposes the condition that the beginning of the trajectory is as close to geodesic as possible, and again the particle is initially at rest in the embedding frame. The rest of the trajectory is constructed as per the algorithm. Note that the initial longest chain may have less than $n_f$ elements, in which case minimisation is over all elements until $n_f$ elements are reached.

A second important point to note about the simulations is scale: the causal set models give rise to the continuum diffusion equation when $1<<n_f<<l$ where $l$ is the trajectory length. In other words, the causal set must be large enough that the trajectory involves many $n_f$ steps, and $n_f$ itself must be much larger than 1. This is quite difficult to obtain. Suppose $n_f=10$ and we want the trajectory length to be $10n_f$ steps (or in the case of Model 3, 100 elements). In 1+1 dimensions this requires a causal set of roughly 10000 elements -- quite manageable. If $n_f = 100$, however, a causal set of roughly $1\times 10^8$ is needed for $l=100n_f$.\footnote{In 3+1 dimensions you would need the absurd number of $1\times 10^{16}$ elements.} This is certainly not manageable with the current code, so it is not really possible to explore realistic values of $n_f$ and trajectory length. Despite this limitation the simulations clearly show diffusion behaviour, as will be demonstrated below.

\subsection{Example trajectories in 1+1 dimensions}

\begin{figure}[p]
\begin{center}
\includegraphics[height=9cm]{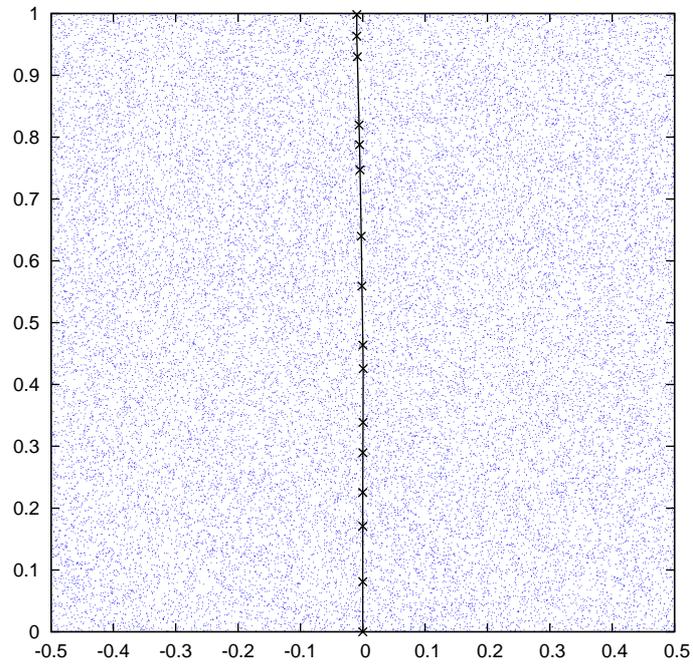}
\caption[An example of a Model 1 particle trajectory in 1+1 dimensions]{Example trajectory using Model 1 in 1+1 dimensions: $N=32768$, $\tau_f = 0.1105$, $l=16$.}\label{f:2Dexamplemom}
\end{center}
\end{figure}

Simply constructing an example trajectory for each of the three models in 1+1 dimensions reveals something of the relationship between the parameters of the model and the diffusion parameter in the continuum. Figure~\ref{f:2Dexamplemom} shows a trajectory constructed in a causal set of size $N=32768$, using the swerves momentum model, Model 1.\footnote{The causal sets are generated by a Poisson process with a mean of $N$, the actual number of elements varies.} For this trajectory $\tau_f = 0.1105$ in embedding coordinates, and the final length of the trajectory is $l=16$ elements. Clearly this trajectory does not have a forgetting time, $\tau_f$, many orders of magnitude greater than the discreteness scale, or a total trajectory length much greater than $\tau_f$. The result is, however, very very close to the expected straight line in the continuum.  

\begin{figure}[p]
\begin{center}
\includegraphics[height=9cm]{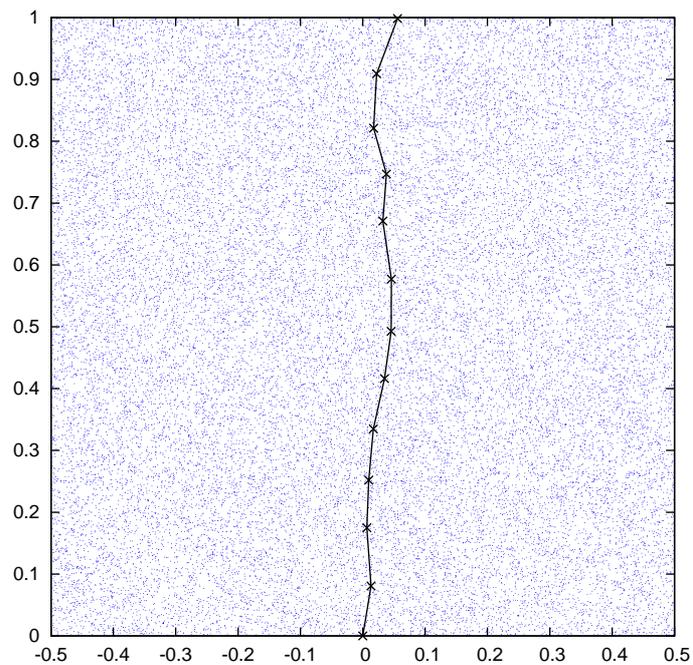}
\caption[An example of a Model 2 particle trajectory in 1+1 dimensions]{Example trajectory using Model 2 in 1+1 dimensions: $N=32768$, $n_f = 20$, $l=13$.}\label{f:2Dexampleintrinsicnf}
\end{center}
\end{figure}

\begin{figure}[p]
\begin{center}
\subfigure[Model 3, $n_f = 20$, $l=163$.]{
\includegraphics[height = 9cm]{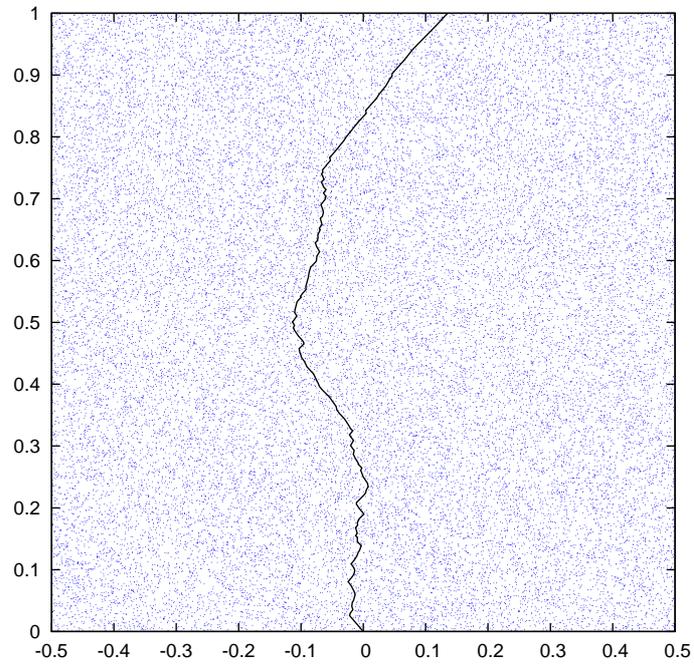}}
\subfigure[Model 3, $n_f = 40$, $l=154$.]{
\includegraphics[height=9cm]{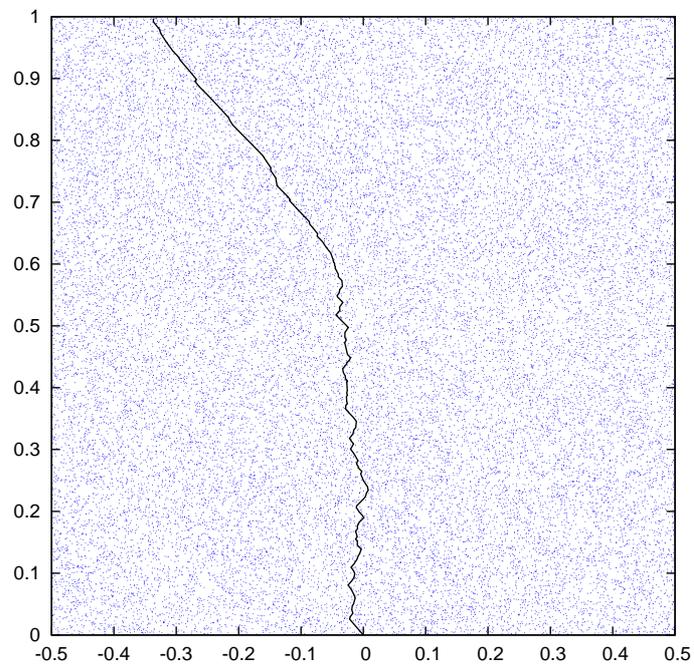}}
\caption[Examples of Model 3 particle trajectories in 1+1 dimensions]{Example Model 3 trajectories in 1+1 dimensions: $N=32768$.}\label{f:2Dexampleintrinsic}
\end{center}
\end{figure}

\begin{figure}[th]
\begin{center}
\includegraphics[]{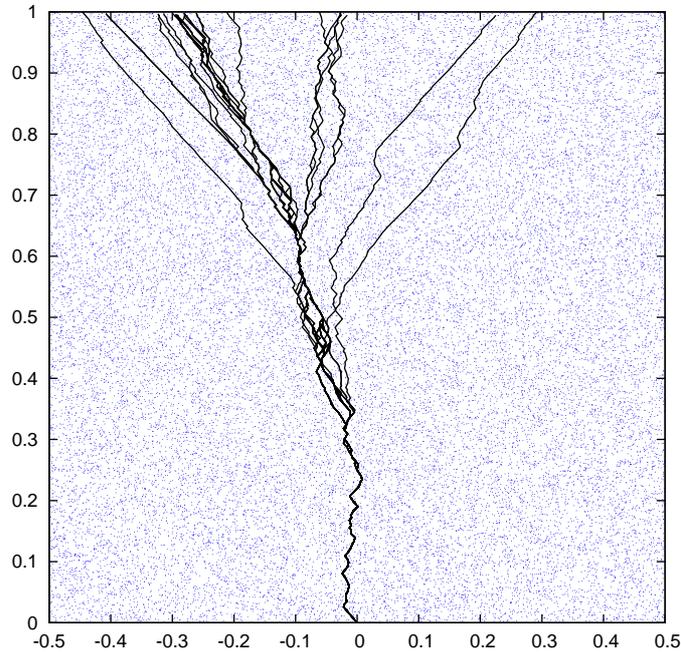}
\caption[Multiple Model 3 trajectories]{30 Model 3 trajectories in 1+1 dimensions: $N=32768$, $n_f = 40$.}\label{f:intrinsicfixedembedding}
\end{center}
\end{figure}

Model 2 also gives a reasonably straight line for comparable parameters. Figure~\ref{f:2Dexampleintrinsicnf} shows a trajectory constructed in a causal set with $N=32768$ and $n_f=20$. Note that $n_f=20$ is approximately equivalent to $\tau_f=0.1105$ for a causal set of size $N=32768$. Recall, however, that Model 2 actually relies on information $2n_f$ into the past of the trajectory: it would be more correct to compare a $n_f=10$ Model 2 trajectory with a $\tau_f=0.1105$ Model 1 trajectory. Regardless, it is clear that with roughly comparable forgetting times Model 2 results in more fluctuations in position and momentum.

A comparison of the two intrinsic models, Model 2 and Model 3, is also interesting. In Figure~\ref{f:2Dexampleintrinsic} two trajectories constructed using Model 3 are shown: one with $n_f=20$ and one with $n_f=40$. As mentioned above, Model 2 relies on information $2n_f$ steps into the past, so we might expect similar trajectories from $n_f=20$ Model 2 and $n_f=40$ Model 3. The trajectories from Model 3, however, swerve much more. Given that the values of $n_f$ and $l$ are not really in the realm $1<<n_f<<l$, where the continuum diffusion is expected to hold, care should be taken in drawing conclusions about the continuum behaviour of Models 2 and 3. Even so, it seems reasonable to conclude that a simple universal relation between the forgetting number of an intrinsic particle model and the diffusion parameter in the continuum does not exist (a concrete relation between the forgetting time of Model 1 and the diffusion parameter will be found in Section~\ref{ss:model1sim}). Of course, the models here are not physically realistic. It seems likely, however, that when physically realistic models are developed, the determination of the diffusion constant $k$ from the `fundamental' parameters of the underlying model on the causal set, will be model dependent. 

The diffusion-like behaviour of the models is most obvious if multiple trajectories are generated. Recall that the intrinsic models do not define unique trajectories -- an element is chosen at random from the eligible elements at each step. For a fixed causal set and initial trajectory, multiple trajectories can thus be generated. Figure~\ref{f:intrinsicfixedembedding} shows 30 (distinct) Model 3 trajectories in a fixed causal set, $N=32768$, $n_f = 40$. Note that from $t=0$ to $t=0.25$ the trajectory is a fixed longest chain for all 30 trajectories.  
%[NOT USEFUL!] Histograms of the final position and momentum can be plotted, and are shown in Figure STILL TO DO!
Although diffusion-like behaviour is clear, it is difficult to determine whether it is exactly the diffusion given by the swerves diffusion equation. The values of $N$ and $n_f$ are nowhere close to the continuum limit, and 30 trajectories is simply insufficient. To compare more directly with the swerves diffusion equation, I will look instead at the non-intrinsic model, Model 1.

\subsection{A close investigation of Model 1}
\label{ss:model1sim}
\subsubsection{Simulating trajectories}

Unlike the intrinsic models, for a given causal set and fixed initial position and momentum Model 1 defines a unique trajectory. To say that Model 1 results in diffusion is, in a sense, saying that the exact underlying causal set is unknown. To simulate this I generate many different sprinklings into the same region of Minkowski spacetime, and calculate the unique trajectory in each. As above, the starting point of each trajectory is fixed at $x=0,\,t=0$, an extra element artificially added to each sprinkling. For each trajectory a final position and momentum can be calculated at $t_{final}=0.95$. This is done by identifying elements $e_n$ and $e_{n+1}$ in the trajectory such that $e_n$ is the last element with $t<t_{final}$ and $e_{n+1}$ is the first element with $t>t_{final}$. Note that there is a vanishingly small probability that any element has $t=t_{final}$. The final position is found by linear interpolation between $e_{n}$ and $e_{n+1}$. The final momentum is simply the usual momentum at $e_{n+1}$, i.e.~it is proportional to the vector between $e_n$ and $e_{n+1}$. 

\begin{figure}[ht]
\begin{center}
%\centering
\includegraphics[height=9cm]{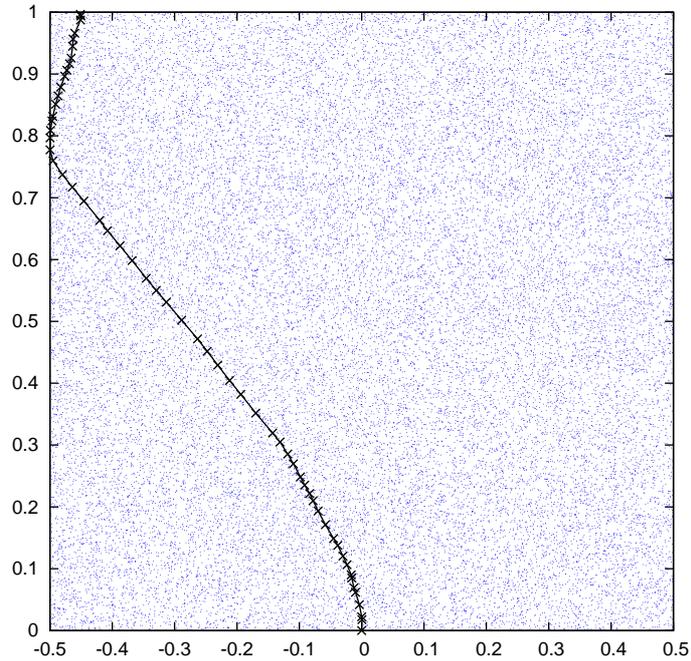}
\caption[An invalid Model 1 trajectory]{An invalid Model 1 trajectory, $N=32768$, $\tau_f=0.02$.}\label{f:invalidtraj}
\end{center}
\end{figure}

The first step in the analysis is to reject some trajectories -- if a trajectory is close to the boundary of the region of Minkowski spacetime at any point, it will `bounce' back and distort the results. An example of such a trajectory is given in Figure~\ref{f:invalidtraj}. Provided only a small fraction of the trajectories are invalid, the invalid ones can be safely rejected without much loss of information about the trajectory distribution. The parameter values investigated in this section are chosen such that there are indeed few, if any, invalid trajectories.

Once the invalid trajectories have been removed, histograms of final position and final momentum can be generated from the remaining trajectories. These histograms clearly show diffusion behaviour, but to truly test the model the results must be compared with the distributions expected from the swerves diffusion equation.

\subsubsection{The swerves equation}

The inhomogeneous, 1+1-dimensional swerves diffusion equation is
%%%%%%%%%%%%%%%%%%%%%%%%%%%%%%%%%%%%%%%%%%%%%%%%%%%%%%%%%%%%%%%%%%%%%%%%%%%%%%%%%%%%%%%%%%%%%%%%%%%%%%%%%%%%%%%%
\begin{equation}
\pd{\rho_t}{t}= \frac{-p}{\sqrt{m^2+p^2}}\pd{\rho_t}{x} + k\pd{}{p}\left(\frac{\sqrt{m^2+p^2}}{m}\pd{\rho_t}{p}\right).\label{e:1dswerves}
\end{equation}
%%%%%%%%%%%%%%%%%%%%%%%%%%%%%%%%%%%%%%%%%%%%%%%%%%%%%%%%%%%%%%%%%%%%%%%%%%%%%%%%%%%%%%%%%%%%%%%%%%%%%%%%%%%%%%%%
As this does not have an exact solution, it must be evolved numerically. For the results shown here this numerical evolution is conducted with \texttt{NDSolve} in \texttt{Mathematica}. The initial distribution of the simulated trajectories is a $\delta$-function: all trajectories begin with $x=0$ and $p=0$. To numerically evolve Equation~\ref{e:1dswerves} with a $\delta$-function initial distribution is not possible -- instead, the initial distribution is taken to be a highly peaked Gaussian
%%%%%%%%%%%%%%%%%%%%%%%%%%%%%%%%%%%%%%%%%%%%%%%%%%%%%%%%%%%%%%%%%%%%%%%%%%%%%%%%%%%%%%%%%%%%%%%%%%%%%%%%%%%%%%%%
\begin{equation}
\rho_t(x,p,0) = \exp\left(-\frac{x^2}{\sigma_x^2}-\frac{p^2}{\sigma_p^2}\right).\label{e:1dinit}
\end{equation}
%%%%%%%%%%%%%%%%%%%%%%%%%%%%%%%%%%%%%%%%%%%%%%%%%%%%%%%%%%%%%%%%%%%%%%%%%%%%%%%%%%%%%%%%%%%%%%%%%%%%%%%%%%%%%%%%
Here, $\sigma_x^2$ and $\sigma_p^2$ are not fixed: they are chosen such that Equation~\ref{e:1dinit} is a good approximation to a histogram of the initial position-momentum data constructed with the same bin size as used for the final data, and such that Equation~\ref{e:1dswerves} evolves smoothly without numerical artifacts. To obtain final position and momentum distributions the evolved $\rho_t(x,p,t_{final})$ is summed over momentum and position, respectively.

\subsubsection{The matter of units}

To compare the simulated trajectories with the swerves diffusion equation a consistent set of units needs to be chosen: Equation~\ref{e:1dswerves}, above, is expressed in terms of Planck units; the simulated trajectories are expressed in terms of `embedding' coordinates. As mentioned earlier in this thesis, the discreteness scale in causal set theory is expected to be of the order of the Planck scale. For concreteness I assume the two are the same, and thus Equation~\ref{e:1dswerves} can be considered as expressed in `discreteness units'. A discreteness length $d_{pl} = \sqrt{V/N}$, where $V$ is the volume of the region of Minkowski spacetime in embedding units and $N$ is the mean causal set size, can be calculated from the simulated causal sets. The position, time, momentum, and mass in the simulations are then re-expressed in terms of discreteness units. 

\subsubsection{Comparing the simulations to the swerves equation}

Equation~\ref{e:1dswerves} is evolved for a range of values of $k$ and the resulting distributions are compared to the position and momentum histograms for given values of the model parameters ($\tau_f$, $N$, and particle mass $m$). A best fit value of $k$ for each set of model parameters is found by minimising the reduced $\chi^2$:
%%%%%%%%%%%%%%%%%%%%%%%%%%%%%%%%%%%%%%%%%%%%%%%%%%%%%%%%%%%%%%%%%%%%%%%%%%%%%%%%%%%%%%%%%%%%%%%%%%%%%%%%%%%%%%%%
\begin{eqnarray}
\chi^2 &=& \sum_i\frac{\left(O_i-E_i\right)^2}{E_i}\,,\\
\chi_{red}^2&=&\chi^2/f,
\end{eqnarray}
%%%%%%%%%%%%%%%%%%%%%%%%%%%%%%%%%%%%%%%%%%%%%%%%%%%%%%%%%%%%%%%%%%%%%%%%%%%%%%%%%%%%%%%%%%%%%%%%%%%%%%%%%%%%%%%%
where $O_i$ is the observed frequency, $E_i$ is the expected frequency (i.e.~that given by the evolution of the diffusion equation), and $f$ is the number of degrees of freedom (here, $f=$~number of data points - 1). The reduced $\chi^2$ is calculated for the distribution in momentum rather than the full momentum-position distribution. The momentum distribution is chosen for comparison as it drives the position diffusion.
Care must be taken when calculating the value of the reduced $\chi^2$ -- if a significant proportion of the frequency values are less than five, $\chi^2$ is no longer a good measure of fit (see, for example~\cite{Sheskin:2004}). In situations where the frequency in a given bin is less than five, multiple bins are combined to resolve this problem.
A reduced $\chi^2$ value of $\chi^2_{red}\sim 1$ is usually taken to indicate a good fit (see, for example~\cite{Taylor:1997}), i.e.~the evolved diffusion equation is a good description of the simulated trajectories.

\subsubsection{Results}

The goal of these simulations is not just to demonstrate that the model gives diffusion behaviour, but also to determine the relationship between the continuum diffusion parameter $k$ and the parameters of the underlying discrete model. The relationship between $k$ and the forgetting time $\tau_f$ can be determined by running simulations for a range of values of $\tau_f$ (with fixed $N$ and $m$) and determining the best fit value of $k$ in each case.

A causal set of size $N=32768$ was used in these simulations. The particle mass in embedding units is taken to be $m=1$. Five hundred trajectories were evolved for each value of $\tau_f$ in $\tau_f=\{0.03,0.035,0.04,0.045,0.05,0.055,0.06,0.07,0.08,0.09,0.1\}$ (in embedding units). The results are summarised in Table~\ref{t:results}. Several things need to be considered when interpreting these results, as discussed below.

\begin{figure}[p]
\begin{center}
\subfigure[$\tau_f=0.04$.]{
\begin{minipage}[]{0.4\linewidth}
\includegraphics[height = 3.5cm]{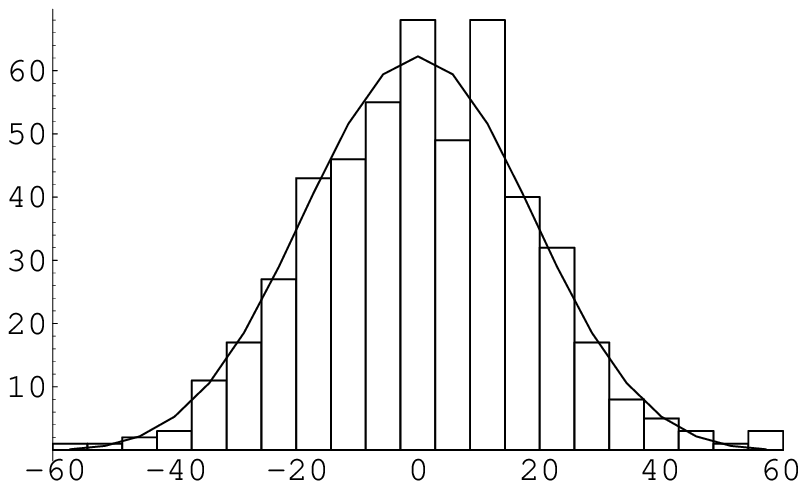}
\end{minipage}
\hspace{4mm}
\begin{minipage}[]{0.4\linewidth}
\includegraphics[height = 3.5cm]{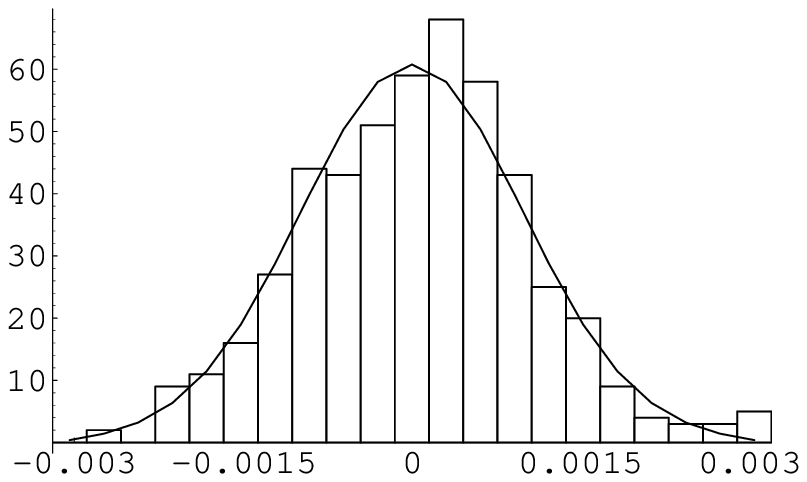}
\end{minipage}}
\subfigure[$\tau_f=0.05$.]{
\begin{minipage}[]{0.4\linewidth}
\includegraphics[height = 3.5cm]{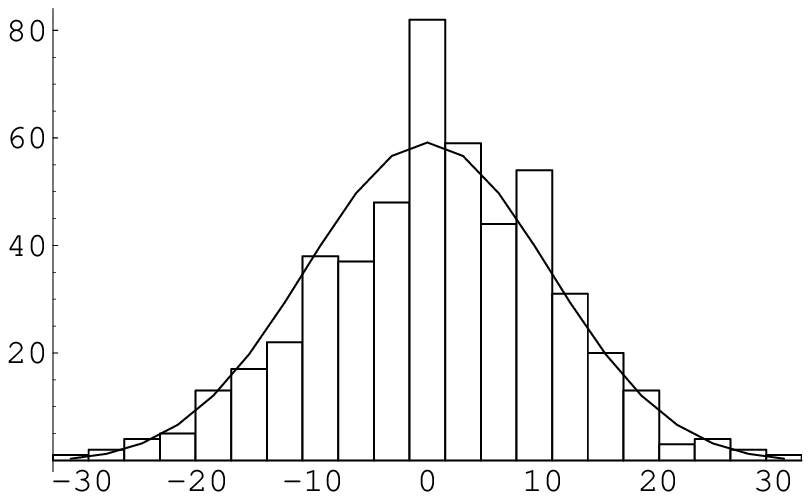}
\end{minipage}
\hspace{4mm}
\begin{minipage}[]{0.4\linewidth}
\includegraphics[height = 3.5cm]{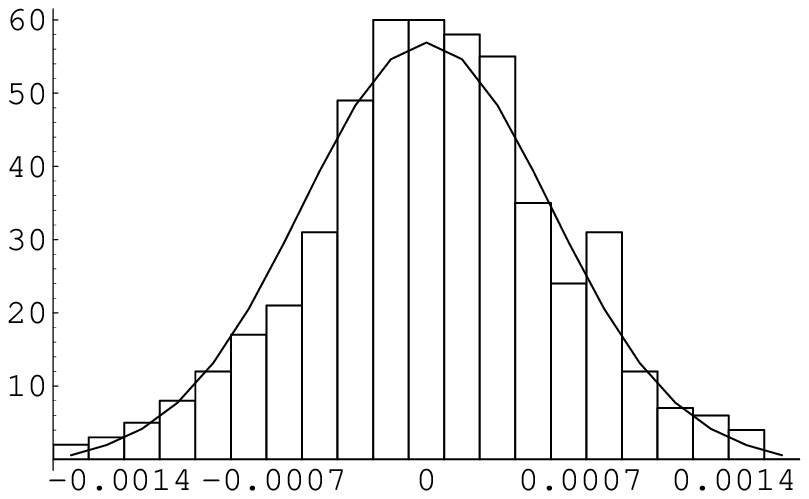}
\end{minipage}}
\subfigure[$\tau_f=0.06$.]{
\begin{minipage}[]{0.4\linewidth}
\includegraphics[height = 3.5cm]{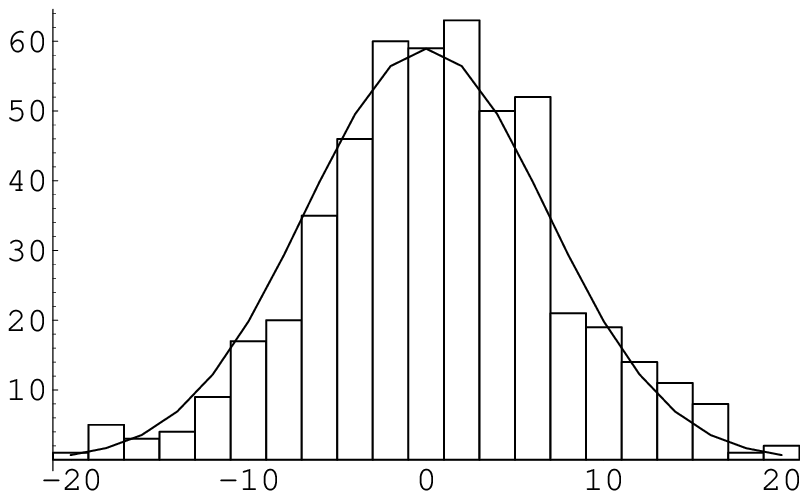}
\end{minipage}
\hspace{4mm}
\begin{minipage}[]{0.4\linewidth}
\includegraphics[height = 3.5cm]{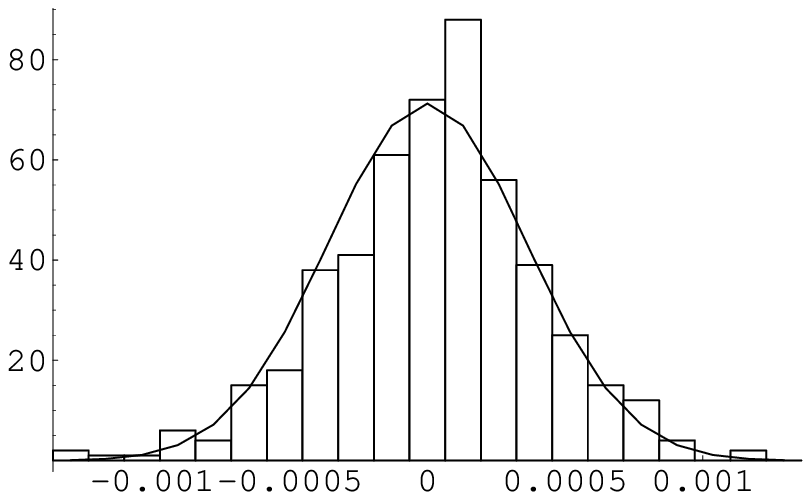}
\end{minipage}}
\subfigure[$\tau_f=0.08$.]{
\begin{minipage}[]{0.4\linewidth}
\includegraphics[height = 3.5cm]{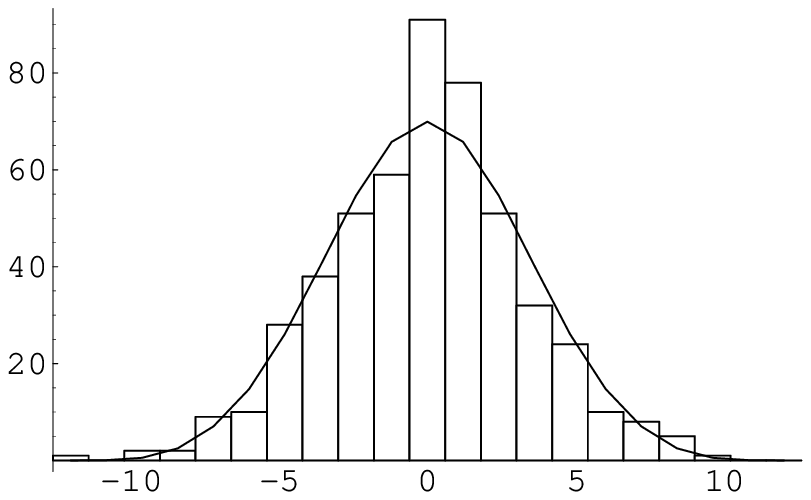}
\end{minipage}
\hspace{4mm}
\begin{minipage}[]{0.4\linewidth}
\includegraphics[height = 3.5cm]{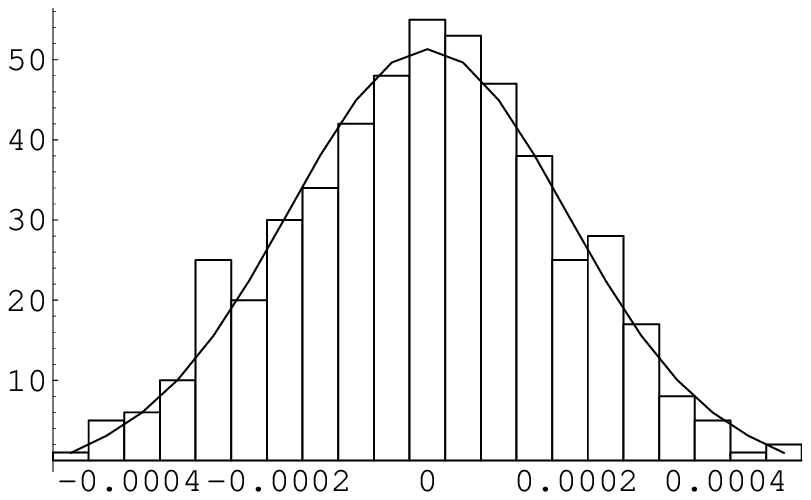}
\end{minipage}}
\caption[Example position and momentum histograms for Model 1]{Example histograms with best fit numerical solutions. The position distribution is shown on the left, momentum on the right.}.\label{f:histograms}
\end{center}
\end{figure}

Table~\ref{t:results} shows the simulated values of $\tau_f$ in both embedding units and multiples of the discreteness length $d_{pl}$. In units of $d_{pl}$, $\tau_f$ varies between $5.4$ and $18$ -- clearly not the many orders of magnitude greater than $d_{pl}$ that is expected in reality. The average trajectory length (this is the number of `steps' in the trajectory rather than the number of elements) is also well out of the realistic range, reaching a maximum of only $48$. Unfortunately realistic values are simply not feasible with the current computing facilities. What is remarkable is that, even with these values of $\tau_f$ and $N$, the distribution of the simulated trajectories is extremely well modelled by the diffusion equation. The best fit values of the diffusion parameter and corresponding $\chi^2_{red}$ values are given in Table~\ref{t:results}. As mentioned earlier, $\chi^2_{red}\sim 1$ is usually considered a good fit. It is easy to see just how good the fit is when the evolved diffusion equation is plotted together with the simulation histograms. Examples of both the position and momentum histograms for four of the values of $\tau_f$ are shown in Figure~\ref{f:histograms}. The particular values of $\tau_f$ shown are not special in any way, and are merely chosen to illustrate the results. 

\begin{table}[t]
\centering
\begin{tabular}[th]{|c|c|c|c|c|c|}
\hline
\multirow{2}{*}{$\tau_f$ } & \multirow{2}{*}{$\tau_f$ ($d_{pl}$)} &  Number of invalid & Average trajectory & \multirow{2}{*}{best fit $k$} &\multirow{2}{*}{$\chi^2_{red}$}\\
& & trajectories & length & &\\
\hline
0.03 & 5.4 & 10 & 48 & $1.2\times 10^{-8}$ & 1.5\\
0.035 & 6.3 & 2 & 42 & $5.5\times 10^{-9}$ & 1.0\\ 
0.04 & 7.2 & 0 & 37 & $2.8\times 10^{-9}$  & 0.59\\
0.045 & 8.1 & 0 & 33 & $1.5\times 10^{-9}$ & 0.64\\
0.05 & 9.1 & 0 & 30 & $9.2\times 10^{-10}$ & 1.0\\
0.055 & 10 & 0 & 27 & $5.0\times 10^{-10}$ & 2.4\\
0.06 & 11 & 0 & 25 & $3.7\times 10^{-10}$ & 1.6 \\
0.07 & 13 & 0 & 22 & $1.7\times 10^{-10}$ & 0.56 \\
0.08 & 14 & 0 & 19 & $9.9\times 10^{-11}$ & 0.55\\
0.09 & 16 & 0 & 17 & $4.7\times 10^{-11}$ & 2.0\\
0.1 & 18 & 0 & 16 & $2.6\times 10^{-11}$ & 2.4\\
\hline
\end{tabular}
\caption{Analysis of Model 1 trajectories for various values of $\tau_f$, with fixed $N$ and $m$.}
\label{t:results}
\end{table}

As Table~\ref{t:results} shows, invalid trajectories are an issue only for the smallest two values of $\tau_f$ and even in the worst case only comprise $2\%$ of the 500 trajectories. As such, they don't influence the results given here. They do, however, become significant for values of $\tau_f$ smaller than $0.03$ lessening the utility of simulations for those values. 
 
\begin{figure}[p]
\begin{center}
\subfigure[$k$ vs.~$\tau_f$.]{
\hspace{-0.038\textwidth}
\includegraphics[width = 0.838\textwidth]{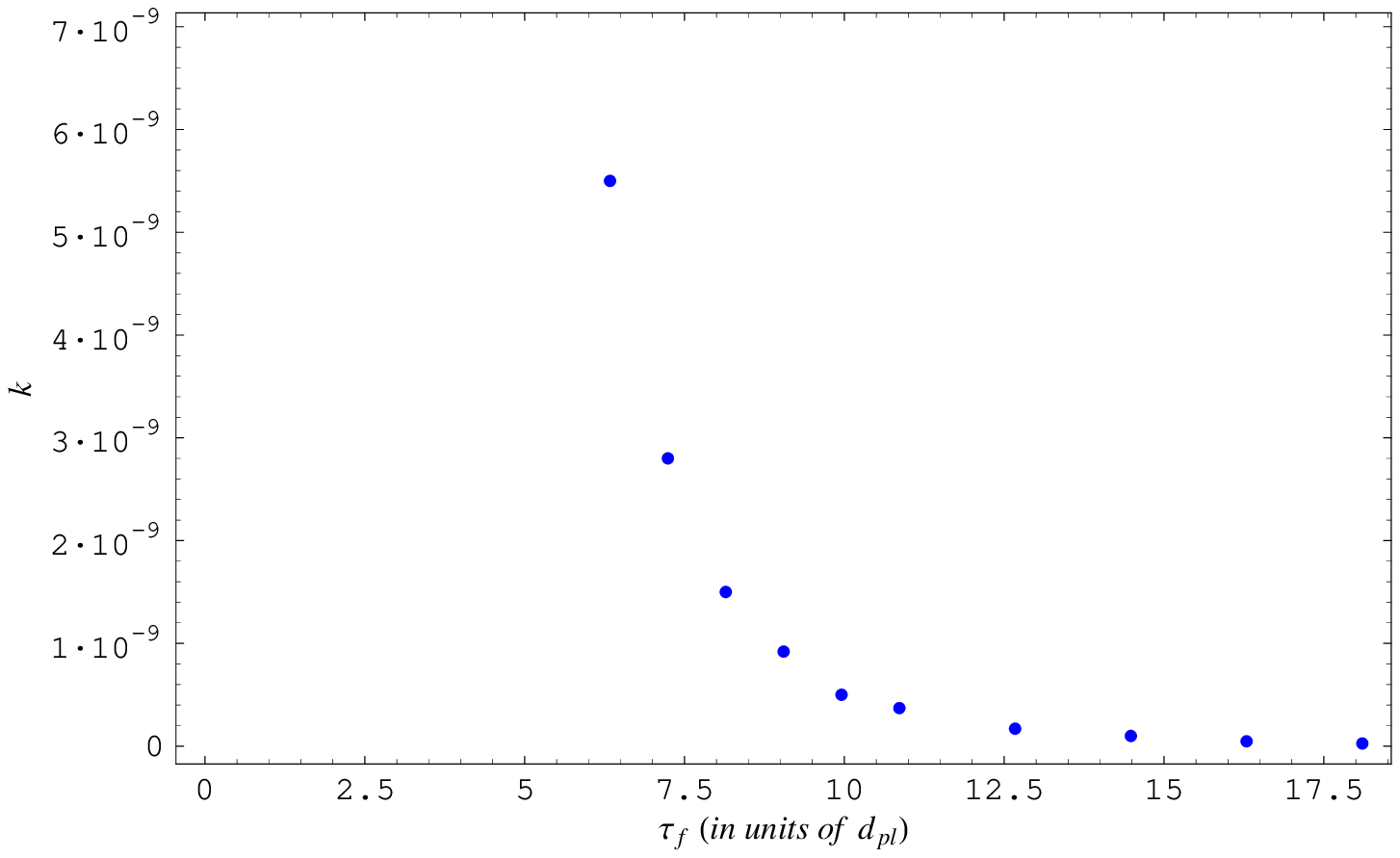}}
\subfigure[$\ln k$ vs.~$\ln\tau_f$ with the best fit line $-5.02\ln\tau_f-9.8$.]{
\includegraphics[width = 0.805\textwidth]{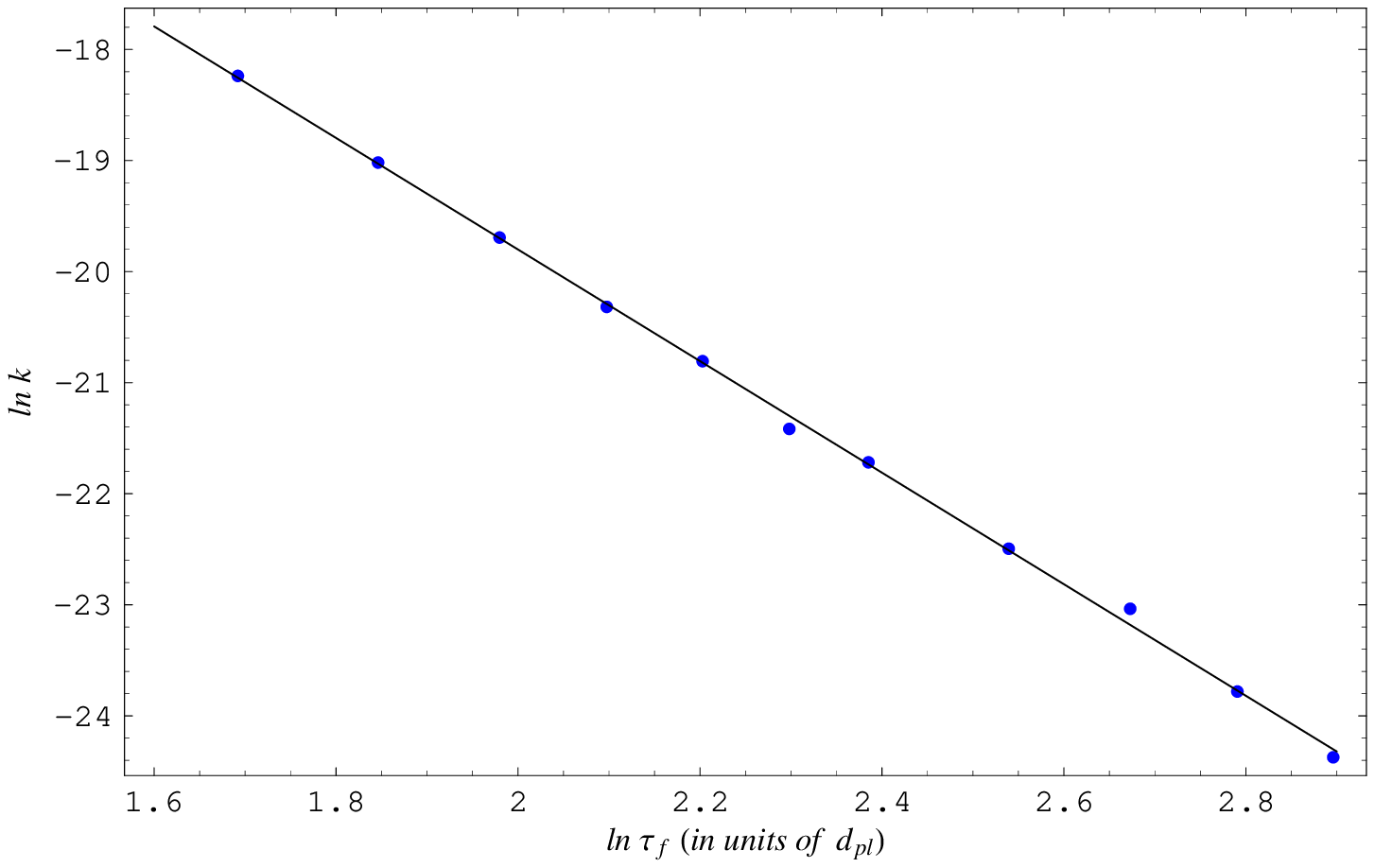}}
\caption[$k$ vs.~$\tau_f$ and $\ln k$ vs.~$\ln\tau_f$]{Comparison of the diffusion parameter $k$ and forgetting time $\tau_f$, for fixed $N$ and $m$.}\label{f:kvstauf}
\end{center}
\end{figure}

Although the best fit values of $k$ and corresponding $\chi^2_{red}$ calculated here are fairly robust, it should be noted that the best fit values of $k$ are determined by calculating $\chi^2_{red}$ for various values of $k$ and choosing the one that minimises $\chi^2_{red}$. A truly systematic search of the parameter space is not undertaken, and a measure of the error in $k$ is not calculated. The exact values of $\chi^2_{red}$ depend slightly on the bin width chosen when generating the histograms. They also vary slightly if $\sigma_x$ and $\sigma_p$ in the initial Gaussian distribution are changed. A further point is that for some $\tau_f$ there are a range of values of $k$ that give $\chi^2_{red}\sim 1$ and thus all could be considered a good fit. For the results that I have obtained, $k$ vs.~$\tau_f$ is plotted in Figure~\ref{f:kvstauf}. The dependence of $k$ on $\tau_f$ is clear in the $\ln-\ln$ plot, where the best fit line to the data, $\ln k = -5.02\ln \tau_f -9.8$ is also shown. Despite the considerations mentioned above, it seems clear that $k\sim 1/\tau_f^5$.

\pagebreak[4]
The diffusion parameter $k$ also depends on the particle mass $m$. In Model 1 the trajectory is independent of the particle mass -- the mass is only used to correctly normalise the momentum. The momentum histogram for trajectories of a particle of mass $m^{\prime} = \alpha m$ can therefore be obtained simply by scaling the original momentum histogram by $\alpha$. Looking at Equation~\ref{e:1dswerves}, if both $m$ and $p$ are scaled by $\alpha$ and $k$ is scaled by $\alpha^2$ the equation remains unchanged. Thus it can be determined, without reanalysing the simulations, that $k\sim m^2$.  

\begin{figure}[p]
\begin{center}
\subfigure[$\ln k$ vs.~$\ln d_{pl}$ with the best fit line $4.0\ln d_{pl} + 13$.]{
\hspace{0.012\textwidth}
\includegraphics[width = 0.798\textwidth]{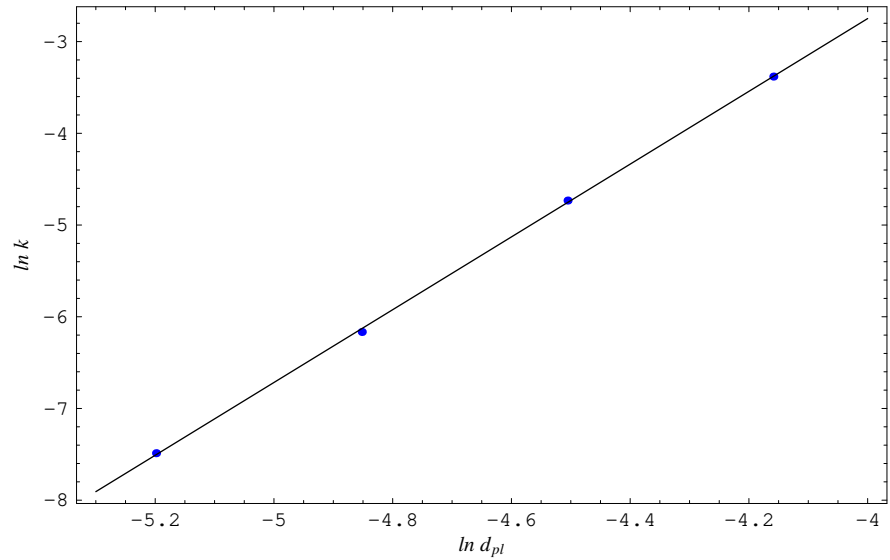}
\label{f:logkvslogdpl}}
\subfigure[$k$ vs.~$m^2/\tau_f^5$ with the best fit line $1.9\,m^2/\tau_f^5-2.5\times 10^{-11}$.]{\hspace{-0.06\textwidth}
\includegraphics[width = 0.865\textwidth]{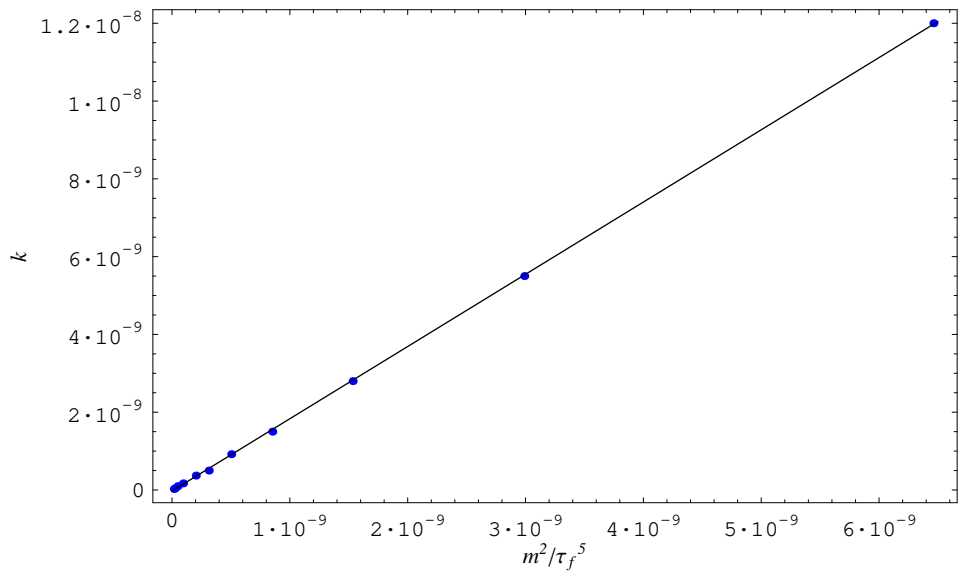}
\label{f:kvsm2tauf5}}
\caption[$\ln k$ vs.~$\ln d_{pl}$ and $k$ vs.~$m^2/\tau_f^5$]{The relationship between the diffusion parameter $k$ and underlying model parameters $d_{pl},\,m,$ and $\tau_f$.}
\end{center}
\end{figure}

There is one remaining parameter that $k$ can depend on: the discreteness length itself. From dimensional analysis of Equation~\ref{e:1dswerves} it can be deduced that $k$ has units of $L^{-3}$ (since Planck units are used, $L = T = M^{-1}$). Since $k\sim m^2\tau_f^{-5}$ the dependence on $d_{pl}$ must be $k\sim d_{pl}^4$. This dependence can be checked through simulations. In this case everything needs to be expressed in terms of embedding units rather than discreteness units.  Five hundred trajectories are calculated for four different values of $N$ (and thus $d_{pl}$) and the same values of forgetting time $\tau_f=0.08$ and mass $m=1$. The value of $\tau_f$ was chosen to avoid too many invalid trajectories in the causal sets considered. As above, the best fit values of $k$ and the corresponding $\chi^2_{red}$ can be calculated. The results are summarised in Table~\ref{t:resultsdpl}. Again, a $\ln-\ln$ plot (see Figure~\ref{f:logkvslogdpl}) allows the dependence of $k$ on $d_{pl}$ to be determined. The result is as expected: $k\sim d_{pl}^4$.

\begin{table}[t]
\centering
\begin{tabular}[t]{|c|c|c|c|c|c|}
\hline
\multirow{2}{*}{$N$} & \multirow{2}{*}{$d_{pl}$} &  Number of invalid & Average trajectory & \multirow{2}{*}{best fit $k$} &\multirow{2}{*}{$\chi^2_{red}$}\\
& & trajectories & length & &\\
\hline
4096 & 0.016 & 12 & 19 & 0.034 & 0.64\\
8192 & 0.011 & 0 & 19 & 0.0088 & 1.0\\ 
16384 & 0.0078 & 0 & 19 & 0.0021 & 0.45\\
32768 & 0.0055 & 0 & 19 & 0.00056 & 0.55\\
\hline
\end{tabular}
\caption{Analysis of Model 1 trajectories for various values of $d_{pl}$ (in embedding units).}
\label{t:resultsdpl}
\end{table}

The analysis of the simulations leads to the relation $k\sim m^2 d_{pl}^4/\tau_f^5$. The constant of proportionality can in fact also be determined. Looking back at the results for varying $\tau_f$ and noting that in terms of discreteness units $m = 0.0055$ and $d_{pl} =1$ the relation $k$ vs.~$m^2/\tau_f^5$ can be plotted (see Figure~\ref{f:kvsm2tauf5}). The best fit line is found to have the equation $k = 1.9\, m^2/\tau_f^5-2.5\times 10^{-11}$. Thus, it can be concluded that 
\begin{equation}
k \approx 2 m^2 d_{pl}^4/\tau_f^5\,.
\label{e:kest}
\end{equation}

Causal set theory contains not only a discreteness scale, but also a hypothesised `nonlocality' scale below which physics cannot remain local~\cite{Sorkin:2007qi,Sorkin:2009bp}. The forgetting parameter that appears in the massive particle models may be a measure of this nonlocality scale. Sorkin~\cite{Sorkin:2007qi} estimates the nonlocality scale to be of the order $10^{20}$ in Planck units. Although the simulations here use forgetting times nowhere near this estimate, and there is no theoretical reason why the forgetting time should be exactly the nonlocality scale, it is interesting to consider the value of the diffusion constant that arises if the above results are extrapolated. Taking the particle to be a proton with mass $m=10^{-20}$, $\tau_f=10^{20}$, and $d_{pl}=1$, Equation~\ref{e:kest} leads to the estimate $k\approx 10^{-140}$. 

As discussed in Section~\ref{s:massivebounds}, tight constraints have already been placed on $k$ through observations. In particular (see Equation~\ref{e:kboundHyd}) $k\leq10^{-56}kg m^2 s^{-3}\sim10^{-102}$ due to the lack of spontaneous heating of hydrogen in the laboratory. The diffusion parameter estimated from the nonlocality scale is thus many orders of magnitude smaller than the current constraint (the factor of two mass difference between the proton and a hydrogen molecule is inconsequential here). This rough calculation should not be given too much weight, but it does suggest that $k$ could be well outside our current observable range.

%% file: SwervesConclusion.tex
\section{Concluding remarks}
\label{s:swervesconclusion}

Following the proposal by Dowker et al.~of the swerves model, and their derivation of the proper time swerves diffusion equation, the question arose of how an underlying model of particle behaviour on a causal set could be formulated without relying on information about the approximating continuum spacetime. The association between longest chains in a causal set and geodesics in the continuum allowed such models to be created, two examples were given in Section~\ref{s:models}. The intrinsic models described in this thesis are by no means the only options, and simply give an idea of the wealth of possibilities available. The models proposed here (including the original swerves model) make no attempt to be physically realistic. A true description of particle motion in causal sets will have to await many further developments in the theory. The first step might involve developing a quantum mechanical rule for the trajectory evolution, or modelling particles as wavepackets. The models proposed do capture the crucial features of causal set theory -- they could, in a sense, be interpreted as a coarse grained description. Just as modelling a gas as an unrealistic collection of hard spheres proved enormously useful, so too do these models prove useful in determining continuum behaviour.

The proper time swerves diffusion equation, first given in~\cite{Dowker:2003hb} and derived in full in Section~\ref{s:massivepropertime}, is the unique continuum diffusion equation given an underlying Markovian, Poincar\'e invariant process causing random fluctuations in momentum. It is independent of the particular underlying models discussed -- those models and many more would result in continuum behaviour described by the swerves diffusion equation. One query that remained, was whether a given model was guaranteed to result in a finite diffusion constant. In Section~\ref{s:finitediffconstant} it was shown analytically that the swerves model does lead to a finite diffusion constant.

The swerves diffusion equation, expressed in terms of proper time, is difficult to compare with observations. Although bounds had been placed on the diffusion parameter by considering the non-relativistic limit, bounds in the fully relativistic case could not formerly be considered. The full inhomogeneous swerves diffusion equation in terms of cosmic time was derived in Section~\ref{s:massivecosmictime}. With this equation complete comparison with observations will be possible.

After exploring swerves analytically, the second task of this chapter was to explore the models numerically. The underlying models give rise to diffusion in the continuum limit. With simulations limited to quite small causal sets (far from the infinite sprinkling density of the continuum limit) it was not expected that the full diffusion behaviour of the models would be apparent. The results were, in fact, far better than anticipated. All three models clearly show diffusion like behaviour. A close investigation of the swerves model shows the model behaviour perfectly matches the swerves diffusion equation. It was even possible to determine an expression for the continuum diffusion parameter in terms of the parameters of the underlying causal set model. That the diffusion behaviour occurs even this far from the continuum limit is very interesting. There are also wider implications for causal set simulations -- the results here show that despite computational limits on causal set sizes, useful information about the continuum appearance of discrete spacetime can be obtained.

%% file: MasslessParticles.tex
\chapter{Massless particles}
\label{c:massless}

\begin{center}
\begin{quote}
\foreignlanguage{germanb}{
Die ganzen 50 Jahre bewusster Gr\"ubelei haben mich der Antwort der Frage "`{Was sind Lickhtquanten}"' nicht n\"aher gebracht. Heute glaubt zwar jeder Lump, er wisse es, aber er t\"auscht sich.}
%\selectlanguage{english}
\end{quote}
\end{center}
\begin{center}
\begin{quote}
A total of fifty years of careful reflection has not brought me any closer to answering the question `what are the quanta of light?'.  It's true, today, that many a fool believes he knows the answer, but he's mistaken. 
\end{quote}
\end{center}
\begin{flushright}
Einstein to Besso, 12th December 1951\footnote{Published in~\cite{Einstein:1972}, English translation from the French translation in~\cite{Einstein:1972} courtesy of Dr.~R.~M.~Pollard.}
\end{flushright}

An underlying spacetime discreteness results in diffusion in momentum and spacetime for massive particles. Does a similar diffusion occur for massless particles? For massive particles concrete models of particle propagation on a causal set were constructed, motivating the derivation of the swerves diffusion equation.  For massless particles such underlying models are unfortunately elusive. Consider a sprinkling into Minkowski spacetime: for any given element, $p$, there will almost surely be no element sprinkled on the future lightcone of $p$. A trajectory constructed, as in the massive case, by jumping from one element $p$ to another $q$, where $p\prec q$, will rarely be close to null. A close analogue to the future lightcone of $p$ can, however, be defined simply as the set of all elements $q$ linked to $p$: $p\link q$. These elements $q$ lie near the hyperboloid with proper time one Planck unit from $p$. This suggests that the simplest analogue of a null ray is a single link. More generally, $p\prec q$ is a good approximation to a section of a null ray if the causal interval between $p$ and $q$ is totally ordered~\cite{Sorkin:2009bp,Levichev:1987}, but it is still hard to see how a discrete Markovian process resulting in a close-to-null trajectory might be defined. It may be that options such as allowing particles to also move backward in time need to be considered. More generally, it is not surprising that attempts to model massless particles as classical point particles run into trouble. It is hoped that the study of massless fields on a causal set~\cite{Sorkin:2007qi,Dowker:2009} and general quantum field theory on a causal set~\cite{Johnston:2008za,Johnston:2009fr} will eventually lead to a model for massless particle propagation at a discrete level.

The current lack of an underlying model does not, fortunately, exclude the investigation of the potential effect of discreteness on photons in the continuum approximation. A massless diffusion equation can be derived in one of two ways: following the stochastic evolution on a manifold of states procedure, as for the massive case; or, simply taking a $m\rightarrow 0$ limit of the massive particle diffusion equation. As will be discussed in the following section, it turns out that the second method gives an incomplete result.

In Section~\ref{s:masslessaffine} the massless particle diffusion equation in terms of affine time is derived. As for massive particles, in order to make comparisons with observations an equation in terms of cosmic time is needed -- such an equation in derived in Section~\ref{s:masslesscosmictime}. 

The resulting massless particle model has two free parameters. In Section~\ref{s:CMBbounds} the effect of massless particle diffusion on the cosmic microwave background is studied, and bounds are placed on the values of the two free parameters. Finally, the consequences of the model for astrophysical spectra are discussed in Section~\ref{s:spectralline}, where it is found that the strong bounds from the cosmic microwave background rule out observable effects. The work in Sections~\ref{s:masslessaffine},~\ref{s:masslesscosmictime},~\ref{s:CMBbounds} was conducted in collaboration with Fay Dowker and Rafael Sorkin, and appears in~\cite{Philpott:2008vd}.

Note that the massless particles considered in this chapter are assumed to have no polarisation. The issue of polarisation is addressed in depth in Chapter~\ref{c:polarisation}.

%% file: MasslessAffineTimeEquation.tex
\section{Diffusion in affine time}
\label{s:masslessaffine}

The state space in the massless case differs from the massive case. For massive particles there is a probability distribution on $\mink\times\Lob$. For massless particles $\Lob$ becomes the light cone in momentum space defined by $p_{\mu}p^{\mu}=0$.
This cone will be denoted $\mathbb{H}_0^3$. Hereafter the momentum of a massless particle will be denoted $k^{\mu}$. Assuming that the photons under consideration are well described in a geometrical optics approximation (and thus have definite spacetime worldlines and momenta), the state space will be $\mathbb{M}^4\times\mathbb{H}_0^3$.

Since proper time vanishes along a lightlike worldline, it is clearly not a suitable time parameter for a massless diffusion process.  Instead, an affine time, $\lambda$, defined by
%%%%%%%%%%%%%%%%%%%%%%%%%%%%%%%%%%%%%%%%%%%%%%%%%%%%%%%%%%%%%%%%%%%%%%%%%%%%%%%%%%%%%%%%%%%%%%%%%%%%%%%%%%%%%%%%
\begin{equation}
dx^\mu = k^\mu d\lambda\,,
\end{equation}
%%%%%%%%%%%%%%%%%%%%%%%%%%%%%%%%%%%%%%%%%%%%%%%%%%%%%%%%%%%%%%%%%%%%%%%%%%%%%%%%%%%%%%%%%%%%%%%%%%%%%%%%%%%%%%%%
is used to parametrise the photon worldline.

In the massive particle case the density of microstates, $n$, was proportional to the determinant of the metric on the state space: $n\propto\sqrt{g}$. In the massless case, the induced metric on $\Lob_0$ is degenerate. A volume element on $\Lob_0$ can, however, still be defined. The four dimensional volume element $d^4 k$ of momentum space, together with the masslessness constraint, $k^\mu k_\mu=0$, gives the invariant volume element $d^4k\,\delta(k^\mu k_\mu)=d^3k/2k^0$ on $\Lob_0$, i.e. $n\propto 1/k^0$ in Cartesian coordinates. It turns out to be more useful, however, to work in polar coordinates on $\Lob_0$: $\{k, \theta, \phi\}$ where $k$ is the magnitude of the three momentum and $\theta$ and $\phi$ are the usual polar angles in momentum space.
In these coordinates, the density of states is $n \propto k \sin\theta$.

As with the massive particle case, the derivation of the massless diffusion equation begins with the current and continuity equations, Equation~\ref{e:current} and Equation~\ref{e:continuity}. The first step is to determine $K^{AB}$ and $u^A$. To begin, note that there is a unique, up to a constant factor, invariant vector field on $\Lob_0$ which is the momentum itself, $k^a$, i.e.~the vector with components $(k, 0, 0)$ in polar coordinates. This is absent in the massive case, where the momentum vector does not lie in the mass shell. Also, although there is no invariant metric on $\Lob_0$, there is
an invariant symmetric two-tensor, $k^a k^b$, again unique up to a constant factor.

From Equation~\ref{e:KABlim}, the spacetime component of $K^{AB}$ is
%%%%%%%%%%%%%%%%%%%%%%%%%%%%%%%%%%%%%%%%%%%%%%%%%%%%%%%%%%%%%%%%%%%%%%%%%%%%%%%%%%%%%%%%%%%%%%%%%%%%%%%%%%%%%%%%
\begin{eqnarray}
K^{\mu\nu} &=& \lim_{\Delta\lambda\rightarrow 0+}\left<\frac{\Delta x^{\mu}\Delta x^{\nu}}{2\Delta\lambda}\right>\nonumber\\
&=& \lim_{\Delta\lambda\rightarrow 0+} k^{\mu}k^{\nu}\Delta\lambda\nonumber\\
&=& 0. 
\end{eqnarray}
%%%%%%%%%%%%%%%%%%%%%%%%%%%%%%%%%%%%%%%%%%%%%%%%%%%%%%%%%%%%%%%%%%%%%%%%%%%%%%%%%%%%%%%%%%%%%%%%%%%%%%%%%%%%%%%%
$K^{AB}$ is positive semidefinite so $K^{\mu a} = 0$. Finally, the requirement that $K^{AB}$ be Lorentz invariant and translation invariant fixes the remaining component $K^{ab}$ to be proportional to $k^ak^b$:
%%%%%%%%%%%%%%%%%%%%%%%%%%%%%%%%%%%%%%%%%%%%%%%%%%%%%%%%%%%%%%%%%%%%%%%%%%%%%%%%%%%%%%%%%%%%%%%%%%%%%%%%%%%%%%%%
\begin{equation}
K^{AB}=
\begin{pmatrix}
0& 0\\
0& d_1k^ak^b
\end{pmatrix}\,,
\end{equation}
%%%%%%%%%%%%%%%%%%%%%%%%%%%%%%%%%%%%%%%%%%%%%%%%%%%%%%%%%%%%%%%%%%%%%%%%%%%%%%%%%%%%%%%%%%%%%%%%%%%%%%%%%%%%%%%%
where $d_1\geq0$ is a constant.

The vector $u^A$ is easily determined by looking individually at the components in
spacetime and momentum space. As in the massive particle case, the spacetime component $u^{\mu} = v^{\mu}$ by Equation~\ref{e:uA}, and $v^{\mu} = k^\mu$ by Equation~\ref{e:vAlim}.
In contrast to the massive case, there can be nonzero components of $u^A$ in the momentum space directions because
the momentum itself is an invariant vector. The momentum direction components are thus given by
$u^{a}=d_2k^a$, where $d_2$ is a constant. Working in polar coordinates the `position' vector $k^a$ on the cone $\Lob_0$
is simply $(k,0,0)$ where $k^2={k_0}^2$. Thus $u^A = (k^0,k^1,k^2,k^3,d_2k,0,0)$ on $\mink\times\Lob_0$.

Substituting the forms for $K^{AB}$ and $u^A$ into Equation~\ref{e:diffequA} gives the massless particle affine time equation:
%%%%%%%%%%%%%%%%%%%%%%%%%%%%%%%%%%%%%%%%%%%%%%%%%%%%%%%%%%%%%%%%%%%%%%%%%%%%%%%%%%%%%%%%%%%%%%%%%%%%%%%%%%%%%%%%
\begin{eqnarray}
\frac{\partial\rho_{\lambda}}{\partial\lambda} &=&
\partial_A\left(K^{AB}n\partial_B\left(\frac{\rho_{\lambda}}{n}\right)-u^A\rho_{\lambda}\right)\nonumber\\
&=& -k^{\mu}\frac{\partial\rho_{\lambda}}{\partial x^{\mu}} +
d_1\frac{\partial}{\partial k}\left(k^3\frac{\partial}{\partial
k}\left(\frac{\rho_{\lambda}}{k}\right)\right)\nonumber\\
&& -
d_2\frac{\partial}{\partial k}\left(k\rho_{\lambda}\right)\label{e:affine}\,.
\end{eqnarray}
%%%%%%%%%%%%%%%%%%%%%%%%%%%%%%%%%%%%%%%%%%%%%%%%%%%%%%%%%%%%%%%%%%%%%%%%%%%%%%%%%%%%%%%%%%%%%%%%%%%%%%%%%%%%%%%%

Lorentz invariance constrains the diffusion process such that, although there is diffusion in the energy of the photon, the direction of the photon is unchanged and it continues to propagate at the speed of light. Note that there are two parameters, $d_1$ and $d_2$, in this phenomenological model, rather than the single parameter of the massive particle model. 
There is not only a diffusion term but an independent drift term, arising from the existence of an invariant vector on $\Lob_0$. It will be shown further below that this extra term leads to the existence of power law equilibrium solutions of the diffusion equation.

It is not unreasonable to expect that a diffusion equation for massless particles could be obtained simply by taking the $m\rightarrow 0$ limit of the massive particle diffusion equation, Equation~\ref{e:mtau}. This, however, results in Equation~\ref{e:affine} with $d_2=0$, as there is no invariant vector in the massive case. It is interesting to note that there is this discontinuity in behaviour between very highly relativistic massive particles, and massless particles. It suggests that massive and massless particles are fundamentally different objects and will need to be described in completely different ways in causal set theory.

%% file: MasslessCosmicTimeEquation.tex
\section{Diffusion in cosmic time}
\label{s:masslesscosmictime}

As with massive particles, the massless particle diffusion equation is more useful for comparison with observations when expressed in terms of cosmic time. The massless cosmic time diffusion equation can be obtained by following the same argument as in the massive case (Section~\ref{s:massivecosmictime}).
It is assumed that the massless particle trajectory begins at $\lambda=0$ and $t=0$. 
The first step is to define a new larger state space incorporating the affine time: $\mathcal{M}^{\prime} = \mink\times\Lob_0\times\mathbb{R}$. In this larger space, a new current component can be defined:
%%%%%%%%%%%%%%%%%%%%%%%%%%%%%%%%%%%%%%%%%%%%%%%%%%%%%%%%%%%%%%%%%%%%%%%%%%%%%%%%%%%%%%%%%%%%%%%%%%%%%%%%%%%%%%%%
\begin{equation}
J^{\lambda}(t,x^i,k^a,\lambda) = \rho_{\lambda}\,.
\label{e:lambda}
\end{equation}
%%%%%%%%%%%%%%%%%%%%%%%%%%%%%%%%%%%%%%%%%%%%%%%%%%%%%%%%%%%%%%%%%%%%%%%%%%%%%%%%%%%%%%%%%%%%%%%%%%%%%%%%%%%%%%%%
The $t$ component of the current $J$ can  be expressed in terms of $J^{\lambda}$, using Equations~\ref{e:current} and~\ref{e:lambda}.
%%%%%%%%%%%%%%%%%%%%%%%%%%%%%%%%%%%%%%%%%%%%%%%%%%%%%%%%%%%%%%%%%%%%%%%%%%%%%%%%%%%%%%%%%%%%%%%%%%%%%%%%%%%%%%%%
\begin{eqnarray}
J^t(t,x^i,k^a,\lambda) &=& k J^{\lambda}\,,\\
\Rightarrow \rho_{\lambda}&=& \frac{J^t}{k}\,.
\end{eqnarray}
%%%%%%%%%%%%%%%%%%%%%%%%%%%%%%%%%%%%%%%%%%%%%%%%%%%%%%%%%%%%%%%%%%%%%%%%%%%%%%%%%%%%%%%%%%%%%%%%%%%%%%%%%%%%%%%%
The remaining components of the current can now be expressed in terms of $J^t$.
%%%%%%%%%%%%%%%%%%%%%%%%%%%%%%%%%%%%%%%%%%%%%%%%%%%%%%%%%%%%%%%%%%%%%%%%%%%%%%%%%%%%%%%%%%%%%%%%%%%%%%%%%%%%%%%%
\begin{eqnarray}
J^{i}\left(t,x^i,k^a,\lambda\right) &=& \frac{k^i}{k}J^t\,,\\
J^{a}\left(t,x^i,k^a,\lambda\right) &=& -d_1 k^a k^b
{k\sin\theta}\,\partial_b\left(\frac{
J^t}{k^2\sin\theta}\right)\nonumber\\&& + \frac{J^t}{k}d_2 k^a\,,
\end{eqnarray}
%%%%%%%%%%%%%%%%%%%%%%%%%%%%%%%%%%%%%%%%%%%%%%%%%%%%%%%%%%%%%%%%%%%%%%%%%%%%%%%%%%%%%%%%%%%%%%%%%%%%%%%%%%%%%%%%
where Equation~\ref{e:altform} has been used.
In polar coordinates the vector $k^a = (k, 0, 0)$,  and thus there is only one nonzero component of $J^{k^a}$
in the radial (energy) direction:
%%%%%%%%%%%%%%%%%%%%%%%%%%%%%%%%%%%%%%%%%%%%%%%%%%%%%%%%%%%%%%%%%%%%%%%%%%%%%%%%%%%%%%%%%%%%%%%%%%%%%%%%%%%%%%%%
\begin{eqnarray}
J^{k}\left(t,x^i,k^a,\lambda\right) &=& -d_1 k\,\frac{\partial
J^t}{\partial k} + \left(2d_1 + d_2\right)J^t\,.
\end{eqnarray}
%%%%%%%%%%%%%%%%%%%%%%%%%%%%%%%%%%%%%%%%%%%%%%%%%%%%%%%%%%%%%%%%%%%%%%%%%%%%%%%%%%%%%%%%%%%%%%%%%%%%%%%%%%%%%%%%
Since the affine time along the particle trajectories is unobservable, it can be integrated over, and a new integrated current, $\bar{J}$, can be defined. Integrating over the $t$ component gives
%%%%%%%%%%%%%%%%%%%%%%%%%%%%%%%%%%%%%%%%%%%%%%%%%%%%%%%%%%%%%%%%%%%%%%%%%%%%%%%%%%%%%%%%%%%%%%%%%%%%%%%%%%%%%%%%
\begin{eqnarray}
\bar{J}^t(t,x^i,k^a) &=& \int_0^\infty{J^t(t,x^i,k^a,\lambda)d\lambda}.
\end{eqnarray}
%%%%%%%%%%%%%%%%%%%%%%%%%%%%%%%%%%%%%%%%%%%%%%%%%%%%%%%%%%%%%%%%%%%%%%%%%%%%%%%%%%%%%%%%%%%%%%%%%%%%%%%%%%%%%%%%
The remaining current components can be integrated and re-expressed in terms of $\bar{J}^t$:
%%%%%%%%%%%%%%%%%%%%%%%%%%%%%%%%%%%%%%%%%%%%%%%%%%%%%%%%%%%%%%%%%%%%%%%%%%%%%%%%%%%%%%%%%%%%%%%%%%%%%%%%%%%%%%%%
\begin{eqnarray}
\bar{J}^{i}(t,x^i,k^a) &=& \frac{k^i}{k}\bar{J}^t\,,\\
\bar{J}^{k}(t,x^i,k^a) &=& -d_1 k\frac{\partial \bar{J}^t}{\partial k}
+ \left(2d_1 + d_2\right)\bar{J}^t\,.
\end{eqnarray}
%%%%%%%%%%%%%%%%%%%%%%%%%%%%%%%%%%%%%%%%%%%%%%%%%%%%%%%%%%%%%%%%%%%%%%%%%%%%%%%%%%%%%%%%%%%%%%%%%%%%%%%%%%%%%%%%
In the extended space $\mathcal{M}^{\prime}$, the continuity equation becomes
%%%%%%%%%%%%%%%%%%%%%%%%%%%%%%%%%%%%%%%%%%%%%%%%%%%%%%%%%%%%%%%%%%%%%%%%%%%%%%%%%%%%%%%%%%%%%%%%%%%%%%%%%%%%%%%%
\begin{equation}
\partial_{\alpha} J^{\alpha} = 0\,,
\end{equation}
%%%%%%%%%%%%%%%%%%%%%%%%%%%%%%%%%%%%%%%%%%%%%%%%%%%%%%%%%%%%%%%%%%%%%%%%%%%%%%%%%%%%%%%%%%%%%%%%%%%%%%%%%%%%%%%%
where $X^{\alpha}=\{X^A,\lambda\}$ are coordinates on $\mathcal{M}\times\mathbb{R}$.
Integrated over $\lambda$ the continuity equation becomes
%%%%%%%%%%%%%%%%%%%%%%%%%%%%%%%%%%%%%%%%%%%%%%%%%%%%%%%%%%%%%%%%%%%%%%%%%%%%%%%%%%%%%%%%%%%%%%%%%%%%%%%%%%%%%%%%
\begin{equation}
[J^{\lambda}]^{\infty}_{0} + \partial_t\bar{J}^t + \partial_i\bar{J}^i + \partial_a\bar{J}^a = 0\,.
\end{equation}
%%%%%%%%%%%%%%%%%%%%%%%%%%%%%%%%%%%%%%%%%%%%%%%%%%%%%%%%%%%%%%%%%%%%%%%%%%%%%%%%%%%%%%%%%%%%%%%%%%%%%%%%%%%%%%%%
$J^{\lambda}|_{\lambda =0}$ is zero for all $t>0$ and $J^{\lambda}$ tends to zero as
$\lambda$ approaches infinity for finite $t$. So for all $t>0$
%%%%%%%%%%%%%%%%%%%%%%%%%%%%%%%%%%%%%%%%%%%%%%%%%%%%%%%%%%%%%%%%%%%%%%%%%%%%%%%%%%%%%%%%%%%%%%%%%%%%%%%%%%%%%%%%
\begin{equation}
\partial_t \bar{J}^t + \partial_i \bar{J}^i + \partial_a \bar{J}^a = 0\,.
\end{equation}
%%%%%%%%%%%%%%%%%%%%%%%%%%%%%%%%%%%%%%%%%%%%%%%%%%%%%%%%%%%%%%%%%%%%%%%%%%%%%%%%%%%%%%%%%%%%%%%%%%%%%%%%%%%%%%%%
If $\bar{J}^t$ is renamed $\rho_t$ the massless particle cosmic time equation can be written:
%%%%%%%%%%%%%%%%%%%%%%%%%%%%%%%%%%%%%%%%%%%%%%%%%%%%%%%%%%%%%%%%%%%%%%%%%%%%%%%%%%%%%%%%%%%%%%%%%%%%%%%%%%%%%%%%
\begin{eqnarray}
\frac{\partial\rho_{t}}{\partial t} &=& -\partial_i \bar{J}^i - \partial_a
\bar{J}^a\nonumber\\  \label{e:masslesscosmictime}
&=& -\frac{k^i}{k}\partial_i\rho_{t} - \left(d_1 +
d_2\right)\frac{\partial \rho_{t}}{\partial k} + d_1 k
\frac{\partial^2 \rho_{t}}{\partial k^2}\;.
\end{eqnarray}
%%%%%%%%%%%%%%%%%%%%%%%%%%%%%%%%%%%%%%%%%%%%%%%%%%%%%%%%%%%%%%%%%%%%%%%%%%%%%%%%%%%%%%%%%%%%%%%%%%%%%%%%%%%%%%%%

\subsection{Power law equilibrium solutions}

It is interesting to note that negative values of the drift parameter $d_2$ allow for power law equilibrium solutions of the massless particle cosmic time diffusion equation. For homogeneous distributions Equation~\ref{e:masslesscosmictime} becomes
\begin{equation}
\frac{\partial\rho_{t}}{\partial t} =  - \left(d_1 +
d_2\right)\frac{\partial \rho_{t}}{\partial k} + d_1 k
\frac{\partial^2 \rho_{t}}{\partial k^2}\;.
\end{equation}
Equilibrium solutions of this equation must satisfy
\begin{equation}
0 =  - \left(d_1 +
d_2\right)\frac{\partial \rho_{t}}{\partial k} + d_1 k
\frac{\partial^2 \rho_{t}}{\partial k^2}\;.\label{e:equil}
\end{equation}
Equation~\ref{e:equil} has a power law solution
\begin{equation}
\rho_{t}\propto k^{\frac{2d_1 + d_2}{d_1}}\;.
\end{equation}
If a small energy cut off $k_{min}$ is assumed, this solution is normalisable when the exponent is less than -1. Since the diffusion parameter $d_1$ is positive this can only occur for negative values of $d_2$. It can be conjectured that if $(2d_1+d_2)/d_1<-1$ (equivalently $|d_2|>3d_1$) any normalised distribution will tend to this power law equilibrium solution at late times.

%% file: CMBBounds.tex
\section{Diffusion and drift in the cosmic microwave background}
\label{s:CMBbounds}

In developing a phenomenological model the aim is to provide a model for currently unexplained observations or to suggest new observations that might be made to test a theory. Before proposing new observations or exotic experiments, however, one should constrain the model as tightly as possible based on what is already known. The massless particle momentum diffusion discussed in the preceding sections has two free parameters: a positive diffusion constant $d_1$ and a `drift' constant $d_2$, which may be either positive or negative. If the photons we encounter in our everyday lives experienced large amounts of diffusion or drift, it would of course have already been noticed. The motivation for the massless momentum diffusion model, however, lies in the underlying discreteness of causal set quantum gravity -- the resulting effects are expected to only be apparent on very small scales, outside the range of current experiments. How then can the parameters be constrained? The key is to look at photons that have been travelling long enough for the diffusion and drift to `accumulate' to an observable level. 

The cosmic microwave background (CMB) seems an ideal testing ground: not only are
its photons the `oldest' we can observe, but its spectrum has been determined with great precision. The photons in the CMB have been `free streaming' since the surface of last scattering at a redshift of $z\sim1100$ -- approximately $13.7$ billion years. When the universe became transparent
at recombination, the photons that are now observed as the CMB would have had a blackbody spectrum with a
temperature of $3000\,K$ (see for example~\cite{Kolb:1990}). As the universe expanded the photons were both stretched and diluted, a process that neatly preserves the blackbody nature of the spectrum and simply lowers the temperature. Current observations of the CMB yield a temperature of $2.728\pm0.004 \,K$ and measure the spectrum to be Planckian (blackbody) over the $2-21cm^{-1}$ frequency range to within a weighted root mean square (rms) deviation of only 50 parts per million (ppm) of the peak brightness~\cite{Fixsen:1996nj}. Any diffusion and drift in energy would have distorted the CMB spectrum -- the fact that the CMB photons have travelled so far but remained so perfectly thermal allows the parameters in the model to be constrained very tightly.

\subsection{Modelling diffusion in the CMB}
\label{ss:noexp}

The derivation of the massless cosmic time diffusion equation assumed
spacetime to be Minkowskian. Although the expansion of the universe can be incorporated into the model (and will be in Section~\ref{ss:expuni} below) a first bound on the parameters will be obtained by simply ignoring the expansion. In a non-expanding universe the temperature of the CMB would have remained constant from the surface of last scattering to today. To allow comparison with data, it will in fact be assumed that the CMB was emitted with the temperature observed today, $T=2.728K$.

If a blackbody spectrum is evolved according to the massless cosmic time diffusion equation, the final distribution will no longer be blackbody. The deviation from blackbody can be determined and, by requiring it to be within the current deviation allowed by observations, bounds can be placed on the drift and diffusion parameters. Specifically, the initial Planckian spectrum, expressed as a number density of
photons per unit spatial volume per unit energy, is
%%%%%%%%%%%%%%%%%%%%%%%%%%%%%%%%%%%%%%%%%%%%%%%%%%%%%%%%%%%%%%%%%%%%%%%%%%%%%%%%%%%%%%%%%%%%%%%%%%%%%%%%%%%%%%%%
\begin{equation}
  \rho(k, t=0) = 8\pi\frac{k^2}{\exp\left(\frac{k}{T}\right) - 1}\,,
\end{equation}
%%%%%%%%%%%%%%%%%%%%%%%%%%%%%%%%%%%%%%%%%%%%%%%%%%%%%%%%%%%%%%%%%%%%%%%%%%%%%%%%%%%%%%%%%%%%%%%%%%%%%%%%%%%%%%%%
with a temperature $T=2.728K$. This distribution is evolved over a time interval equal to that since the
surface of last scattering ($1\times 10^{60}$ in Planck units) via the homogeneous massless cosmic time diffusion equation
%%%%%%%%%%%%%%%%%%%%%%%%%%%%%%%%%%%%%%%%%%%%%%%%%%%%%%%%%%%%%%%%%%%%%%%%%%%%%%%%%%%%%%%%%%%%%%%%%%%%%%%%%%%%%%%%
\begin{equation} \label{e:homcos}
  \frac{\partial\rho_t}{\partial t} = -
  \left(d_1+d_2\right)\frac{\partial\rho_{t}}{\partial k} +
  d_1k\frac{\partial^2\rho_{t}}{\partial k^2}  \ ,
\end{equation}
%%%%%%%%%%%%%%%%%%%%%%%%%%%%%%%%%%%%%%%%%%%%%%%%%%%%%%%%%%%%%%%%%%%%%%%%%%%%%%%%%%%%%%%%%%%%%%%%%%%%%%%%%%%%%%%%
using the \texttt{MATLAB} numerical pde solver \texttt{pdepe}. 

\subsubsection{Boundary conditions}

Some questions must be addressed when choosing a boundary condition at $k=0$ for this integration. What happens to a photon as its momentum approaches zero? Do photons leak away through the tip of the null cone in momentum space? Physically, the photon concept employed by the model breaks down as the wavelength tends to infinity, because the geometrical optics approximation fails. This suggests that an `absorbing boundary condition',
$\rho(E)=0$, is appropriate at $E=0$. As it happens, the current is
%%%%%%%%%%%%%%%%%%%%%%%%%%%%%%%%%%%%%%%%%%%%%%%%%%%%%%%%%%%%%%%%%%%%%%%%%%%%%%%%%%%%%%%%%%%%%%%%%%%%%%%%%%%%%%%%
\begin{equation}
  J = (2d_1+d_2)\rho_t - d_1 k \pd{\rho_t}{k}\,,
\end{equation}
%%%%%%%%%%%%%%%%%%%%%%%%%%%%%%%%%%%%%%%%%%%%%%%%%%%%%%%%%%%%%%%%%%%%%%%%%%%%%%%%%%%%%%%%%%%%%%%%%%%%%%%%%%%%%%%%
and thus (so long as $\partial\rho_t/\partial k$ remains finite) the absorbing boundary condition $\rho_t=0$ is equivalent to the `reflecting boundary condition', $J=0$, at $E=0$. In the simulations that follow, a reflecting boundary condition is used at the upper boundary $k=k_{max}$.

\subsubsection{Calculating the deviation}

The evolved spectrum was converted from a number density per unit volume per unit frequency to
a spectral radiance (energy per unit area per unit time per unit frequency per steradian) as used in the analysis of the COBE FIRAS data. This allowed the deviation from Planckian to be compared with the quoted 50ppm of the peak brightness.
The first step was to fit a Planck spectrum to the evolved spectral radiance using the least squares method. Specifically, the  \texttt{MATLAB} function \texttt{fminsearch} was used to find the temperature that minimised the sum of the squared differences between the evolved distribution and the fitted Planck spectrum. Looking for the best fit Planck spectrum rather than comparing with the initial $2.728K$ spectrum, allowed for
the possibility that the diffusion changes the temperature of the CMB in a way that may be reconciled with observation. As it happens, it was found that the temperature of the best fit Planck spectrum was very close to the initial temperature in cases where the deviation is within the allowed
tolerance. For example the choice of parameters $d_1 = 5\times 10^{-97}$ and $d_2=1\times 10^{-96}$ gives a best fit temperature of $2.7281 K$, indistinguishable from the current observed temperature of $2.728\pm0.004 K$.

Finally, the rms deviation between the fitted Planckian spectrum and the evolved spectrum
%%%%%%%%%%%%%%%%%%%%%%%%%%%%%%%%%%%%%%%%%%%%%%%%%%%%%%%%%%%%%%%%%%%%%%%%%%%%%%%%%%%%%%%%%%%%%%%%%%%%%%%%%%%%%%%%
\begin{equation}
\textrm{rms} = \sqrt{\frac{\sum_i(s_i-f_i)^2}{N}}\;,
\end{equation}
%%%%%%%%%%%%%%%%%%%%%%%%%%%%%%%%%%%%%%%%%%%%%%%%%%%%%%%%%%%%%%%%%%%%%%%%%%%%%%%%%%%%%%%%%%%%%%%%%%%%%%%%%%%%%%%%
where $s_i$ is the value of the evolved spectrum at a point $i$, $f_i$ is the value of the fitted spectrum, and $N$ is the total number of data points, was calculated over the frequency range
$2$$-$$21cm^{-1}$  (energy range $4\times10^{-23}-4\times10^{-22}J$) with all points weighted equally. After calculating the peak brightness of the evolved spectrum, this rms deviation can be compared  to the allowed tolerance of 50 parts per million of the peak brightness. This process
was repeated for a range of values of the parameters $d_1$ and $d_2$.

\subsubsection{Results}

\begin{figure}[p]
\begin{center}
\subfigure[The initial Planck spectrum at $2.728\,K$ (dotted line), evolved spectrum (dashed line) and best fit Planck spectrum at $2.851\,K$ (solid line).]{
\includegraphics[width=0.8\textwidth]{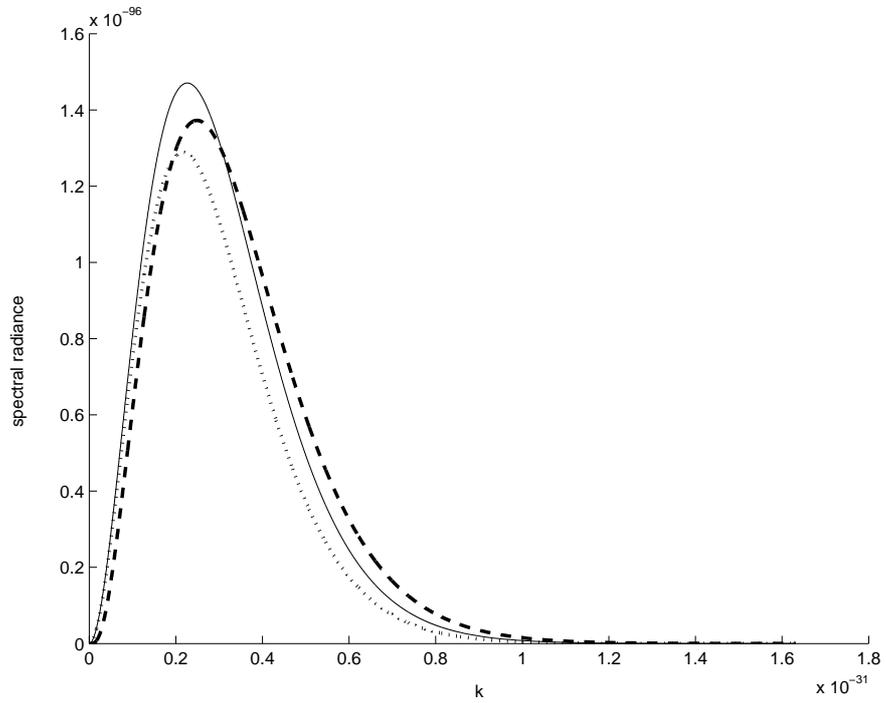}}
\hspace{10mm}
\subfigure[The difference between the initial spectrum and the evolved spectrum (dotted line), and the best fit spectrum and the evolved spectrum (dashed line).]{
\includegraphics[width=0.8\textwidth]{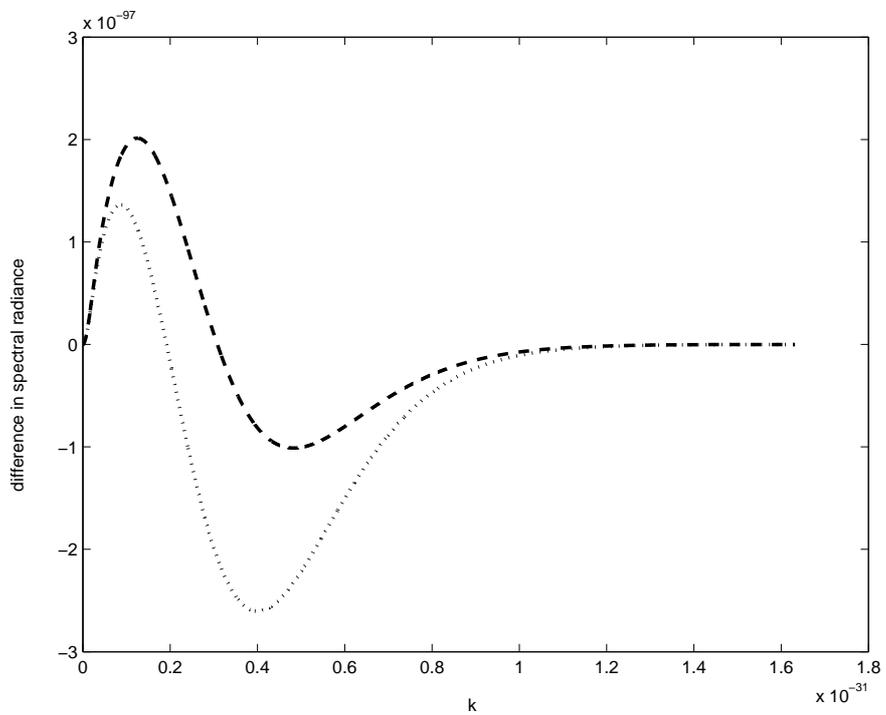}}
\caption[Example Planck spectrum evolved according to the diffusion equation]{Comparison of evolved and best fit Planck spectrum for $d_1 = 1\times 10^{-93},\,d_2 = 1\times 10^{-93}$. Energy $k$ and spectral radiance are in Planck units.}\label{f:egevolspec}
\end{center}
\end{figure}

For the evolved spectrum to be close enough to Planckian to be within the allowed tolerance, the deviation must be essentially invisible to the eye.  Before placing bounds on $d_1$ and $d_2$ it is useful therefore to get a general feel for how a blackbody spectrum evolves under Equation~\ref{e:homcos} by looking at values of $d_1$ and $d_2$ that result in deviations well outside the allowed tolerance. Figure~\ref{f:egevolspec} shows the evolved distribution and best fit Planck spectrum for $d_1 = 1\times 10^{-93},\,d_2 = 1\times 10^{-93}$. The positive drift parameter $d_2$ clearly shifts the spectrum to the right.  With these large values of $d_1$ and $d_2$ the difference between the best fit Planck spectrum ($2.851\,K$) and the initial Planck spectrum ($2.728\,K$) is quite considerable. However, no amount of temperature change will disguise the translation of the Planck spectrum -- the diffusion and drift are clearly visible.

To place bounds on the diffusion and drift constants they were first examined separately,
varying $d_1$ with $d_2 = 0$ and varying $d_2$ with $d_1 = 0$.
When $d_1=0$ Equation~\ref{e:homcos} can be solved exactly:
%%%%%%%%%%%%%%%%%%%%%%%%%%%%%%%%%%%%%%%%%%%%%%%%%%%%%%%%%%%%%%%%%%%%%%%%%%%%%%%%%%%%%%%%%%%%%%%%%%%%%%%%%%%%%%%%
\begin{equation}
  \rho_t(k, t) = \rho_0(k - d_2 t)\,,
\end{equation}
%%%%%%%%%%%%%%%%%%%%%%%%%%%%%%%%%%%%%%%%%%%%%%%%%%%%%%%%%%%%%%%%%%%%%%%%%%%%%%%%%%%%%%%%%%%%%%%%%%%%%%%%%%%%%%%%
i.e.~the spectrum just translates at a constant speed. For $d_2$ negative, this is inconsistent with the boundary condition $\rho_t=0$ at $k=0$.  However, in this case one can implement an absorbing boundary condition trivially: simply cut off the translated distribution at $k=0$.
The exact solution is also useful as a method of checking the level of numerical error in the simulations. Since the deviations from Planckian that are being investigated are so small there is the possibility that the rms deviation calculated is significantly affected by numerical error. Fortunately, this is not the case -- the difference between the exact solution and the numerically evolved solution for $d_1=0$ is several orders of magnitude smaller than the deviation from Planckian.\footnote{Rerunning all the simulations in this section using the \texttt{Mathematica} numerical solver \texttt{NDSolve} yields identical results further suggesting that numerical error and integration step size do not have a significant effect on the bounds.} For example, if $d_2 = \pm4\times 10^{-96}$ the rms deviation from the best fit Planck spectrum is $5\times 10^{-101}$ ($5\times 10^{-5}$ peak brightness) for both the exact and the numerical solution. The rms deviation
\textit{between} the exact and numerical solutions is $4\times 10^{-104}$. This also demonstrates that the $\rho=0$ boundary condition imposed on the numerical solution, although inconsistent with the
exact solution when $d_2<0$, does not introduce noticeable errors for the values of $d_2$ that are of concern here.

\begin{figure}[t]
\begin{center}
\includegraphics[width=0.6\textwidth]{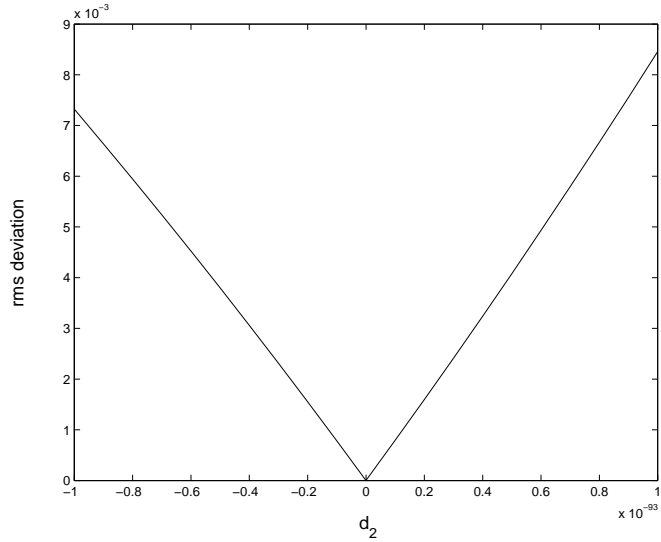}
\caption[Varying $d_2$ with $d_1=0$]{The rms deviation of the simulated spectrum from
Planckian as a proportion of the spectrum peak, varying $d_2$, $d_1=0$.}
\label{f:d2only}
\end{center}
\end{figure}

The rms deviation over a range of values of $d_2$ with $d_1=0$ is shown in Figure~\ref{f:d2only} (plotted from the exact solution).
\begin{figure}[t]
\begin{center}
\includegraphics[width=0.6\textwidth]{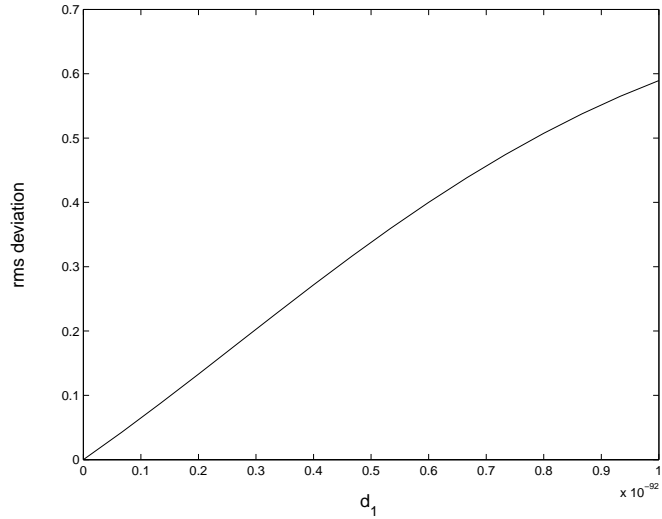}
\caption[Varying $d_1$ with $d_2=0$]{The rms deviation of the simulated spectrum from
Planckian as a proportion of the spectrum peak, varying $d_1$, $d_2=0$.}
\label{f:d1only}
\end{center}
\end{figure}

For $d_2=0$ and $d_1>0$ the equation can only be solved numerically. The results in this case are shown in Figure~\ref{f:d1only}. 
In both cases the deviation from Planckian increases approximately linearly with increasing magnitude
of the parameters. The simulations suggest that for the deviation from Planckian of the CMB to be within the allowed $5\times 10^{-5}$ of the peak brightness the diffusion constant $d_1$ must be less than
approximately $7\times10^{-97}$ if $d_2 = 0$, and the drift parameter $d_2$ must fall within the range
$-5\times10^{-96}<d_2<5\times10^{-96}$ if $d_1 = 0$. Converting to SI units this gives the bounds:
%%%%%%%%%%%%%%%%%%%%%%%%%%%%%%%%%%%%%%%%%%%%%%%%%%%%%%%%%%%%%%%%%%%%%%%%%%%%%%%%%%%%%%%%%%%%%%%%%%%%%%%%%%%%%%%%
\begin{eqnarray}
  &d_1&< 3\times10^{-44}kgm^2s^{-3}\,,\\
  -2\times10^{-43}<&d_2&<2\times10^{-43}kgm^2s^{-3} .
\end{eqnarray}
%%%%%%%%%%%%%%%%%%%%%%%%%%%%%%%%%%%%%%%%%%%%%%%%%%%%%%%%%%%%%%%%%%%%%%%%%%%%%%%%%%%%%%%%%%%%%%%%%%%%%%%%%%%%%%%%

\begin{figure}[t]
\begin{center}
\includegraphics[width=0.6\textwidth]{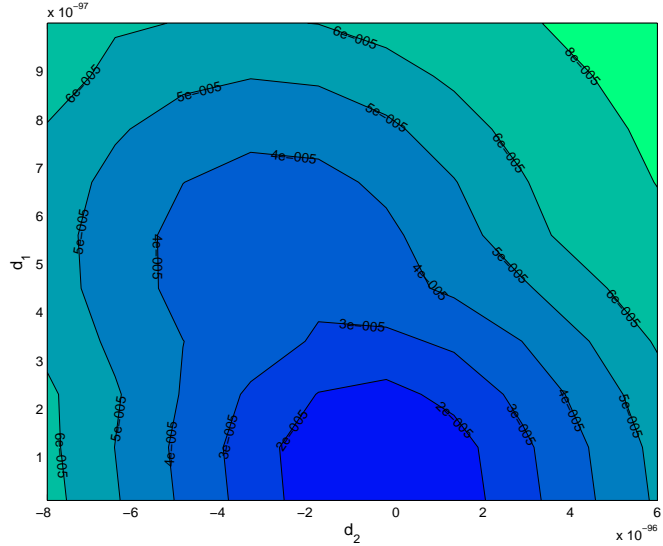}
\caption[Varying both $d_1$ and $d_2$]{The rms deviation of the simulated spectrum from
Planckian as a proportion of the spectrum peak, varying both
$d_1$ and $d_2$. Values of $d_1$ and $d_2$ within the
$5e$$-$$005$ contour give a spectrum that is Planckian to within 50ppm of the
peak brightness.}
\label{f:d1d22pt7K}
\end{center}
\end{figure}

If both $d_1$ and $d_2$ are nonzero the picture is somewhat more complicated, but similar bounds result. The results in this general situation are shown in Figure~\ref{f:d1d22pt7K}. The allowed region (within the $5e-005$ contour) is seen to be not quite symmetric in the drift parameter $d_2$. Placing concrete bounds on $d_1$ and $d_2$ from Figure~\ref{f:d1d22pt7K} is difficult, since the allowed values of $d_1$ depend on the value of $d_2$ and vice versa -- the bounds are, however, of the order of magnitude given above.

\subsection{The expanding universe}
\label{ss:expuni}
 
To determine the bounds on the drift and diffusion parameters, given in Section~\ref{ss:noexp} above, the effect of the expansion of the universe on the CMB was ignored -- it was assumed that the CMB had remained at a temperature of $\sim2.7\,K$ from the surface of last scattering up to today.  This is of course not the
case. At the surface of last scattering the CMB had a temperature of about $3000\,K$. As the universe expanded the individual photons were stretched along with the space, and correspondingly diluted, leaving the
$2.7K$ spectrum observed today. It turns out that the expansion has essentially no effect on the model: the
distribution in the expanding universe can be deduced easily from the nonexpanding one and
the bounds derived from the nonexpanding simulation change only slightly.
 
Before including the effect of cosmological expansion in the diffusion equation, a differential equation that describes the cooling of the CMB needs to be found; the following is a standard result in cosmology. Consider the usual continuity equation
\begin{equation}
\pd{\rho}{t} = -\partial_i(\rho v^i)\,,
\end{equation}
for flux $\rho v^i$. In this case we are interested in the flux in momentum space of the CMB photons and thus $v$ has a single component in the energy direction $v = \partial k/\partial t$. This gives the continuity equation
\begin{equation}
\pd{\rho_t}{t} = -\pd{}{k}\left(\rho_t\pd{k}{t}\right)\,.
\end{equation}
Under expansion the CMB photons are stretched such that $k\sim 1/a$, where $a$ is the scale factor. Thus, $\partial k/\partial t = -\dot{a}/a^2 = -k\dot{a}/a$, and the continuity equation becomes
\begin{eqnarray}
\pd{\rho_t}{t} &=& -\pd{}{k}\left(-\rho_t k \frac{\dot{a}}{a}\right)\nonumber\\
&=& \frac{\dot{a}}{a}\pd{}{k}\left(\rho_t k\right)\,.
\label{e:expeqn}
\end{eqnarray}

Consider the distribution
\begin{equation}
\rho_0(k,t) = \frac{8\pi}{a_0^3}\frac{a^3k^2}{\exp\left(\frac{a k}{a_0 T_0}\right)-1}\,,
\label{e:planckexp}
\end{equation}
where $a_0$ is the scale factor of the universe today. Photons in the CMB are not only stretched, but also diluted by the expansion of the universe. The distribution in Equation~\ref{e:planckexp} can be seen as a Planck spectrum where the normalisation depends on the scale factor $a$, to take into account the photon dilution. Equation~\ref{e:planckexp} is indeed a solution to the expansion equation, Equation~\ref{e:expeqn}. This confirms, as mentioned earlier, that the blackbody nature of the CMB is preserved during the expansion of the universe. 

If the expansion effect is included in the diffusion equation, Equation~\ref{e:homcos}, the equation becomes
\begin{equation}
\pd{\rho_t}{t} = -(d_1+d_2)\pd{\rho_t}{k} +d_1 k\pdsq{\rho_t}{k}+\frac{\dot{a}}{a}\pd{}{k}\left(\rho_t k\right)\,.
\label{e:expdiff}
\end{equation}
Now consider defining a new variable $\tilde{k}=k\, a/a_0 $ and a new density function $\tilde{\rho}(\tilde{k})=a_0\rho_t(k)/a$. Expressed in these variables the spectrum, Equation~\ref{e:planckexp}, becomes
\begin{equation}
\tilde{\rho}_0(\tilde{k},t) = 8\pi\frac{\tilde{k}^2}{\exp{\left(\frac{\tilde{k}}{T_0}\right)-1}}\,,
\end{equation} 
which is constant in time.

Expressing Equation~\ref{e:expdiff} in terms of $\tilde{\rho}$ and $\tilde{k}$ gives
\begin{eqnarray}
LHS &=& \left(\pd{}{t} + \pd{\tilde{k}}{t}\pd{}{\tilde{k}}\right)\left(\frac{a}{a_0}\tilde{\rho}\right)\nonumber\\
&=&  \left(\pd{}{t} + \frac{\dot{a}}{a}\tilde{k}\pd{}{\tilde{k}}\right)\left(\frac{a}{a_0}\tilde{\rho}\right)\nonumber\\
&=& \frac{\dot{a}}{a}\tilde{\rho} + \frac{a}{a_0}\pd{\tilde{\rho}}{t} +\frac{\dot{a}}{a_0}\tilde{k}\pd{\tilde{\rho}}{\tilde{k}}\nonumber\,,\\
RHS &=& -(d_1+d_2)\frac{a^2}{a_0^2}\pd{\tilde{\rho}}{\tilde{k}} + d_1\tilde{k}\frac{a^2}{a_0^2}\pdsq{\tilde{\rho}}{\tilde{k}}+\frac{\dot{a}}{a_0}\pd{}{\tilde{k}}\left(\tilde{k}\tilde{\rho}\right)\nonumber\,,\\
\Rightarrow \pd{\tilde{\rho}}{t} &=& -(d_1+d_2)\frac{a}{a_0}\pd{\tilde{\rho}}{\tilde{k}} + d_1\frac{a}{a_0}\tilde{k}\pdsq{\tilde{\rho}}{\tilde{k}}\,.
\end{eqnarray}

Defining a new time variable $t^{\prime}$ such that $dt^{\prime}/dt = a/a_0$ reduces this to
\begin{equation}
\pd{\tilde{\rho}}{t^{\prime}} = -(d_1+d_2)\pd{\tilde{\rho}}{\tilde{k}} + d_1\tilde{k}\pdsq{\tilde{\rho}}{\tilde{k}}\,,
\end{equation}  
exactly the same form as the non-expanding diffusion equation, Equation~\ref{e:homcos}.

\begin{figure}[t]
\begin{center}
\includegraphics[width=0.6\textwidth]{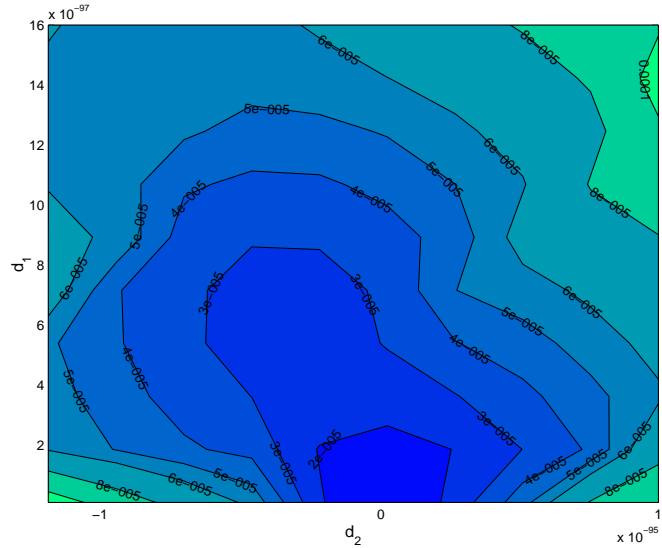}
\caption[Varying both $d_1$ and $d_2$ for an expanding universe]{The rms deviation of the simulated spectrum from
Planckian as a proportion of the spectrum peak, varying both
$d_1$ and $d_2$ in the expanding universe. Values of $d_1$ and $d_2$ within the
$5e$$-$$005$ contour give a spectrum that is Planckian to within 50ppm of the
peak brightness.}
\label{f:d1d2exp}
\end{center}
\end{figure}

For a matter dominated FRW universe $a\sim t^{2/3}$, i.e. $a(t) = a_0 t^{2/3}/t_0^{2/3}$, where $t_0$ is the current value of $t$. Since $dt^{\prime}/dt = a/a_0$ integration gives
\begin{equation}
t^{\prime} = \frac{3}{5}\frac{t^{5/3}}{t_0^{2/3}} + \textrm{const}\,.
\end{equation}
Since the time since the surface of last scattering ($t=10^{60}$) is of the same order as the age of the universe $t_0$, this effectively gives $t^{\prime}\sim 3/5 t$. In other words, including expansion is essentially the same as running the non-expanding simulations for $3/5$ of the time. As such, the bounds found in Section~\ref{ss:noexp} would not be expected to change greatly. For completeness the simulations have been run for the expanding case, and the results are shown in Figure~\ref{f:d1d2exp} -- the order of magnitude of the bounds is indeed the same.

%% file: SpectralLineShift.tex
\section{The shifting of spectral lines}
\label{s:spectralline}

The blackbody nature of the cosmic microwave background placed strong bounds on the magnitude of the diffusion and drift parameters. It is worthwhile, however, looking at other potential astrophysical consequences of the diffusion and drift: are observable effects ruled out by our current bounds, or do other sources allow even stronger bounds to be derived?
Here the effect of drift and diffusion on the spectral lines of distant sources will be investigated. 

Diffusion will result in a broadening of spectral lines. The energy dependence of the diffusion term in Equation~\ref{e:masslesscosmictime} implies higher energy lines will experience greater broadening than low energy lines. The difficulty is, however, that spectral lines from distant sources will experience line broadening due to a number of other effects. For example, the varying velocities of atoms in a gas cloud source will result in thermal Doppler broadening of the spectral lines. A calculation of the exact amount of broadening due to diffusion would not be trivial. Given the very tight bound on the diffusion parameter $d_1<7\times10^{-97}$, together with broadening due to other effects and the inherent spectral resolution (discussed further below), it seems unlikely that broadening due to diffusion would be observable. In this section I will instead concentrate on drift.

The drift term will shift all lines in a spectrum by a fixed amount. Spectral lines from astrophysical sources are, of course, already shifted with respect to laboratory spectra due to the expansion of the universe. The expansion of the universe results in lines shifted by a fixed ratio, the redshift $z$:
\begin{equation}
1+z = \frac{\lambda_{obs}}{\lambda_{e}}\,,
\end{equation}
where $\lambda_{obs}$ is the wavelength we observe, and $\lambda_{e}$ is the wavelength at which the light was emitted. Since the drift is independent of wavelength, this effect could be easily distinguished from redshifting. In fact, if drift occurs, the calculated redshift for a source would depend on which spectral line was used. 
Assuming that the diffusion constant is zero, the drift will shift spectral lines by an energy $\Delta k = d_2 t$, where $t$ is the time of travel of the photons. If the line has also been redshifted the observed line will have energy
\begin{eqnarray}
E_{obs} &=& E_{redshift}+d_2 t\nonumber\\
&=&\frac{E_{emit}}{1+z}+d_2t\,.
\end{eqnarray}
If drift occurs, but is not taken into account, observations would lead to the calculation of an apparent redshift $z^{\prime}$ that differed from the actual redshift:
\begin{eqnarray}
1+z^{\prime}&=&\frac{E_{emit}}{E_{obs}}\nonumber\\
&=&(1+z)\frac{E_{emit}}{E_{emit}+(1+z)d_2 t}\,.\label{e:newredshift}
\end{eqnarray}

\begin{figure}[t]
\begin{center}
\begin{minipage}{\textwidth}
\includegraphics[width=\textwidth]{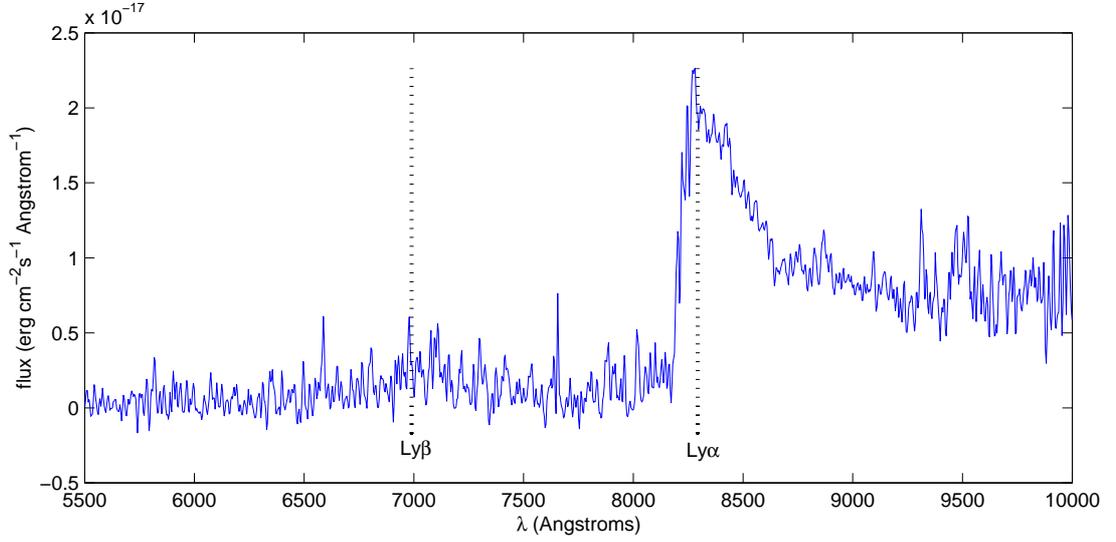}
\caption[Example quasar spectrum]{The spectrum of quasar SDSSp J083643.85+005453.3 at redshift $z=5.82$.\footnotemark}
\label{f:spectrum}
\end{minipage}
\end{center}
\end{figure}
\footnotetext{Many thanks to Rick White (Space Telescope Science Institute) and Bob Becker (University of California Davis) for providing the data to plot this spectrum.}

As an example, consider the quasar SDSSp J083643.85+005453.3 with redshift $z=5.82\pm0.02$~\cite{Fan:2001ff}, whose spectrum is shown in Figure~\ref{f:spectrum}. A prominent line in the spectra of quasars is the hydrogen Lyman-$\alpha$ line, which has a wavelength of $121.6nm$. For a quasar at $z=5.82$ this line is redshifted to $829.3nm$, or working in energies
\begin{eqnarray}
E_{emit} &=& 1.634\times10^{-18}J,\\
E_{redshift} &=& 2.395\times10^{-19}J\,.
\end{eqnarray}

To determine how much the Lyman-$\alpha$ line would be shifted by the drift, it is necessary to know the time it has taken for the light to travel from the quasar. The cosmic time-redshift relation
\begin{equation}
t_u = \frac{2}{3H_0\Omega_m^{1/2}(1+z)^{3/2}}\,,
\end{equation}
(see for example~\cite{Bergstrom:2006}) gives the approximate age of the universe when photons at a redshift of $z$ were emitted. Take the Hubble parameter to be $H_0 = 70.5kms^{-1}Mpc^{-1}$, the matter density to be $\Omega_M = 0.274$ and the current age of the universe to be $t_0 = 13.72\times10^9$ years~\cite{Hinshaw:2008kr}. The age of the universe when the photons from the quasar were emitted is $t_u = 9.64\times 10^8$ years ($2.25\times10^{59}$ in Planck units), and thus they have been travelling for $t_0-t_u = 1.28\times 10^{10}$ years ($2.98\times 10^{60}$ in Planck units). 
In Section~\ref{s:CMBbounds} the drift parameter was constrained to be $-5\times 10^{-96}<d_2<5\times 10^{-96}$. These values give a drift of magnitude 
\begin{eqnarray}
\Delta E &=& d_2 t\nonumber\\
&=& 1.49\times 10^{-35}\nonumber\\
&=& 7.30\times 10^{-26}J\,.
\end{eqnarray}
This drift is seven orders of magnitude smaller than the redshifted energy $2.395\times10^{-19}J$. 
When measuring the spectra of the quasar, Fan et al.~\cite{Fan:2001ff} note they have a spectral resolution of about $20$\AA. A remeasurement of the spectra~\cite{Becker:2001ee} gives a much better resolution of about $2$\AA. If the measured Lyman-$\alpha$ line is $8293\pm2$\AA, this corresponds in energy to approximately $2.395\times10^{-19}\pm1\times10^{-22}J$. The drift is thus much smaller than even the improved spectral resolution. 

If $z=5.82$ is the true redshift of the quasar, the measured redshift from Equation~\ref{e:newredshift} due to the shift in the Lyman-$\alpha$ line would be imperceptibly different. If one were to also consider the Lyman-$\beta$ line at $102.5nm$, the difference between the redshifts calculated using Lyman-$\alpha$ and Lyman-$\beta$ would be of the order $10^{-7}$, much smaller than the stated uncertainty of $0.02$.

The effect of drift would be more apparent for more distant objects. The calculations given here for the high redshift quasar at $z=5.82$ do however seem to rule out any observation of the drift in spectra in the near future.
It is certainly clear that the bounds placed on the drift parameter through consideration of the CMB are much stronger than any that could currently be obtained from spectral lines. This is of course hardly surprising given the combination of the very accurate measurements of the CMB and the redshift of $z\sim 1000$.

%% file: MasslessConclusion.tex
\section{Concluding remarks}
\label{s:masslessconclusion}

The assumption of Lorentz invariance is very powerful: despite the lack of an underlying microscopic model for massless particles, Lorentz invariance allows a concrete diffusion model to be derived in the continuum limit. Such models were derived in Sections~\ref{s:masslessaffine} and~\ref{s:masslesscosmictime} in terms of affine time and cosmic time respectively. For massless particles without internal degrees of freedom, the diffusion model is found to have two free parameters -- making the phenomenological model, although powerful, slightly less powerful than the massive particle case. 

A distribution of massless particles initially peaked at some energy will undergo a diffusion in energy at a rate governed by the diffusion parameter $d_1$, and also drift to a higher or lower energies depending on the parameter $d_2$. The direction of the photon is unchanged, and light still travels at the speed of light. 

The values of the two parameters will depend on properties of the, as yet unknown, microscopic model. For massive particles the diffusion parameter depended on the forgetting time, $\tau_f$ -- for massless particles it is likely there will also be some dependence on a non-locality scale, such as that discussed in~\cite{Sorkin:2007qi}. The parameters could also depend on properties of the wave packet associated with the photon. Future work on scalar field propagation may lead to progress in this area. In the meantime, constraints can be placed on the values of the parameters by comparison with observations.

The CMB, consisting as it does of the `oldest' photons, provides the perfect laboratory for constraining the drift and diffusion. As shown in Section~\ref{s:CMBbounds}, the blackbody nature of the CMB constrains $d_1$ and $d_2$ very strongly. The tightness of the constraints is due not only to time the photons have been propagating, but also to the accuracy with which the CMB spectrum has been measured. Conceivably, future measurements could narrow the deviation from blackbody even further and thus result in even tighter bounds on $d_1$ and $d_2$. 

The drift and diffusion in energy of massless particles could have other observable consequences. One possibility, as discussed in Section~\ref{s:spectralline},  is the broadening and shift of spectral lines. The tight constraints from the CMB do, however, appear to rule out the detection of the effect in spectral lines with current technology.

As mentioned earlier, the work up to this point has considered only particles with no internal degrees of freedom. The ideas here can, however, be applied more generally. In particular, as will be seen in the next chapter, the effect of an underlying discreteness on photon polarisation can be modelled.

%% file: Polarization.tex
\chapter{Polarisation}
\label{c:polarisation}
\vspace{-5mm}
In the previous chapter the effect of spacetime discreteness on the energy of massless particles was discussed, but the possible polarisation of the particles was neglected. Here that omission is remedied. 

The study of polarisation is of particular relevance at present -- the polarisation of the cosmic microwave background (CMB) was first detected as recently as 2002 by DASI. It has since been observed by a number of experiments. Current and future experiments, such as the recently launched Planck, will hopefully reveal even further information. As such, the CMB polarisation is an ideal testing ground for quantum gravity phenomenology at the moment. 

If spacetime is discrete it is expected that the polarisation of a photon travelling through spacetime will be subject to some fluctuations. Unfortunately there is, again, no underlying model for polarisation on a causal set. It is hoped that such a model will be developed in the future. In the meantime, the expected continuum behaviour of a polarised photon in an underlying discrete spacetime can be found by following the stochastic evolution on a manifold of states procedure.

Before deriving a diffusion equation for polarisation it is necessary to choose, from the multitude of possibilities, a suitable way to describe the polarisation. This choice is discussed in Section~\ref{s:polarstatespace}.  In Section~\ref{s:polaraffinetime} the affine time polarisation diffusion equation is derived. It is found that restricting to linear polarisation is necessary for useful concrete phenomenology. The cosmic time linear polarisation diffusion equation is derived in Section~\ref{s:polarcosmictime}, and it is found that polarised light will experience both a suppression in polarisation fraction and a rotation in polarisation angle.

Section~\ref{s:polarobs} discusses possible ways to constrain the free parameters of the model. Rough constraints from radio galaxy polarisation data are discussed, as is the effect of rotation and suppression on the CMB.

\nopagebreak
The majority of the work in this chapter appears in~\cite{Contaldi:2010fh}.

%% file: PolarizationStateSpace.tex
\section{The polarisation state space}
\label{s:polarstatespace}

The polarisation of a massless particle with momentum $k^{\mu}$ can be described by a complex four-vector $a^{\mu}$, that satisfies $k_{\mu}a^{\mu}=0$ and $a_{\mu}^{*}a^{\mu}=1$. This description is not, however, gauge invariant: if $a^{\mu}$ satisfies these conditions then ${a^{\prime}}^{\mu}=a^{\mu}+\lambda k^{\mu}$, for any complex number $\lambda$, will describe the same polarisation state. To avoid this problem, the polarisation can be described in terms of a complex two-form $P_{\mu\nu} = k_{\mu}a_{\nu}-a_{\mu}k_{\nu}$. This $P_{\mu\nu}$ satisfies the following Lorentz invariant conditions:
\begin{enumerate}
\item $P_{\mu\nu}=-P_{\nu\mu}\,$;
\item $P_{\mu\nu}P^{\mu\nu}=0$ and $P^{*}_{\mu\nu}P^{\mu\nu}=0\,$;
\item $P^{\mu\nu}k_{\mu}=0\,$;
\item $P^{\mu\nu*}P_{\mu\sigma}=k^{\nu}k_{\sigma}\,$.
\end{enumerate}
Note that it is also the case that any $P_{\mu\nu}$ satisfying these conditions can be expressed in terms of some $a^{\mu}$ in the form above.
The antisymmetry condition $(1)$ reduces $P_{\mu\nu}$ to six complex (12 real) degrees of freedom. The constraints $(2)$ reduce it further to 10 real degrees of freedom. $P^{\mu\nu}k_{\mu}=0$ and $P^{\mu\nu*}k_{\mu}=0$ give six independent constraints, leaving only four degrees of freedom. Finally, the constraint $(4)$ is only one new condition, leaving $P_{\mu\nu}$ with three real degrees of freedom.

Consider the specific case $k^{\mu}=s^{\mu}$ where $s^{\mu}=(1,0,0,1)$. The condition $k_{\mu}a^{\mu}=0$ requires $a^{\mu}=(a^0,a^1,a^2,-a^0)$, and thus:
\begin{eqnarray}
P_{\mu\nu}&=&s_{\mu}a_{\nu}-a_{\mu}s_{\nu}\nonumber\\
&=&\begin{pmatrix}
0& -a_1 & -a_2 & 0\\
a_1 & 0 & 0 & -a_1\\
a_2 & 0 & 0 & -a_2\\
0 & a_1 & a_2 & 0
\end{pmatrix}\,,
\label{e:Pmatrix}
\end{eqnarray}
where $|a_1|^2 + |a_2|^2 = 1$. The overall phase of the complex vector $v=(a_1,a_2)\in\mathbb{C}^2$ is not physically relevant for the description of the polarisation state of a single photon (see for example~\cite{Weinberg:1995}), leaving a state space with only two real dimensions. This state space is the Bloch (or Poincar\'e) sphere, which will be denoted $\mathcal{B}$. The diffusion equations in the following sections will be derived in terms of the angles on the Bloch sphere rather than the polarisation two-form $P_{\mu\nu}$. The resulting equations are of course still Lorentz invariant, despite this not necessarily being immediately apparent. For further discussion of the explicitly Lorentz invariant $P_{\mu\nu}$ formulation see Appendix~\ref{a:polarLI}.

\begin{figure}[t]
%\begin{center}
\includegraphics[width=0.8\textwidth]{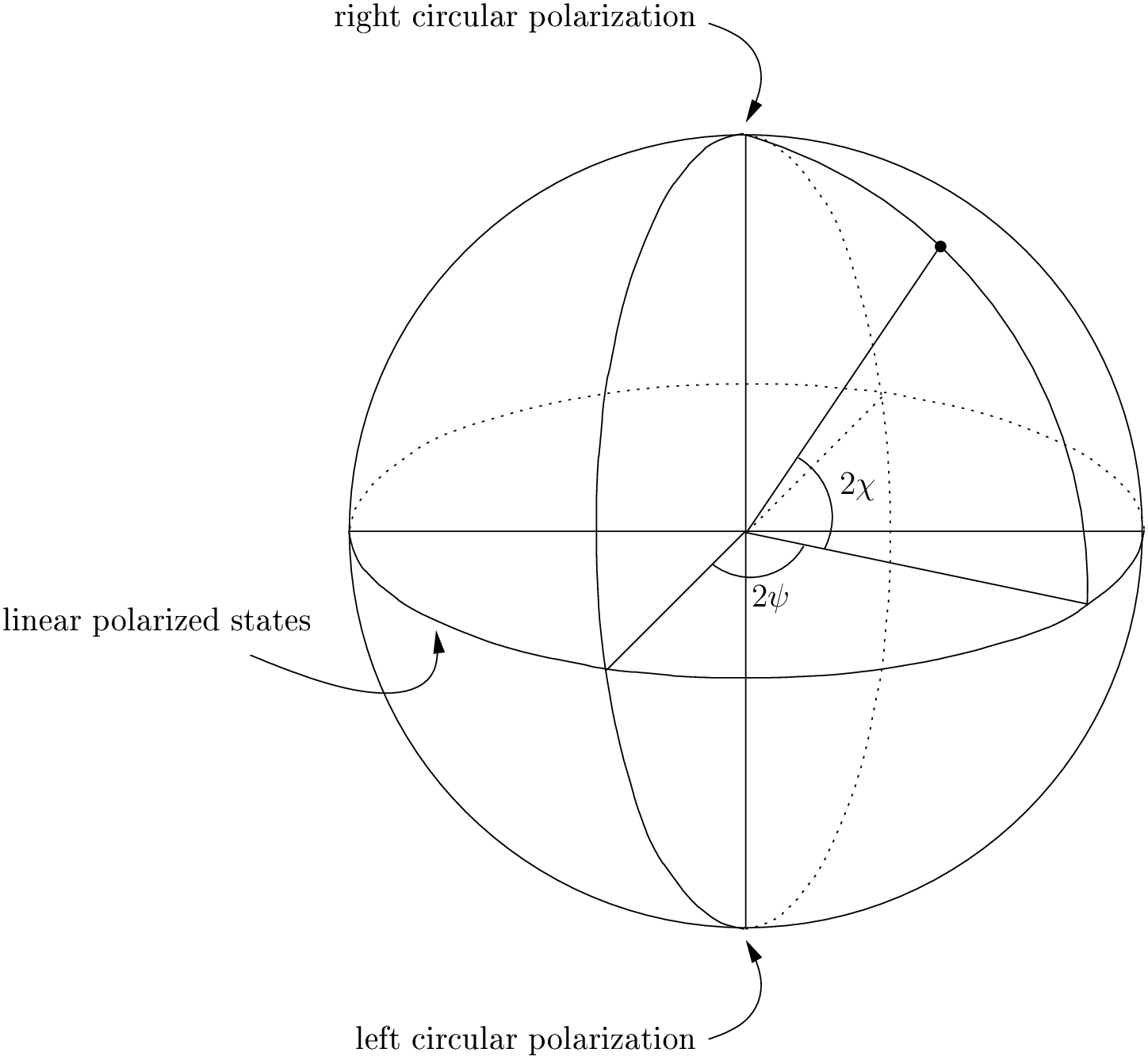}
\caption[The Bloch sphere]{The Bloch/Poincar\'e sphere description of polarisation.}
\label{f:blochsphere}
%\end{center}
\end{figure}

The next step is to show that the state space given by $v=(a_1,a_2)$ is indeed the Bloch sphere. Let $m_+ = \frac{1}{\sqrt{2}}(1,i)$ and $m_- = \frac{1}{\sqrt{2}}(1,-i)$, corresponding to left and right circularly polarised states, respectively. The vector $v=(a_1,a_2)$ can be expanded in terms of $\{m_+,m_-\}$, and an overall phase extracted to leave the coefficient of $m_+$ real and positive:
\begin{eqnarray}
v &=& e^{i\gamma}\left(\cos\left(\chi+\frac{\pi}{4}\right)m_+ + e^{2i\psi}\sin\left(\chi+\frac{\pi}{4}\right)m_-\right)\,,\\
\Rightarrow 
a_1&=&\frac{e^{i\gamma}}{\sqrt{2}}\left[\cos\left(\chi+\frac{\pi}{4}\right) + e^{2i\psi}\sin\left(\chi+\frac{\pi}{4}\right)\right]\,,\label{e:a1}\\
a_2&=&\frac{i e^{i\gamma}}{\sqrt{2}}\left[\cos\left(\chi+\frac{\pi}{4}\right) - e^{2i\psi}\sin\left(\chi+\frac{\pi}{4}\right)\right]\,\label{e:a2},
\end{eqnarray}
where $\chi\in[-\pi/4,\pi/4]$ and $\psi\in[0,\pi]$. Note that this slightly unnatural choice of angles is made to match the standard Poincar\'e sphere description of polarisation shown in Figure~\ref{f:blochsphere} (see, for example~\cite{Born:1980}).

The north and south poles ($\chi=\pm\pi/4$) of the Bloch sphere correspond to
\begin{eqnarray}
v\left(\chi=\frac{\pi}{4}\right)&=& \frac{e^{i\gamma+2i\psi}}{\sqrt{2}}(1,-i)\,,\\
v\left(\chi=-\frac{\pi}{4}\right)&=& \frac{e^{i\gamma}}{\sqrt{2}}(1,i)\,,
\end{eqnarray}
right and left circularly polarised states, respectively. States on the equator ($\chi=0$) have
\begin{eqnarray}
a_1 &=& \frac{e^{i\gamma}}{\sqrt{2}}\left(\frac{1}{\sqrt{2}}+e^{2i\psi}\frac{1}{\sqrt{2}}\right)\nonumber\\
&=& e^{i\psi+i\gamma}\cos\psi\,,\\
a_2 &=& i\frac{e^{i\gamma}}{\sqrt{2}}\left(\frac{1}{\sqrt{2}}-e^{2i\psi}\frac{1}{\sqrt{2}}\right)\nonumber\\
&=& e^{i\psi+i\gamma}\sin\psi\,,
\end{eqnarray}
and thus the equator consists of the linear polarisation states. The remainder of the sphere describes elliptical polarisation.

\subsubsection{General $k^{\mu}$}

The above calculations showed that for the specific case of $k^{\mu}=s^{\mu}=(1,0,0,1)$ the polarisation state can be described by the angles $\chi$ and $\psi$ on the Bloch sphere. It remains to show that for a general $k^{\mu}$ the polarisation state $P_{\mu\nu}$ can be expressed in terms of the Bloch sphere. This can be done by transforming $P_{\mu\nu}$ by the Lorentz transformation that takes $k^{\mu}$ to $s^{\mu}$ and then simply using the Bloch sphere description for the special $s^{\mu}$ case. In general the Lorentz transformation mapping $k^{\mu}$ to $s^{\mu}$ is not unique. A \textit{standard Lorentz transformation} taking $s^{\mu}$ to $k^{\mu}$ can, however, be defined by (see, for example~\cite{Weinberg:1995}):
\begin{equation}
L(k):= R(\hat{\underbar{k}})B_z(|\underbar{k}|)\,,
\end{equation}
where $B_z(|\underbar{k}|)$ is the boost along the $z$-direction that takes $s^{\mu}$ to $(|\underbar{k}|,0,0,|\underbar{k}|)$, and $R(\hat{\underbar{k}})$ is the \textit{standard rotation} that takes the $z$-axis into the direction of the unit vector $\hat{\underbar{k}}$. If $\hat{\underbar{k}}$ has spherical polar coordinates $\theta\,,\phi$ then the \textit{standard rotation} is defined by
\begin{equation}
R(\hat{\underbar{k}}):=R_z(\phi)R_y(\theta)\,,
\end{equation}
i.e.~a rotation by angle $\theta$ around the $y$-axis followed by a rotation by $\phi$ around the $z$-axis.
The transformation $L(k)^{-1}$ takes $k^{\mu}$ to $s^{\mu}$. Under the same transformation the polarisation state, $P_{\mu\nu}$, gives
\begin{equation}
P(k)_{\mu\nu}:=(L(k)^{-1})^{\rho}_{\mu}(L(k)^{-1})^{\sigma}_{\nu}P_{\rho\sigma}\,.
\label{e:Pk}
\end{equation}
$P(k)$ now has the same form as Equation~\ref{e:Pmatrix}, with components $P(k)_{\mu\nu}=s_{\mu}a_{\nu}-a_{\mu}s_{\nu}$, where $a_1=P(k)_{10}$ and $a_2=P(k)_{20}$. Equation~\ref{e:Pk} will henceforth be written $P(k) = L(k)^{-1}P$ for simplicity.

Thus, a general state $(k^{\mu},\,P_{\mu\nu})$ is specified by coordinates $(k^{\mu},\,\chi,\,\psi)$, where the angles $\chi$ and $\psi$ are determined by first Lorentz transforming $P_{\mu\nu}$ into the form of Equation~\ref{e:Pmatrix}, and then using Equations~\ref{e:a1} and~\ref{e:a2}.

\subsection{Polarisation under Lorentz transformation}

Before moving onto the derivation of the diffusion equation, it is necessary to know how the polarisation state $(\chi,\psi)$ changes under Lorentz transformation. The photon state $(k^{\mu},\,P_{\mu\nu})$ transforms as usual under a Lorentz transformation $\Lambda$: $(k^{\mu},\,P_{\mu\nu})\rightarrow({k^{\prime}}^{\mu},\,{P^{\prime}}_{\mu\nu})=(\Lambda_{\sigma}^{\mu}k^{\sigma},\,\Lambda_{\mu}^{\rho}\Lambda_{\nu}^{\sigma}P_{\rho\sigma})$. For simplicity this will be written $(k^{\prime},\,P^{\prime})=(\Lambda k,\,\Lambda P)$. Under Lorentz transformation $(a_1,\,a_2)\rightarrow(a^{\prime}_1,\,a^{\prime}_2)=(P^{\prime}(k^{\prime})_{10},\,P^{\prime}(k^{\prime})_{20})$ and
\begin{eqnarray}
P^{\prime}(k^{\prime})&=&L(k^{\prime})^{-1}P^{\prime}\nonumber\\
&=& L(\Lambda k)^{-1}\Lambda P\nonumber\\
&=& L(\Lambda k)^{-1}\Lambda L(k)P(k)\nonumber\\
&=& W(\Lambda,k)P(k)\,,
\end{eqnarray}
where $W(\Lambda,k)=L(\Lambda k)^{-1}\Lambda L(k)$ is an element of the little group of the Lorentz transformations that leaves $s^{\mu}$ fixed.

The little group is isomorphic to the Euclidean group $ISO(2)$: the set of rotations and translations of the plane. Following Weinberg~\cite{Weinberg:1995}, an element of the little group can be written
\begin{equation}
W(\theta,\,\alpha,\,\beta) = S(\alpha,\,\beta)R(\theta)\,,
\end{equation}
where $S$ is the `translation'
\begin{equation}
S^{\mu}_{\nu}(\alpha,\,\beta) = 
\begin{pmatrix}
1+\zeta & \alpha & \beta & -\zeta\\
\alpha & 1 & 0 & -\alpha\\
\beta & 0 & 1 & -\beta\\
\zeta & \alpha & \beta & 1-\zeta
\end{pmatrix}\,,
\end{equation}
$\alpha$ and $\beta$ are real numbers and $\zeta = (\alpha^2+\beta^2)/2$. $R(\theta)$ is the usual rotation
\begin{equation}
R^{\mu}_{\nu}(\theta) = 
\begin{pmatrix}
1 & 0 & 0 & 0\\
0 & \cos\theta & -\sin\theta & 0\\
0 & \sin\theta & \cos\theta & 0\\
0 & 0 & 0 & 1
\end{pmatrix}\,.
\end{equation}
To see how the translations $S(\alpha,\,\beta)$ act on the polarisation consider the infinitesimal transformation
\begin{equation}
\Delta S^{\mu}_{\nu}(\alpha,\,\beta) =
\begin{pmatrix}
1 & \alpha & \beta & 0 \\
\alpha & 1 & 0 & -\alpha \\
\beta & 0 & 1 & -\beta\\
0 & \alpha & \beta & 1
\end{pmatrix}\,.
\end{equation}
From Equation~\ref{e:Pmatrix} it is clear that $\Delta S$ can be written
\begin{equation}
\Delta S^{\mu}_{\nu}(\alpha,\,\beta) = \delta^{\mu}_{\nu} + s^{\mu}b_{\nu}-b^{\mu}s_{\nu}\,,
\end{equation}
where $s^{\mu}b_{\mu}=0$ and thus $b^{\mu} = (b^0,\,\alpha,\,\beta,\,-b^0)$. Recall $P(k)_{\mu\nu}=s_{\mu}a_{\nu}-a_{\mu}s_{\nu}$. Thus
\begin{eqnarray}
\Delta S^{\mu^{\prime}}_{\mu}\Delta S^{\nu^{\prime}}_{\nu} P(k)_{\mu^{\prime}\nu^{\prime}} &=& \left(\delta^{\mu^{\prime}}_{\mu} + s^{\mu^{\prime}}b_{\mu}-b^{\mu^{\prime}}s_{\mu}\right)\left(\delta^{\nu^{\prime}}_{\nu} + s^{\nu^{\prime}}b_{\nu}-b^{\nu^{\prime}}s_{\nu}\right)\nonumber\\
&&\times\left(s_{\mu^{\prime}}a_{\nu^{\prime}}-a_{\mu^{\prime}}s_{\nu^{\prime}}\right)\nonumber\\
&=& s_{\mu}a_{\nu}-a_{\mu}s_{\nu}\,,
\end{eqnarray}
using the relations $s^{\mu}s_{\mu}=0,\,s^{\mu}a_{\mu}=0$ and $s^{\mu}b_{\mu}=0$. The `translations' therefore leave the polarisation unchanged. 

The rotation $R$ acts as usual, thus under a Lorentz transformation
\begin{equation}
(a_1,\,a_2)\rightarrow(a^{\prime}_1,\,a^{\prime}_2)=(a_1\cos\delta - a_2\sin\delta,\,a_1\sin\delta + a_2\cos\delta)\,,
\end{equation}
where $\delta = \delta(\Lambda,k^{\mu})$.
The transformation of the Bloch sphere angles can be determined from Equations~\ref{e:a1},~\ref{e:a2}:
\begin{eqnarray}
a^{\prime}_1 &=& \frac{e^{i\gamma}}{\sqrt{2}}\left[\cos\left(\chi+\frac{\pi}{4}\right) + e^{2i\psi}\sin\left(\chi+\frac{\pi}{4}\right)\right]\cos\delta\nonumber\\
&& + \frac{i e^{i\gamma}}{\sqrt{2}}\left[\cos\left(\chi+\frac{\pi}{4}\right) - e^{2i\psi}\sin\left(\chi+\frac{\pi}{4}\right)\right]\sin\delta\nonumber\\
&=& \frac{e^{i\gamma}}{\sqrt{2}}\left[\cos\left(\chi+\frac{\pi}{4}\right) + e^{2i\psi}\sin\left(\chi+\frac{\pi}{4}\right)\right]\frac{1}{2}\left(e^{i\delta}+e^{-i\delta}\right)\nonumber\\
&& + \frac{i e^{i\gamma}}{\sqrt{2}}\left[\cos\left(\chi+\frac{\pi}{4}\right) - e^{2i\psi}\sin\left(\chi+\frac{\pi}{4}\right)\right]\frac{-i}{2}\left(e^{i\delta}-e^{-i\delta}\right)\nonumber\\
&=& \frac{e^{i\gamma+i\delta}}{\sqrt{2}}\left[\cos\left(\chi+\frac{\pi}{4}\right) + e^{2i(\psi+\delta)}\sin\left(\chi+\frac{\pi}{4}\right)\right]\,,\\
a^{\prime}_2 &=& \frac{e^{i\gamma}}{\sqrt{2}}\left[\cos\left(\chi+\frac{\pi}{4}\right) + e^{2i\psi}\sin\left(\chi+\frac{\pi}{4}\right)\right]\sin\delta\nonumber\\
&&+ \frac{i e^{i\gamma}}{\sqrt{2}}\left[\cos\left(\chi+\frac{\pi}{4}\right) - e^{2i\psi}\sin\left(\chi+\frac{\pi}{4}\right)\right]\cos\delta\nonumber\\
&=& \frac{e^{i\gamma}}{\sqrt{2}}\left[\cos\left(\chi+\frac{\pi}{4}\right) + e^{2i\psi}\sin\left(\chi+\frac{\pi}{4}\right)\right]\frac{-i}{2}\left(e^{i\delta}-e^{-i\delta}\right)\nonumber\\
&&+ \frac{i e^{i\gamma}}{\sqrt{2}}\left[\cos\left(\chi+\frac{\pi}{4}\right) - e^{2i\psi}\sin\left(\chi+\frac{\pi}{4}\right)\right]\frac{1}{2}\left(e^{i\delta}+e^{-i\delta}\right)\nonumber\\
&=& \frac{i e^{i\gamma+i\delta}}{\sqrt{2}}\left[\cos\left(\chi+\frac{\pi}{4}\right) - e^{2i(\psi+\delta)}\sin\left(\chi+\frac{\pi}{4}\right)\right]\,,
\end{eqnarray}
thus
\begin{equation}
(\chi,\,\psi)\rightarrow(\chi^{\prime},\,\psi^{\prime})=(\chi,\psi+\delta)\,,
\end{equation}
i.e.~Lorentz transformations act on the Bloch sphere as rotations around the north-south polar axis.

\subsection{Stokes parameters}

The Stokes parameters are a common method of describing polarisation, particularly in cosmology and astrophysics. They have the advantage of being quantities that can be directly measured by experiments. As such, it is useful to understand the relationship between the Bloch sphere coordinates $\chi\,,\psi$ and the Stokes parameters $I\,,Q\,,U\,,V$. 

The first Stokes parameter, $I$, is the total intensity of the light. The Bloch sphere, in fact, only parametrises polarisation of a fixed intensity (or more correctly a fixed polarisation intensity $I p$ where $p$ is the polarisation fraction). The parameters $Q$ and $U$ describe linear polarisation. Consider a set of orthogonal axes $\{x,y\}$ in the plane perpendicular to the direction of propagation of the light, and another set of orthogonal axes $\{a,b\}$ at $45^{\circ}$ in the plane to $\{x,y\}$. $Q$ is the difference between the intensity transmitted by a polariser that accepts light polarised in the $x$-direction, and the intensity transmitted by one that accepts light polarised in the $y$-direction. $U$ is given by the difference between the intensity transmitted by a polariser that accepts light in the $a$-direction, and the intensity transmitted by one that accepts light in the $b$-direction. Note that the $\{x,y\}$ can be chosen freely within the plane perpendicular to the propagation of the light. In practice there are several standard coordinate conventions, for example the International Astronomical Union coordinate convention defines the $z$-axis as pointing along the line of sight towards the observer, the $x$-axis points north and the $y$-axis, east. The remaining Stokes parameter, $V$, measures circular polarisation. 

The Stokes parameters obey the inequality
\begin{equation}
I^2\geq Q^2+U^2+V^2\,.
\end{equation}
The polarisation fraction $p$ is defined by
\begin{equation}
p=\frac{\sqrt{Q^2+U^2+V^2}}{I}\,.
\end{equation}

Consider a perfectly polarised beam with fixed intensity normalised to one. The relationship between the Bloch sphere angles and the Stokes parameters can be found by calculating the density matrix $S= v v^{\dagger}$ from the vector $v=(a_1,a_2)$. Equating
\begin{eqnarray}
S&=& \frac{1}{2}
\begin{pmatrix}
I + Q & U-iV\\
U+iV & I-Q
\end{pmatrix}\,,
\end{eqnarray}
 (see, e.~g.~\cite{Kosowsky:1994cy}), gives
\begin{eqnarray}
Q &=& \cos2\psi\cos2\chi\,,\\
U &=& \sin2\psi\cos2\chi\,,\\
V &=& \sin2\chi\,.
\end{eqnarray}

More generally, if a collection of photons has a distribution of polarisations given by a probability density $\rho(\chi,\psi)$ on the Bloch sphere, $\mathcal{B}$, the Stokes parameters are
\begin{eqnarray}
Q &=& \int_{\mathcal{B}}\rho(\chi,\psi)\cos2\psi\cos2\chi\,d\chi\,d\psi\,,\\
U &=& \int_{\mathcal{B}}\rho(\chi,\psi)\sin2\psi\cos2\chi\,d\chi\,d\psi\,,\\
V &=& \int_{\mathcal{B}}\rho(\chi,\psi)\sin2\chi d\chi d\psi\,.
\end{eqnarray}
Note that a given set of Stokes parameters $Q\,,U\,,V$ does not correspond to a unique distribution. For example, unpolarised light $Q=U=V=0$ could be modelled as a uniform distribution of linearly polarised states, or a uniform distribution on the two circularly polarised states alone -- or any other distribution uniform in $\psi$ and symmetric in $\chi$. 

%% file: PolarizationAffineTimeEquation.tex
\section{Polarisation diffusion in affine time}
\label{s:polaraffinetime}

The state space of a photon with polarisation is $\mathcal{M}=\mink\times\Lob_0\times\mathcal{B}$.  The affine time $\lambda$ is again the suitable time parameter for the trajectories. 

Recall that the stochastic process on $\mathcal{M}$ can be described by the equation (Equation~\ref{e:diffequA}):
\begin{equation}
\pd{\rho_{\lambda}}{\lambda} = \partial_A\left(K^{AB}n\,\partial_B\!\left(\frac{\rho_{\lambda}}{n}\right)-u^A\rho_{\lambda}\right)\,,
\label{e:diffequAPolar}
\end{equation}
and thus to derive the affine time diffusion equation the Lorentz invariant, symmetric positive semi-definite matrix $K^{AB}$ and the Lorentz invariant vector $u^A$ need to be determined. The microscopic density of states, $n$, on $\mathcal{M}$ must also be known.

Unfortunately there are just too many choices when it comes to determining invariant objects on the Bloch sphere. As discussed earlier, Lorentz transformations act on the Bloch sphere as polar rotations. Thus, any tensor, vector, or scalar that does not depend on $\psi$ is Lorentz invariant. For example, $K$ could be the metric on a sphere, or an oblate or prolate spheroid, or indeed any other such strange shape generated by rotating a curve around the polar axis. It need not even be invariant under reflection in the equatorial plane. The invariant vector, $u$, can be any linear combination of $f(\chi)\partial/\partial\chi$ and $g(\chi)\partial/\partial\psi$, where $f$ and $g$ are arbitrary functions of $\chi$. All these free functions mean the model is not much use for phenomenology. Predictive power can be regained if the model is restricted to linear polarisation only. For astrophysical and cosmological applications this is precisely the area of interest -- the CMB, for example, is linearly polarised. The matter is not quite so simple, however, as assuming the initial distribution has no circular polarisation. An initially linearly polarised distribution could become elliptically (or circularly) polarised if $K$ or $g(\chi)$ are not invariant under reflection in the equatorial plane. To allow progress to be made, it will be assumed from here on that circular polarisation $\textit{can}$ be neglected. It is hoped that it will be possible to determine if this assumption is justified when the microscopic physics is better understood. In the meantime, useful results can still be obtained. 

Restricting to linear polarisation, the polarisation state space is just the equator of the Bloch sphere: the unit circle, with coordinate $\psi$. In a slight abuse of notation the state space with linear polarisation only will also be denoted $\mathcal{M}$.  There is one Lorentz invariant vector, $\partial/\partial\psi$, on the space of linear polarised states. The Lorentz invariant density of states $n$ is constant on the circle. As in Chapter~\ref{c:massless}, it is convenient to work in polar coordinates $(k,\theta,\phi)$ on the momentum space $\Lob_0$. The coordinates on the state space $\mathcal{M}$ will be denoted $X^A=(x^{\mu},\,k,\,\theta,\,\phi,\,\psi)$. As shown Section~\ref{s:masslessaffine} there is the invariant vector field $k^a=(k,0,0)$, and invariant two-tensor $k^ak^b$ on $\Lob_0$. The density of states on $\Lob_0$ is $k\sin\theta$. We also know from Section~\ref{s:masslessaffine} that the spacetime components of $K^{AB}$ are zero, and the spacetime component of $u^A$ is simply $k^{\mu}$.

Let $v_1$ and $v_2$ be the invariant vectors $v_1^A=(\mathbf{0},k,0,0,0)$ and $v_2^A=(\mathbf{0},0,0,0,1)$ on $\mathcal{M}$.  The  most general symmetric, positive semi-definite invariant two-tensor on the full state space $\mathcal{M}$ is 
\begin{eqnarray}
K^{AB}&=&av_1^Av_2^A + b(v_1^Av_2^B+v_2^Av_1^B) + cv_2^Av_2^B\nonumber\\
&=&\bordermatrix{
              & x^{\mu} & k     &\theta &\phi&\psi\cr
      x^{\mu} & \mathbf{0}       & \mathbf{0}     & \mathbf{0}     & \mathbf{0}  & \mathbf{0} \cr
           k  & \mathbf{0}       & a k^2 & 0     & 0  & b k\cr
      \theta  & \mathbf{0}       & 0     & 0     & 0  & 0  \cr
      \phi    & \mathbf{0}       & 0     & 0     & 0  & 0  \cr
      \psi   & \mathbf{0}       & bk    & 0     & 0  & c  \cr
}\,; 
\end{eqnarray}
the invariant vector is
\begin{equation}
u^A = (k^{\mu},dk,0,0,e)\,.
\end{equation}
Here $a\geq0,\,c\geq0,\,b,\,d,\,e$ are real constants. In order for $K^{AB}$ to be positive semi-definite the constants must also satisfy $ac-b^2\geq0$. On $\mathcal{M}$ the density of states is simply $n=k\sin\theta$. 

Substituting these forms for $K^{AB}$ and $u^A$ into Equation~\ref{e:diffequAPolar} above gives the affine time linear polarisation diffusion equation
\begin{eqnarray}
\pd{\rho_{\lambda}}{\lambda}&=&\partial_A\left(K^{AB}n\partial_B\left(\frac{\rho_{\lambda}}{n}\right)-u^A\rho_{\lambda}\right)\nonumber\\
&=& -k^{\mu}\pd{\rho_{\lambda}}{x^{\mu}} + ak^2\pdsq{\rho_{\lambda}}{k}+(a-d)k\pd{\rho_{\lambda}}{k}-(a+d)\rho_{\lambda}\nonumber\\
&& + c\pdsq{\rho_{\lambda}}{\psi} + 2bk\frac{\partial^2\rho_{\lambda}}{\partial\psi\partial k} - e\pd{\rho_{\lambda}}{\psi}\,.\label{e:polarfull}
\end{eqnarray}
At first glance this equation appears unwieldy, with still too many free parameters to be useful for phenomenology. Recall, however, the massless particle affine time diffusion equation, Equation~\ref{e:affine}:
\begin{eqnarray}
\pd{\rho_{\lambda}}{\lambda}&=&-k^{\mu}\pd{\rho_{\lambda}}{x^{\mu}}+d_1\pd{}{k}\left(k^3\pd{}{k}\left(\frac{\rho_{\lambda}}{k}\right)\right) - d_2\pd{}{k}\left(k\rho_\lambda\right)\nonumber\\
&=& -k^{\mu}\pd{\rho_{\lambda}}{x^{\mu}} + d_1k^2\pdsq{\rho_{\lambda}}{k} + (d_1-d_2)k\pd{\rho_{\lambda}}{k} - (d_1+d_2)\rho_{\lambda}\,.
\end{eqnarray}
The parameters $a$ and $d$ in Equation~\ref{e:polarfull} are just $d_1$ and $d_2$ renamed. In Section~\ref{s:CMBbounds} very tight constraints were placed on the values of $d_1$ and $d_2$. Because $d_1$ and $d_2$ are constrained to be so small, they can be neglected in this discussion of linear polarisation. Taking $a=d=0$ further requires $b=0$. Finally, the homogeneous polarisation affine time equation reduces to
\begin{equation}
\pd{\rho_{\lambda}}{\lambda} = c\pdsq{\rho_{\lambda}}{\psi} - e\pd{\rho_{\lambda}}{\psi}\,.
\label{e:affinepolar}
\end{equation}
A linearly polarised photon in an underlying discrete spacetime therefore experiences a diffusion, determined by the parameter $c$, and a drift, given by $e$, in its angle of polarisation.
Although Equation~\ref{e:affinepolar} is Lorentz invariant, the Lorentz invariance is not immediately obvious. In Appendix~\ref{a:polarLI} the linear polarisation diffusion equation is formulated in an explicitly Lorentz invariant way, using the polarisation two-form $P_{\mu\nu}$. The resulting equation, although equivalent to the equation above, provides less immediate physical insight.

%% file: PolarizationCosmicTimeEquation.tex
\section{Polarisation diffusion in cosmic time}
\label{s:polarcosmictime}

Again, to compare with observations the affine time polarisation diffusion equation must be re-expressed in terms of cosmic time. Following the arguments given in Sections~\ref{s:massivecosmictime} and~\ref{s:masslesscosmictime} a larger state space incorporating the affine time is defined: $\mathcal{M}^{\prime}=\mathcal{M}\times\mathbb{R}$. The probability density $\rho_{\lambda}$ can be written as a component of the current in $\mathcal{M}^{\prime}$:
\begin{equation}
J^{\lambda}(t,x^i,k^a,\psi,\lambda)=\rho_{\lambda}\,.
\end{equation}
The $t$ component of the current is
\begin{equation}
J^t(t,x^i,k^a,\psi,\lambda) = kJ^{\lambda}\,,
\end{equation}
where Equation~\ref{e:current} has been used. The remaining components of the current, expressed in terms of $J^{t}$ are:
\begin{eqnarray}
J^i(t,x^i,k^a,\psi,\lambda) &=& \frac{k^i}{k}J^t\,,\\
J^k(t,x^i,k^a,\psi,\lambda) &=& -ak^3\pd{}{k}\left(\frac{J^t}{k^2}\right) - b\pd{J^t}{\psi} + dJ^t\,,\\
J^{\psi}(t,x^i,k^a,\psi,\lambda) &=& -\frac{c}{k}\pd{J^t}{\psi}-bk^2\pd{}{k}\left(\frac{J^t}{k^2}\right)+\frac{e}{k}J^t\,.
\end{eqnarray}

The unobservable affine time can be integrated over, and a new integrated current $\bar{J}$ defined:
\begin{eqnarray}
\bar{J}^t(t,x^i,k^a,\psi)&=& \int_0^{\infty}J^t(t,x^i,k^a,\psi,\lambda)d\lambda\,,\\
\bar{J}^i(t,x^i,k^a,\psi) &=& \frac{k^i}{k}\bar{J}^t\,,\\
\bar{J}^k(t,x^i,k^a,\psi) &=& -ak^3\pd{}{k}\left(\frac{\bar{J}^t}{k^2}\right) - b\pd{\bar{J}^t}{\psi} + d\bar{J}^t\,,\\
\bar{J}^{\psi}(t,x^i,k^a,\psi) &=& -\frac{c}{k}\pd{\bar{J}^t}{\psi}-bk^2\pd{}{k}\left(\frac{\bar{J}^t}{k^2}\right)+\frac{e}{k}\bar{J}^t\,.
\end{eqnarray}
The continuity equation on $\mathcal{M}^{\prime}$ can also be integrated over $\lambda$ to give:
\begin{equation}
\left[J^\lambda\right]^{\infty}_0 + \partial_t\bar{J}^t + \partial_i\bar{J}^i+\partial_k\bar{J}^k + \partial_{\psi}\bar{J}^{\psi} = 0\,.
\end{equation}
$J^{\lambda}|_{\lambda =0}$ is zero for all $t>0$ and $J^{\lambda}$ tends toward zero as
$\lambda$ approaches infinity for finite $t$. Thus, renaming $\bar{J}^t$ as $\rho_t$, the polarisation cosmic time equation is
\begin{eqnarray}
\pd{\rho_t}{t}&=&-\partial_i\bar{J}^i-\partial_k\bar{J}^k-\partial_{\psi}\bar{J}^{\psi}\nonumber\\
&=& -\frac{k^i}{k}\pd{\rho_t}{x^i} + a k \pdsq{\rho_t}{k} - (a+d)\pd{\rho_t}{k} + 2b\frac{\partial^2\rho_t}{\partial k\partial \psi}\nonumber\\
&& + \frac{c}{k}\pdsq{\rho_t}{\psi} - \frac{2b}{k}\pd{\rho_t}{\psi} - \frac{e}{k}\pd{\rho_t}{\psi}\,.
\end{eqnarray}
Neglecting the energy drift and diffusion as discussed in Section~\ref{s:polaraffinetime}~($a=d=0\Rightarrow b=0$), and looking at the homogeneous case gives
\begin{equation}
\pd{\rho_t}{t} = \frac{\delta_1}{k}\pdsq{\rho_t}{\psi} - \frac{\delta_2}{k}\pd{\rho_t}{\psi}\,,
\label{e:poldiff}
\end{equation}
where, for convenience, the diffusion parameter, $c$, and drift parameter, $e$, have been renamed $\delta_1$ and $\delta_2$ respectively.
From this equation we see that the rates of drift and diffusion in linear polarisation angle are energy dependent: the higher the energy of the photon, the slower the rate of drift and diffusion. This result is particularly interesting because, although there are other models that predict drift in linear polarisation angle, most do not have this particular energy dependence. This will be discussed further in Section~\ref{s:polarobs}.

\subsection{The solution to the diffusion equation}

Equation~\ref{e:poldiff} is a standard equation describing diffusion with drift on a circle. Solutions to this type of equation are well known, see for example~\cite{Dym:1972}. Here I will solve the equation using Fourier series. First, note that the diffusion occurs on a half circle rather than a full circle: $\psi\in[0,\pi]$. Physically, a polarisation angle $\psi$ and a polarisation angle $\psi +\pi$ are equivalent. For simplicity I define a new variable $x = \psi/\pi$, $0\leq x< 1$. The distribution $\rho_t(t,\psi)$ is a scalar density, rather than a scalar, and thus under this rescaling transforms as
\begin{eqnarray}
\bar{\rho}_t(t,x) &=& \frac{d\psi}{dx}\rho_t(t,\psi)\nonumber\\
&=&\frac{1}{\pi}\rho_t(t,\psi)\,.
\end{eqnarray}
Equation~\ref{e:poldiff} becomes
\begin{eqnarray}
\frac{1}{\pi}\pd{\bar{\rho}_t}{t} &=& \frac{\delta_1}{k\pi^3}\pdsq{\bar{\rho}_t}{x} - \frac{\delta_2}{k\pi^2}\pd{\bar{\rho}_t}{x}\,,\\
\Rightarrow \pd{\bar{\rho}_t}{t} &=& \frac{\delta_1}{k\pi^2}\pdsq{\bar{\rho}_t}{x} - \frac{\delta_2}{k\pi}\pd{\bar{\rho}_t}{x}.
\end{eqnarray}
Expand $\bar{\rho}_t$ in a Fourier series:
\begin{eqnarray}
\bar{\rho}_t\left(t,x\right)&=&\sum_{n=-\infty}^{\infty}{c_n(t)e^{2\pi inx}}\,,\\
c_n(t) &=& \int^{1}_{0}{\bar{\rho}_t(t,x)e^{-2\pi inx}dx}.
\end{eqnarray}
The diffusion equation can thereby be written as an ordinary differential equation
\begin{eqnarray}
\dot{c}_n &=& \int^{1}_{0}{\pd{\bar{\rho}_t}{t}e^{-2\pi inx}dx}\nonumber\\
&=& \int^{1}_{0}{\left(\frac{\delta_1}{k\pi^2}\pdsq{\bar{\rho}_t}{x} - \frac{\delta_2}{k\pi}\pd{\bar{\rho}_t}{x}\right)e^{-2\pi inx}dx}.
\end{eqnarray}
Integrating by parts and recalling $\bar{\rho}_t(x=1)=\bar{\rho}_t(x=0)$ and $e^{-2\pi in} = 1$ gives
\begin{eqnarray}
\dot{c}_n &=& \int^{1}_{0}{\left(\frac{\delta_1}{k\pi^2}\rho_t\pdsq{(e^{-2\pi inx})}{x} + \frac{\delta_2}{k\pi}\rho_t\pd{(e^{-2\pi inx})}{x}\right)dx}\nonumber\\
&=& \left(-\frac{4\delta_1n^2}{k}- \frac{2i\delta_2n}{k}\right)c_n\,,\\
\Rightarrow c_n(t) &=&c_n(0)\exp\left[-\left(\frac{4\delta_1n^2}{k}+ \frac{2i\delta_2n}{k}\right)t\right]\,.
\end{eqnarray}
Let the initial distribution be $\bar{f}(x) = \bar{\rho}_t(0,x)$, then
\begin{eqnarray}
c_n(0) &=& \int^{1}_{0}{\bar{f}(y)e^{-2\pi iny}dy}\,,\\
\Rightarrow \bar{\rho}_t(t,x) &=&\sum_{n=-\infty}^{\infty}{c_n(t)e^{2\pi inx}}\nonumber\\
&=&\sum_{n=-\infty}^{\infty}{\int^{1}_{0}{f(y)e^{-2\pi iny}dy}\exp\left[-\left(\frac{4\delta_1n^2}{k}+ \frac{2i\delta_2n}{k}\right)t\right]e^{2\pi inx}}\nonumber\\
&=&\int^{1}_{0}\sum_{n=-\infty}^{\infty}\exp\left[-\left(\frac{4\delta_1 n^2}{k}+ \frac{2i\delta_2 n}{k}\right)t\right]e^{2\pi in(x-y)}\bar{f}(y)dy\nonumber\\
&=& \bar{p}\circ \bar{f}\,,
\end{eqnarray}
where 
\begin{equation}
\bar{p}(x) = \sum_{n=-\infty}^{\infty}\exp\left[-\left(\frac{4\delta_1n^2}{k}+ \frac{2i\delta_2n}{k}\right)t\right]e^{2\pi inx}\,,
\label{e:fundsol}
\end{equation}
is the fundamental solution (see for example~\cite{Dym:1972}). To be consistent with the notation that appears in other papers (e.g.~\cite{Klimov:1997,Luan:2004}) $\bar{p}$ can be expressed in a slightly different way. Making use of the Jacobi identity
\begin{equation}
\sum_{n=-\infty}^{\infty}\exp\left[-\frac{(x-n)^2}{2t}\right] = \sqrt{2\pi t}\sum_{n=-\infty}^{\infty}\exp\left[-2\pi^2n^2t\right]e^{2\pi inx},
\end{equation}
$\bar{p}$ can be written
\begin{eqnarray}
\bar{p}(x) &=&  \sum_{n=-\infty}^{\infty}\exp\left[-2\pi^2 n^2\frac{2\delta_1t}{k\pi^2}\right]\exp\left[2\pi in\left(x-\frac{\delta_2t}{k\pi}\right)\right]\nonumber\\
 &=& \sqrt{\frac{\pi k}{4\delta_1 t}}\sum_{n=-\infty}^{\infty}\exp\left[-\frac{\left(x-\frac{\delta_2t}{k\pi}-n\right)^2}{\frac{4\delta_1t}{k\pi^2}}\right]\,,\\
 p(\psi) &=& \frac{1}{\pi}\bar{p}(x)\nonumber\\
 &=& \sqrt{\frac{k}{4\pi \delta_1 t}}\sum_{n=-\infty}^{\infty}\exp\left[-\frac{\left(\psi-\frac{\delta_2t}{k}-\pi n\right)^2}{\frac{4\delta_1t}{k}}\right]\,.
\end{eqnarray}
The full solution to Equation~\ref{e:poldiff} is thus
\begin{eqnarray}
\bar{\rho}_t(t,x) &=&\sqrt{\frac{\pi k}{4\delta_1 t}}\int^{1}_{0}\sum_{n=-\infty}^{\infty}\exp\left[-\frac{\left(\pi(x-y)-\frac{\delta_2t}{k}-n\pi\right)^2}{\frac{4\delta_1t}{k}}\right]\bar{f}(y)dy\nonumber\\
&=&\sqrt{\frac{\pi k}{4\delta_1 t}}\sum_{n=-\infty}^{\infty}\int^{n+1}_{n}\exp\left[-\frac{\left(\pi(x-y)-\frac{\delta_2t}{k}\right)^2}{\frac{4\delta_1t}{k}}\right]\bar{f}(y)dy\nonumber\\
&=&\int^{\infty}_{-\infty}\sqrt{\frac{\pi k}{4\delta_1 t}}\exp\left[-\frac{\left(\pi(x-y)-\frac{dt}{k}\right)^2}{\frac{4\delta_1t}{k}}\right]\bar{f}(y)dy\,,\\
\Rightarrow \rho_t(t,\psi) &=& \frac{1}{\pi}\bar{\rho}_t(t,x)\nonumber\\
&=& \int^{\infty}_{-\infty}\sqrt{\frac{k}{4\pi \delta_1 t}}\exp\left[-\frac{\left(\psi-\psi^{\prime}-\frac{\delta_2t}{k}\right)^2}{\frac{4\delta_1t}{k}}\right]f(\psi^{\prime})d\psi^{\prime}\,.
\end{eqnarray}

\pagebreak[4]
In the case of zero diffusion, i.e.~$\delta_1=0$, the fundamental solution, Equation~\ref{e:fundsol}, becomes

\begin{equation}
\bar{p}(x) = \sum_{n=-\infty}^{\infty}\exp\left[2\pi in\left(x-\frac{\delta_2t}{\pi k}\right)\right],
\end{equation}

\noindent i.e.~an initial distribution of $\delta(x)$ will become rotated by $\delta_2t/\pi k$, corresponding to an angle of $\delta_2t/k$.
For an arbitrary initial distribution $\bar{f}(x)$, this gives a solution

\begin{eqnarray}
\bar{\rho}_t(t,x) &=&\int^{1}_{0}\sum_{n=-\infty}^{\infty}\exp\left[2\pi in\left((x-y)-\frac{\delta_2t}{\pi k}\right)\right]\bar{f}(y)dy\,,\\
\rho_t(t,\psi) &=&\int^{\pi}_{0}\sum_{n=-\infty}^{\infty}\exp\left[2 in\left((\psi-\psi^{\prime})-\frac{\delta_2t}{k}\right)\right]f(\psi^{\prime})d\psi^{\prime}\,.
\end{eqnarray}

\vspace{1cm}
Thus, the drift term in the linear polarisation diffusion equation simply rotates the plane of polarisation of light by an angle $\alpha$: $\psi\rightarrow\psi^{\prime}=\psi+\alpha$, where $\alpha = \delta_2t/k$. In terms of the Stokes parameters, $V=0$ since there is no circular polarisation and $Q$ and $U$ become:
\begin{eqnarray}
Q\rightarrow Q^{\prime}&=& Q\cos 2\alpha - U\sin 2\alpha\,,\\
U\rightarrow U^{\prime}&=& U\cos 2\alpha + Q\sin 2\alpha\,.
\end{eqnarray}

The diffusion term acts as an overall suppression of polarisation. To see this, first note that given a collection of photons with a linear polarisation distribution $\rho(\psi)$, the Stokes parameters are
\begin{eqnarray}
Q &=& \int^{\pi}_0 \rho(\psi) \cos 2\psi  d\psi\,,\\
U &=& \int^{\pi}_0 \rho(\psi) \sin 2\psi  d\psi\,.
\end{eqnarray}

\pagebreak
Consider a source with an initial distribution $\rho_i(\psi) = \delta(\psi)$. This implies $Q_i=1,\,U_i=0$, and the light is fully polarised, i.e.~the polarisation fraction is $p=1$. Suppose $\rho_i$ evolves under some process, such as the diffusion process, to a distribution $p(\psi)$, the fundamental solution of the process. The Stokes parameters are then
%\pagebreak[3]
\begin{eqnarray}
Q_f &=& \int^{\pi}_0 p(\psi) \cos 2\psi\,d\psi\,,\\
U_f &=& \int^{\pi}_0 p(\psi) \sin 2\psi\,d\psi\,.
\end{eqnarray}
The polarisation fraction is now
\begin{eqnarray}
p^2 &=& Q_f^2 + U_f^2\nonumber\\
&=& \left(\int^{\pi}_0 p(\psi) \cos 2\psi\,d\psi\right)^2 + \left(\int^{\pi}_0 p(\psi) \sin 2\psi\,d\psi\right)^2\nonumber\\
&\leq& 1\,.
\end{eqnarray}

Assuming that the process is homogeneous, i.e.~$\delta(\psi-\psi^{\prime})\rightarrow p(\psi-\psi^{\prime})$, an arbitrary initial distribution $\rho_t(0,\psi)$ will evolve to $\rho_t(t,\psi) = \int^{\pi}_0\rho_t(0,\psi^{\prime})p(\psi-\psi^{\prime})d\psi^{\prime}\,$. The Stokes parameters are therefore
\begin{eqnarray}
Q_i &=& \int^{\pi}_0 \rho_t(0,\psi) \cos 2\psi\,d\psi\,,\\
U_i &=& \int^{\pi}_0 \rho_t(0,\psi) \sin 2\psi\, d\psi\,,\\
Q(t) &=& \int^{\pi}_0\int^{\pi}_0 \rho_t(0,\psi^{\prime})p(\psi-\psi^{\prime})\cos 2\psi\, d\psi^{\prime} d\psi\,,\\
U(t) &=& \int^{\pi}_0\int^{\pi}_0 \rho_t(0,\psi^{\prime})p(\psi-\psi^{\prime})\sin 2\psi\, d\psi^{\prime} d\psi\,.
\end{eqnarray}

Consider the linear polarisation diffusion equation with drift $\delta_2=0$. 
For zero drift the fundamental solution, $p(\psi)$, is given by Equation~\ref{e:fundsol} with $\delta_2 = 0$:
\begin{equation}
p(\psi) = \frac{1}{\pi}\sum_{n=-\infty}^{\infty}\exp\left(\frac{-4\delta_1n^2t}{k}\right)e^{2in\psi}\,.
\end{equation}
The $Q$ Stokes parameter is therefore
\begin{eqnarray}
Q(t) &=& \int^{\pi}_0\int^{\pi}_0 \rho_t(0,\psi^{\prime})\frac{1}{\pi}\sum_{n=-\infty}^{\infty}\exp\left(\frac{-4\delta_1n^2t}{k}\right)e^{2in(\psi-\psi^{\prime})}\cos 2\psi\, d\psi^{\prime} d\psi\nonumber\\
&=& \frac{1}{\pi}\sum_{n=-\infty}^{\infty}\exp\left(\frac{-4\delta_1n^2t}{k}\right)\int^{\pi}_0\int^{\pi}_0 \rho_t(0,\psi^{\prime})\cos 2n\!\left(\psi-\psi^{\prime}\right)\cos 2\psi\, d\psi^{\prime} d\psi\nonumber\\
&=&
\frac{1}{\pi}\sum_{n=-\infty}^{\infty}\exp\left(\frac{-4\delta_1n^2t}{k}\right)\nonumber\\
&&
\times\int^{\pi}_0\int^{\pi}_0 \rho_t(0,\psi^{\prime})\left[\cos 2n\psi\cos 2n\psi^{\prime}+\sin 2n\psi\sin 2n\psi^{\prime}\right]\cos 2\psi\,d\psi^{\prime} d\psi\nonumber\\
&=&
\frac{1}{\pi}\sum_{n=-\infty}^{\infty}\exp\left(\frac{-4\delta_1n^2t}{k}\right)
\int^{\pi}_0\cos 2n\psi\cos 2\psi\, d\psi
\int^{\pi}_0 \rho_t(0,\psi^{\prime})\cos 2n\psi^{\prime}d\psi^{\prime}\nonumber\\
&&
+\frac{1}{\pi}\sum_{n=-\infty}^{\infty}\exp\left(\frac{-4\delta_1n^2t}{k}\right)
\int^{\pi}_0\sin 2n\psi\cos 2\psi\, d\psi
\int^{\pi}_0 \rho_t(0,\psi^{\prime})\sin 2n\psi^{\prime}d\psi^{\prime}\,.\nonumber\\
\end{eqnarray}
The second sum vanishes and only the $n=\pm1$ terms in the first sum contribute, leaving
\begin{eqnarray}
Q(t)&=&
\exp\left(\frac{-4\delta_1 t}{k}\right)\int^{\pi}_0 \rho_t(0,\psi^{\prime})\cos 2\psi^{\prime}d\psi^{\prime}\nonumber\\
&=&
\exp\left(\frac{-4\delta_1 t}{k}\right)Q_i\,.
\end{eqnarray}
Similarly, the $U$ Stokes parameter is
\begin{eqnarray}
U(t) &=& \int^{\pi}_0\int^{\pi}_0 \rho_t(0,\psi^{\prime})\frac{1}{\pi}\sum_{n=-\infty}^{\infty}\exp\left(\frac{-4\delta_1n^2t}{k}\right)e^{2in(\psi-\psi^{\prime})}\sin 2\psi\, d\psi^{\prime} d\psi\nonumber\\
&=& \frac{1}{\pi}\sum_{n=-\infty}^{\infty}\exp\left(\frac{-4\delta_1n^2t}{k}\right)\int^{\pi}_0\int^{\pi}_0 \rho_t(0,\psi^{\prime})\cos 2n\!\left(\psi-\psi^{\prime}\right)\sin 2\psi\, d\psi^{\prime} d\psi\nonumber\\
&=&
\frac{1}{\pi}\sum_{n=-\infty}^{\infty}\exp\left(\frac{-4\delta_1n^2t}{k}\right)\nonumber\\
&&
\times\int^{\pi}_0\int^{\pi}_0 \rho_t(0,\psi^{\prime})\left[\cos 2n\psi\cos 2n\psi^{\prime}+\sin 2n\psi\sin 2n\psi^{\prime}\right]\sin 2\psi\, d\psi^{\prime} d\psi\nonumber\\
&=&
\frac{1}{\pi}\sum_{n=-\infty}^{\infty}\exp\left(\frac{-4\delta_1n^2t}{k}\right)
\int^{\pi}_0\cos 2n\psi\sin 2\psi\, d\psi
\int^{\pi}_0 \rho_t(0,\psi^{\prime})\cos 2n\psi^{\prime}d\psi^{\prime}\nonumber\\
&&
+\frac{1}{\pi}\sum_{n=-\infty}^{\infty}\exp\left(\frac{-4\delta_1n^2t}{k}\right)
\int^{\pi}_0\sin 2n\psi\sin 2\psi\, d\psi
\int^{\pi}_0 \rho_t(0,\psi^{\prime})\sin 2n\psi^{\prime}d\psi^{\prime}\,.\nonumber\\
\end{eqnarray}
Here the first sum vanishes and the $n=\pm1$ terms of the second sum give
\begin{eqnarray}
U(t)&=&
\exp\left(\frac{-4\delta_1 t}{k}\right)\int^{\pi}_0 \rho_t(0,\psi^{\prime})\sin 2\psi^{\prime}d\psi^{\prime}\nonumber\\
&=&
\exp\left(\frac{-4\delta_1 t}{k}\right)U_i\,.
\end{eqnarray}
The polarisation suppression due to the diffusion is thus
\begin{eqnarray}
p^2&=&\frac{p_f^2}{p_i^2}\nonumber\\
&=&\frac{Q(t)^2+U(t)^2}{Q_i^2+U_i^2}\,,\\
\Rightarrow p&=&\exp\left(\frac{-4\delta_1 t}{k}\right)\,.
\end{eqnarray}

Both the suppression and rotation are independent of the current polarisation angle, and thus in the general case of nonzero $\delta_1$ and $\delta_2$ a linearly polarised beam in a spacetime with an underlying discreteness will experience a rotation in polarisation angle and a suppression of its polarisation fraction.

%% file: ObservationalConstraints.tex
\section{Observational constraints}
\label{s:polarobs}

The model discussed here is by no means the only mechanism for producing rotation of the plane of linear polarisation. Such rotations, when discussed in the context of the polarisation of astrophysical or cosmological sources, are usually referred to as `cosmological birefringence'. Rotation of polarisation is not merely hypothetical -- the polarisation from cosmological sources is well known to be subject to rotation, namely the Faraday effect. Light travelling through the interstellar medium is subject to rotation of the plane of polarisation, because the presence of a magnetic field results in different refractive indices for left and right circularly polarised modes. The angle of rotation, $\alpha$, is dependent on the wavelength: $\alpha\propto\lambda^2$.

Proposed methods for generating additional cosmological birefringence include a loop quantum gravity motivated modification of the dispersion relation~\cite{Gambini:1998it,Gleiser:2001rm}. This modification leads to circular polarisation modes travelling at different velocities and thus can give birefringence. In this case the rotation angle $\alpha\propto 1/\lambda^2\propto k^2$. Quantum gravity motivated birefringence proportional to $k^2$ is also discussed in~\cite{Maccione:2008tq} and~\cite{Gubitosi:2009eu}, where polarisation data from the Crab Nebula and CMB respectively are used to constrain the model parameter.

Birefringence also arises in discussions of dark energy, specifically the quintessence approach. The coupling of the quintessence scalar field to the pseudoscalar of electromagnetism leads to birefringence independent of wavelength~\cite{Carroll:1998zi,Liu:2006uh,Lue:1998mq}. Violating CPT symmetry through the addition of a Chern-Simons term to the Maxwell Lagrangian also results in rotation of the plane of polarisation independent of wavelength~\cite{Li:2008tma,Carroll:1989vb,Feng:2004mq}.
Constraints on such polarisation rotation come from radio galaxy polarisation data and CMB polarisation. Methods of detecting polarisation rotation through radio galaxy data and the CMB will be discussed further below.

None of the proposed effects discussed above have the $\alpha\propto 1/k$ dependence of the polarisation diffusion model. One source of a $1/k$ rotation has been suggested previously: according to Prasanna and Mohanty~\cite{Prasanna:2001xk}, polarisation rotation could be induced by a gravitational wave propagating in the same direction as the photons. This effect would only be present where there is a source of strong gravitational radiation, such as a binary pulsar, and thus it is unlikely to be confused with the effect studied here. 
Note also that most models that predict birefringence violate Lorentz invariance as well as parity. The polarisation diffusion model of this chapter is particularly interesting because it preserves Lorentz invariance. 

Before discussing the bounds on the model parameters, it should be noted that special care needs to be taken when searching for changes in polarisation. Unless the method of generation of polarisation in a source is completely understood, natural variations in polarisation at the source could be mistaken for effects occurring over the travel time of the light to us. 

\subsection{Constraining the drift parameter with radio galaxy data}
\label{ss:radiogalaxy}

\begin{figure}[t]
\begin{center}
\includegraphics[width=0.6\textwidth]{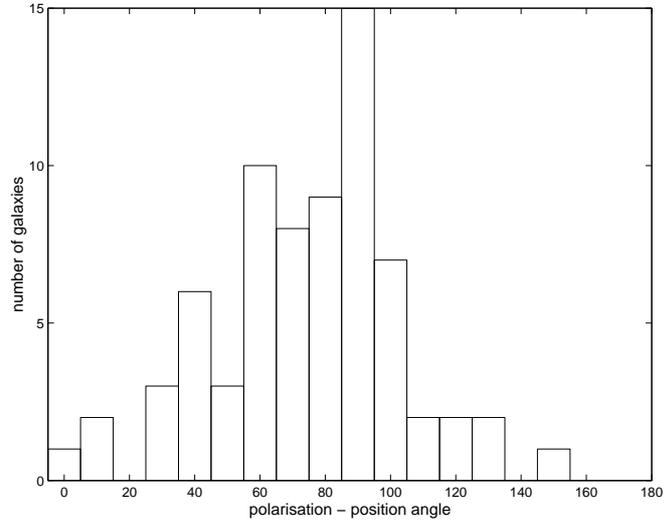}
\caption[Polarisation of radio galaxies]{The correlation between radio galaxy axis and polarisation angle. Data from 71 galaxies with redshift $z>0.3$ is shown here. This data is taken from~\cite{Carroll:1989vb,Carroll:1997tc}.}
\label{f:polarisationdata}
\end{center}
\end{figure}

Radio galaxies emit linearly polarised light and often have an elongated structure defining a natural axis against which their polarisation angle can be measured. After subtracting the effect of Faraday rotation it is found that the polarisation angle for radio galaxies is highly correlated with the observed galaxy axis (radio axis), peaking at $90^{\circ}$ to the radio axis (see, for example,~\cite{Carroll:1997tc, Carroll:1989vb, Cimatti:1994yc} and Figure~\ref{f:polarisationdata}). The linear polarisation of radio galaxies is due to the synchrotron process. Although the cause of the correlation between polarisation angle and radio axis does not seem to be fully understood it is likely there is a physical basis for the alignment of the magnetic field. The correlation holds over a range of redshifts and thus provides a method of constraining a birefringence effect. If it is assumed that the light has a polarisation of $90^{\circ}$ at the source, the difference from $90^{\circ}$ in the measured polarisation angle constrains the amount of rotation, and in the case of the model discussed here, the magnitude of the drift parameter. Data also exists for polarisation fraction for radio galaxies. Unfortunately there seems to be less understanding of what the polarisation fraction is at the source and thus the change in polarisation fraction (and the diffusion parameter) cannot be constrained.

In this section I will use data from one particular galaxy to place an order of magnitude bound on the drift parameter. In~\cite{Cimatti:1994yc} the galaxy MRC 2025-218 was studied. Cimatti et al.~deduced a polarisation angle (with respect to the galaxy axis) of $\theta = 93\pm8^{\circ}$ and a polarisation fraction of $p=8.3\pm2.3\%$. They concluded that there is ``strong evidence that the plane of polarisation is not rotated by more than 10 degrees when the radiation travels from $z=2.63$ to us''. Recall from Section~\ref{s:polarcosmictime} that the angle of rotation is related to the drift parameter by $\alpha = \delta_2 t/k$. The time of travel $t$ can be calculated from the redshift, $z=2.63$, of the galaxy. As mentioned in Section~\ref{s:spectralline} the \mbox{cosmic time - redshift relation},
\begin{equation}
t_u = \frac{2}{3H_0\Omega_m^{1/2}(1+z)^{3/2}}\,,
\end{equation}
gives the approximate age of the universe when photons at a redshift of $z$ were emitted. Taking the Hubble parameter to be $H_0=70.5 kms^{-1}Mpc^{-1}$, the matter density to be $\Omega_M=0.274$, and the current age of the universe to be $t_0=13.72\times 10^9$ years (see~\cite{Hinshaw:2008kr}), gives $t_u=2.55\times 10^9$ years, and thus the photons have been travelling for a cosmic time of $t=11.17\times10^9$ years.

Cimatti et al.~observed the galaxy MRC 1015-218 in the $R$ band, corresponding to a rest frame spectral region (i.e.~non-redshifted wavelength) of $\Delta\lambda_{rest}=1600$ to $2100$\AA. In the calculation below, I will assume the photons were emitted with a wavelength of $\lambda=1850$\AA. As the polarisation rotation angle, $\alpha$, depends on the wavelength, some error is introduced by taking $\lambda$ to be a fixed value within the bandwidth. The variation of $\alpha$ over this wavelength range is not expected to be significant however, and should not affect the order of magnitude bound obtained.

The polarisation diffusion model assumes Minkowksi spacetime, and thus to correctly bound the drift parameter the change in energy of the photons from the galaxy due to the expansion of the universe must be taken into account. The inclusion of expansion into the model will be discussed more generally in Section~\ref{ss:polarCMBbounds}, here I will only note that the wavelength of photons in an expanding universe is given by
\begin{equation}
\lambda(t) = \lambda_{e} \frac{a(t)}{a_{e}}\,,
\end{equation}
where $\lambda_{e}$ is the wavelength the photons were emitted at, $a(t)$ is the scale factor, and $a_{e}$ is the scale factor at the time when the photons were emitted. Assuming a matter dominated FRW universe, the scale factor goes as $a\sim t^{2/3}$, i.e.~$a(t) = t^{2/3}/t_0^{2/3}$ where $t_0$ is the current age of the universe and the scale factor today is taken to be $a_0 = 1$. 
The total angle of rotation of the polarization is thus
\begin{eqnarray}
\Delta\alpha &=& \int_{t_e}^{t_0}{\delta_2 \lambda(t) dt}\nonumber\\
&=& \int_{t_e}^{t_0}{\delta_2 \lambda_{e}\frac{a(t)}{a_{e}}dt}\nonumber\\
&=& \int_{t_e}^{t_0}{\delta_2\lambda_{e}\frac{t^{2/3}}{t_{e}^{2/3}} dt}\nonumber\\
&=& \frac{3\delta_2\lambda_{e}}{5}\left(\frac{t_0^{5/3}}{t_{e}^{2/3}}  - t_{e}\right)\,.
\end{eqnarray}
The drift parameter can be constrained to be
\begin{eqnarray}
\delta_2 
 &<& 7\times 10^{-90}\nonumber\\
 &<& 2\times 10^{-37}kgm^2s^{-3}.
\end{eqnarray}

\subsection{Polarisation diffusion and drift in the cosmic microwave background}
\label{ss:polarCMBbounds}

The effect of polarisation rotation on CMB data has been well studied. Suppression of polarisation, however, seems to have been little considered. Before discussing the CMB we need to fully understand how the expansion of the universe changes the diffusion equation. In fact, an equation that includes expansion can be determined in a similar manner to that discussed in Section~\ref{s:CMBbounds}.

\subsubsection{Including the expansion of the universe}

The redshifting of the photons is given by
\begin{equation}
k(t) = k_{e} \frac{a_{e}}{a(t)},
\end{equation}
where $k_{e}$ is the emitted frequency, $a(t)$ is the scale factor, and $a_{e}$ is the scale factor at the time the photons were emitted. 
Note that the expansion of the universe also dilutes the photons, but this dilution does not change the diffusion equation.
As mentioned in the previous section, if a matter dominated FRW universe is assumed, $a(t) = t^{2/3}/t_0^{2/3}$, where $t_0$ is the current age of the universe and the scale factor today is taken to be $a_0=1$.
If the form for $k$ is substituted into the diffusion equation, Equation~\ref{e:poldiff}, it becomes
\begin{eqnarray}
\pd{\rho_t}{t}&=& \frac{\delta_1 a(t)}{k_{e} a_{e}}\pdsq{\rho_t}{\psi}-\frac{\delta_2 a(t)}{k_{e} a_{e}}\pd{\rho_t}{\psi}\nonumber\\
&=&\frac{t^{2/3}}{t_0^{2/3}}\left(\frac{\delta_1}{k_{e} a_{e}}\pdsq{\rho_t}{\alpha}-\frac{\delta_2}{k_{e} a_{e}}\pd{\rho_t}{\psi}\right).
\end{eqnarray}
A new time variable $t^{\prime}$ can be defined by
\begin{equation}
\frac{dt^{\prime}}{dt} = a(t),
\end{equation}
and thus
\begin{eqnarray}
\pd{\rho_t}{t^{\prime}}&=&\frac{t^{2/3}}{t_0^{2/3}}\pd{\rho_t}{t}\,,\\
\Rightarrow \pd{\rho_t}{t}&=& \pd{\rho_t}{t^{\prime}}\frac{dt^{\prime}}{dt}\nonumber\\
&=&\frac{t^{2/3}}{t_0^{2/3}}\pd{\rho_t}{t^{\prime}}\,,\\
\Rightarrow \pd{\rho_t}{t^{\prime}}&=& \frac{\delta_1}{k_{e} a_{e}}\pdsq{\rho_t}{\psi}-\frac{\delta_2}{k_{e} a_{e}}\pd{\rho_t}{\psi}.
\end{eqnarray}
If we further note $k_{e}/k_0 = a_0/a_{e}\,$, where $k_0$ is the energy observed today, the equation becomes simply
\begin{equation}
\pd{\rho_t}{t^{\prime}} =  \frac{\delta_1}{k_0}\pdsq{\rho_t}{\psi}-\frac{\delta_2}{k_0}\pd{\rho_t}{\psi}.
\end{equation}
In other words, the change in energy due to expansion can be compensated for by fixing the energy at the observed energy and simply transforming to a new time variable
\begin{equation}
t^{\prime} = \frac{3}{5}\frac{t^{5/3}}{t_0^{2/3}}+\textrm{const}.
\end{equation}
For analysis of the CMB this is approximately equivalent simply to evolving the equation for 3/5 the time since $t$ is so close to $t_0$.

\subsubsection{The polarisation of the CMB}

The cosmic microwave background is expected to be partially polarised due to Thomson scattering off free electrons at the surface of last scattering. Thomson scattering can produce only linear polarisation (and only do so when there is a quadrupole anisotropy in the incident radiation) and thus the CMB is expected to have no circular polarisation. The CMB polarisation has now been detected by multiple experiments, including WMAP~\cite{Dunkley:2008mk}, BICEP~\cite{Chiang:2009xsa}, and QUaD~\cite{Brown:2009uy}.

The polarisation of the CMB is commonly discussed not in terms of Stokes parameters, or angles on the Bloch/Poincar\'e sphere, but rather in terms of $E$ and $B$ modes (see, e.~g.~\cite{Dodelson:2003}). Instead of describing the polarisation of a single photon, $E$ and $B$ modes describe how the polarisation varies over a small area. The names `$E$' and `$B$' mode are drawn from an analogy with electric and magnetic fields: the polarisation pattern on the sky (direction and intensity) is decomposed into a curl-free component ($E$ mode) and a divergence free component ($B$ mode). The exact method by which these components are calculated is not important for the current discussion. As mentioned above, a quadrupole anisotropy in the radiation at last scattering is necessary for there to be polarisation in the CMB. This anisotropy can arise from scalar, vector, or tensor perturbations. Scalar perturbations, due to density fluctuations in the plasma, give rise to $E$ mode polarisation; vector perturbations, due to vorticity in the plasma, are expected to be negligible; tensor perturbations can be caused by gravitational waves, and would give rise to $B$ mode polarisation. 

\begin{figure}[t]
\begin{center}
\subfigure[TT.]{
\includegraphics[width = 0.48\textwidth]{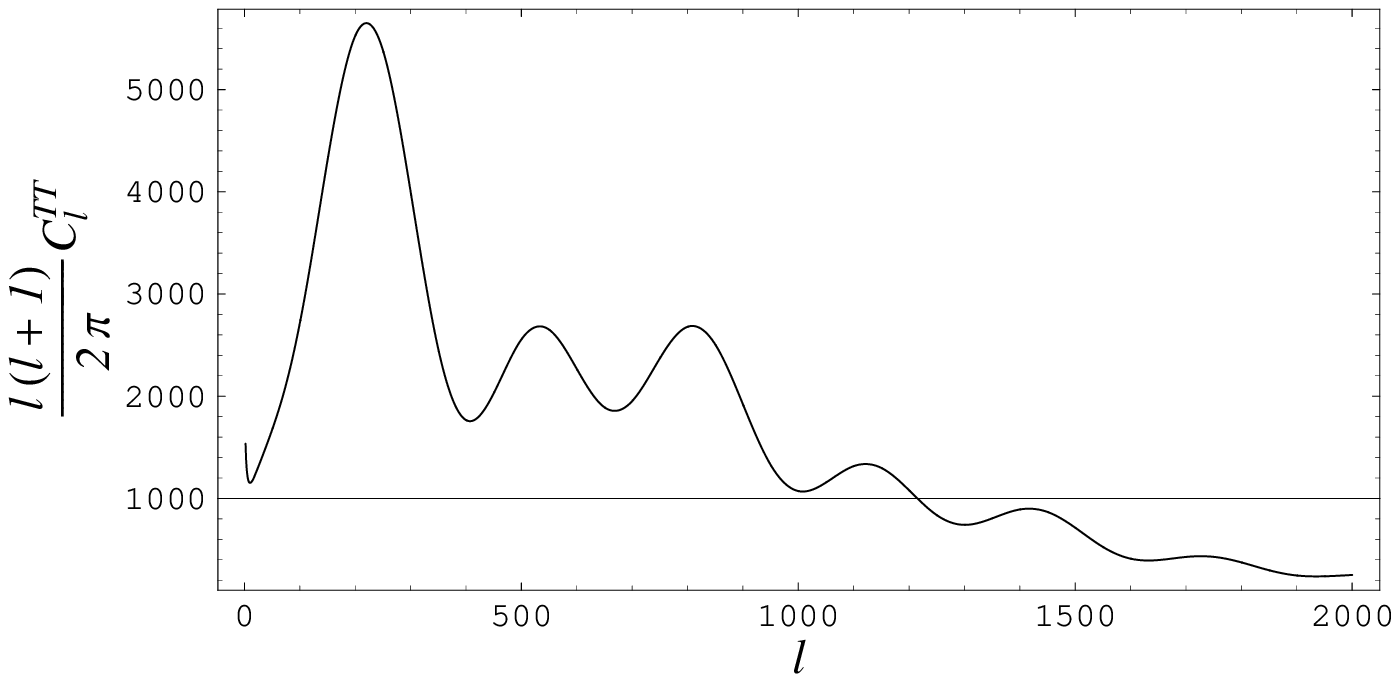}}
\subfigure[TE.]{
\includegraphics[width = 0.48\textwidth]{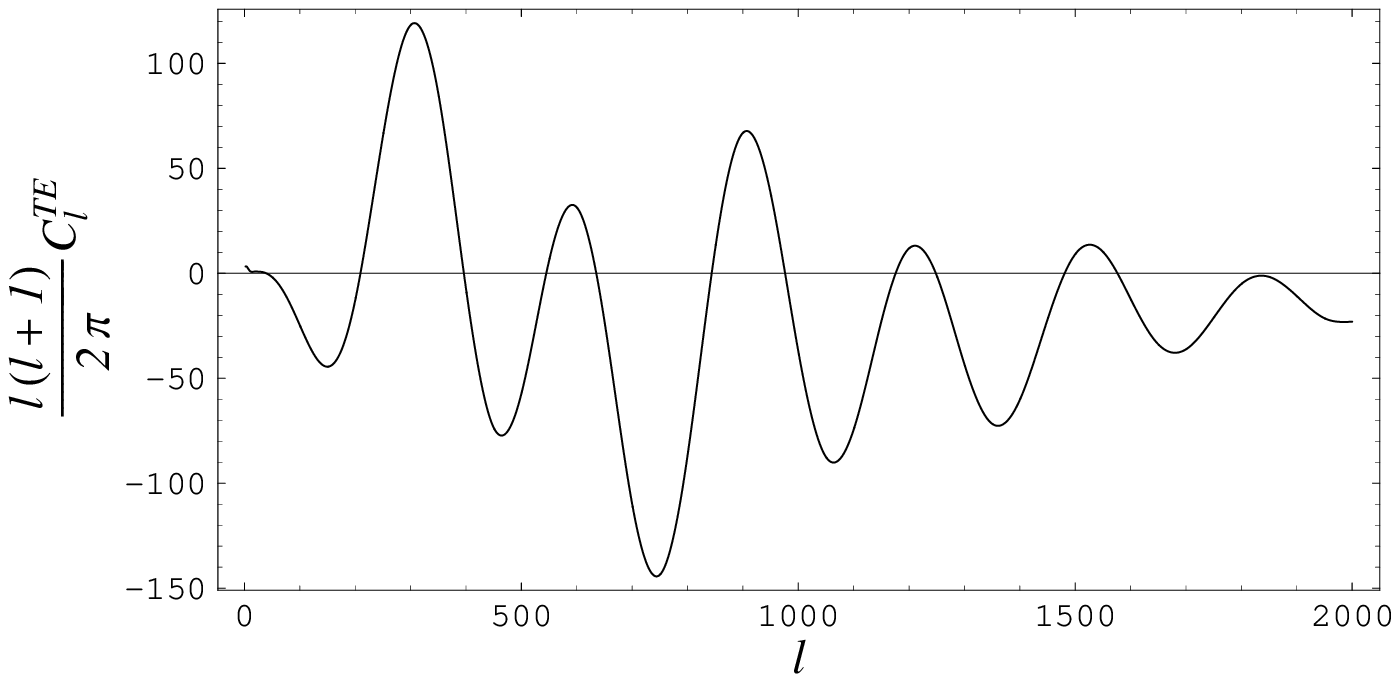}}
\subfigure[EE.]{
\includegraphics[width = 0.48\textwidth]{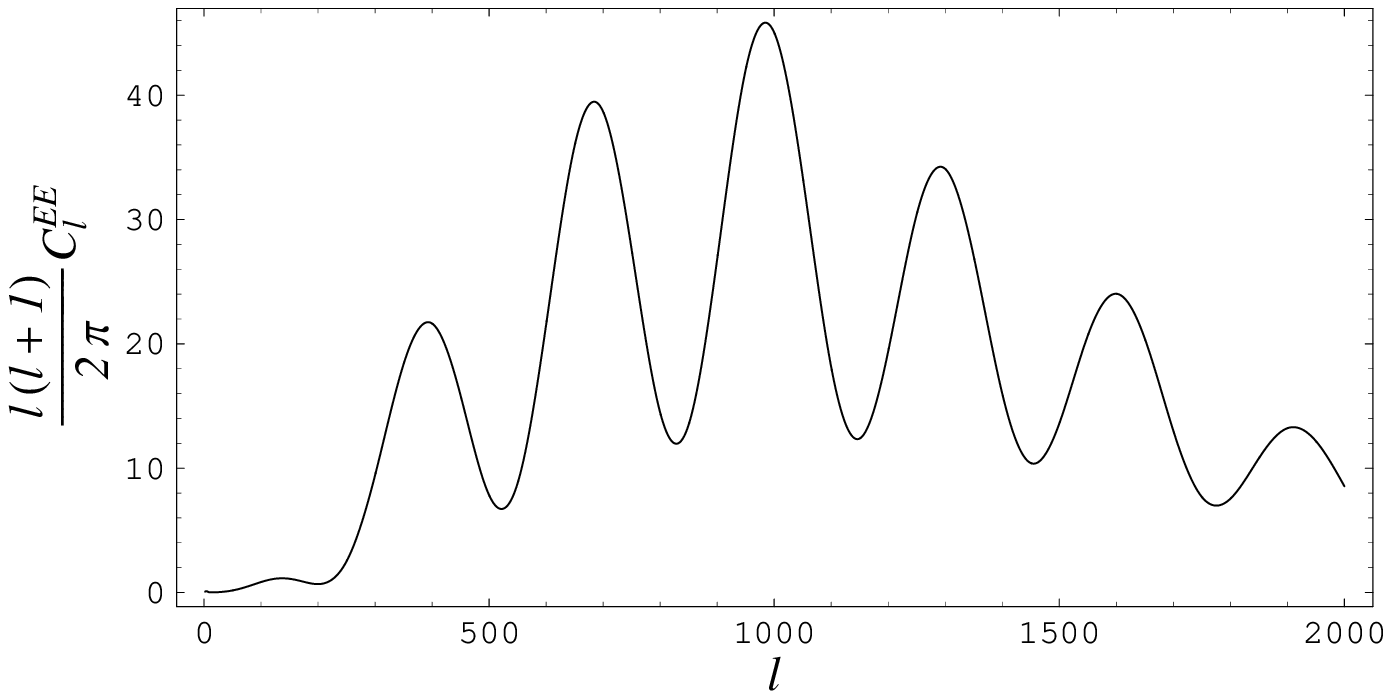}}
\subfigure[BB.]{
\includegraphics[width = 0.48\textwidth]{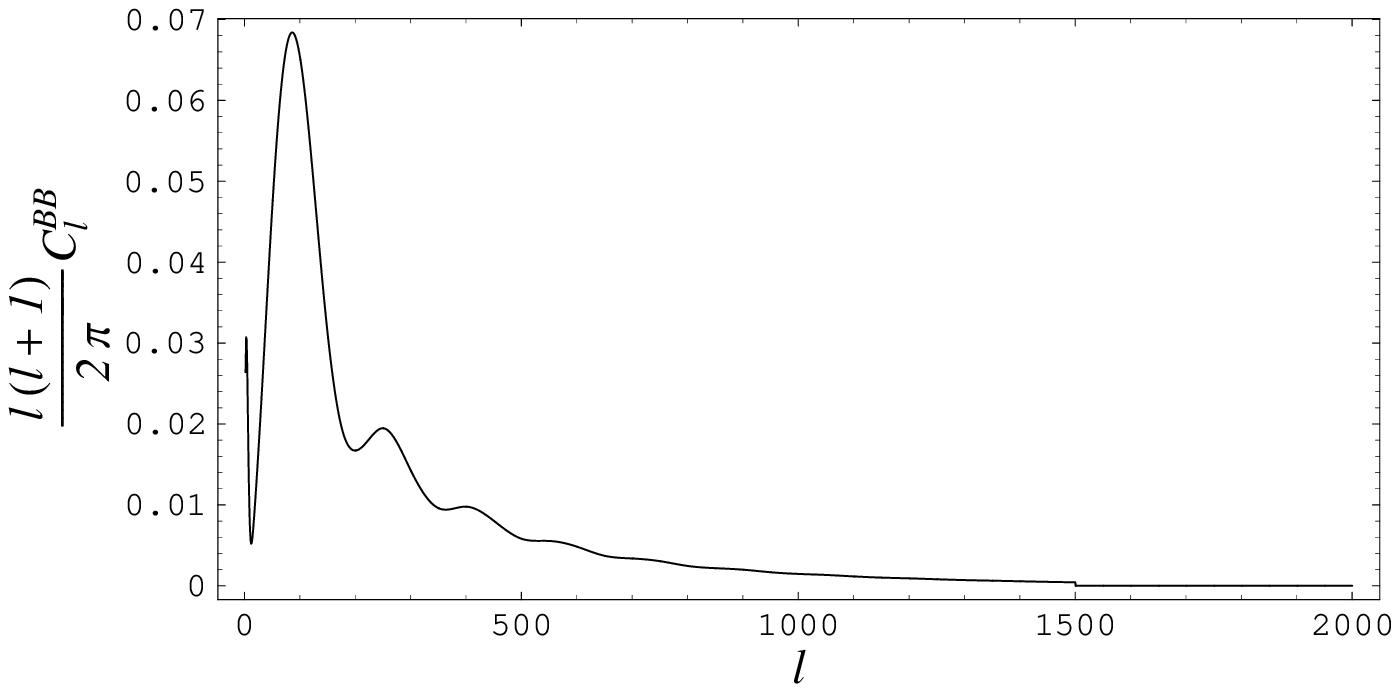}}
\caption[CMB correlation spectra]{CMB $TT,\,TE,\,EE$ and $BB$ correlation spectra produced using CAMB with standard parameters. $l$ is the multipole moment.}
\label{f:stdspectra}
\end{center}
\end{figure} 

Analysis of the CMB is usually in terms of correlations between $E$ mode, $B$ mode, and temperature $T$. In all, there are six cross-correlation spectra that can be examined: $TT,\,TE,\,TB,\,EE,\,EB,\,BB$. In the absence of a parity violating effect, the correlators $TB$ and $EB$ are expected to vanish. The rotation of the polarisation angle can thus be detected through the presence of a nonzero $TB$ and $EB$. More specifically, for an angle of rotation $\Delta\alpha$ the exact change in the correlators can be written as (see, for example~\cite{Lue:1998mq,Feng:2006dp,Li:2008tma}):
\begin{eqnarray}
{C^{\prime}}_l^{TB} &=& C_l^{TE}\sin 2\Delta\alpha\,,\\
{C^{\prime}}_l^{EB} &=& \frac{1}{2}\left(C_l^{EE} - C_l^{BB}\right)\sin 4\Delta\alpha\,,\\
{C^{\prime}}_l^{TE} &=& C_l^{TE}\cos 2\Delta\alpha\,,\\
{C^{\prime}}_l^{EE} &=& C_l^{EE}\cos^2 2\Delta\alpha + C_l^{BB}\sin^2 2\Delta\alpha\,,\\
{C^{\prime}}_l^{BB} &=& C_l^{BB}\cos^2 2\Delta\alpha + C_l^{EE}\sin^2 2\Delta\alpha\,, 
\end{eqnarray}
where $C_l$ are the spectra when there is no rotation. Note that the $TT$ correlation is unchanged by rotation.
For the polarisation diffusion and drift model discussed here the angle $\Delta\alpha$ is dependent on the energy, and thus the expected spectra above will depend on the frequency band one is working in. 

\begin{figure}[t]
\begin{center}
\subfigure[TB.]{
\includegraphics[width = 0.6\textwidth]{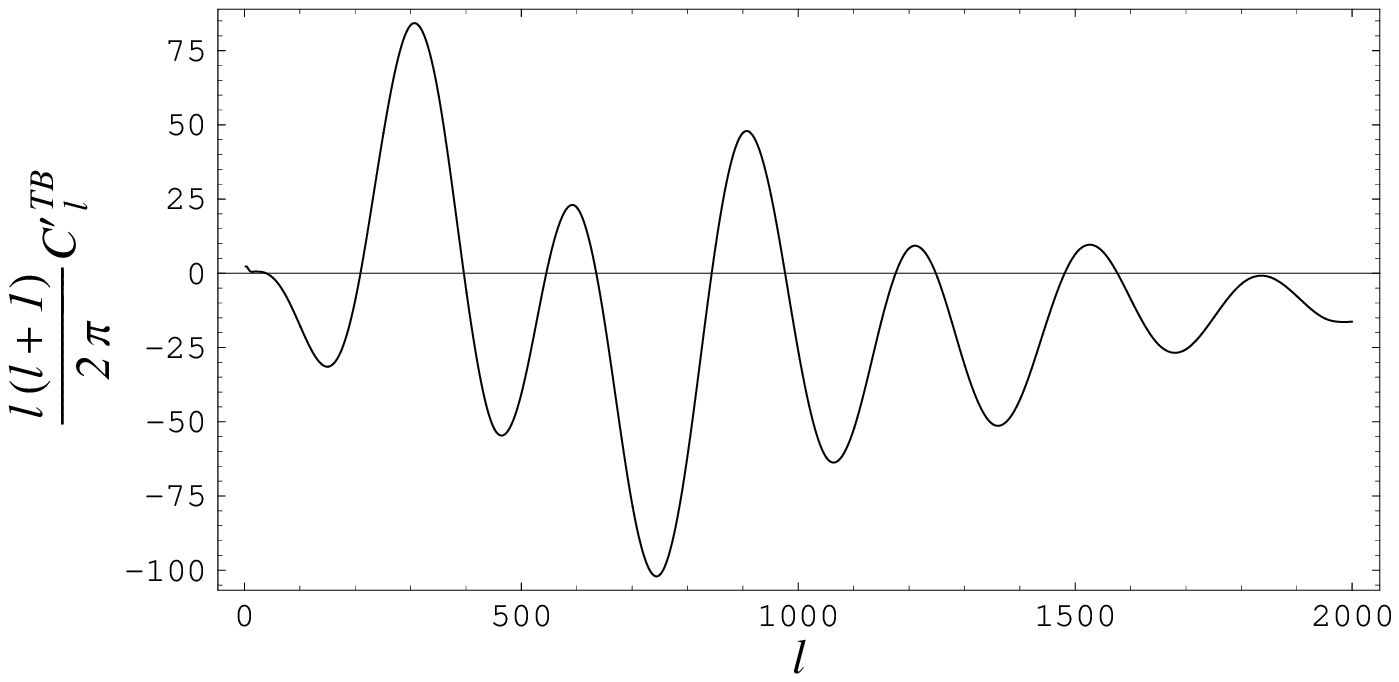}}
\subfigure[EB.]{
\includegraphics[width = 0.6\textwidth]{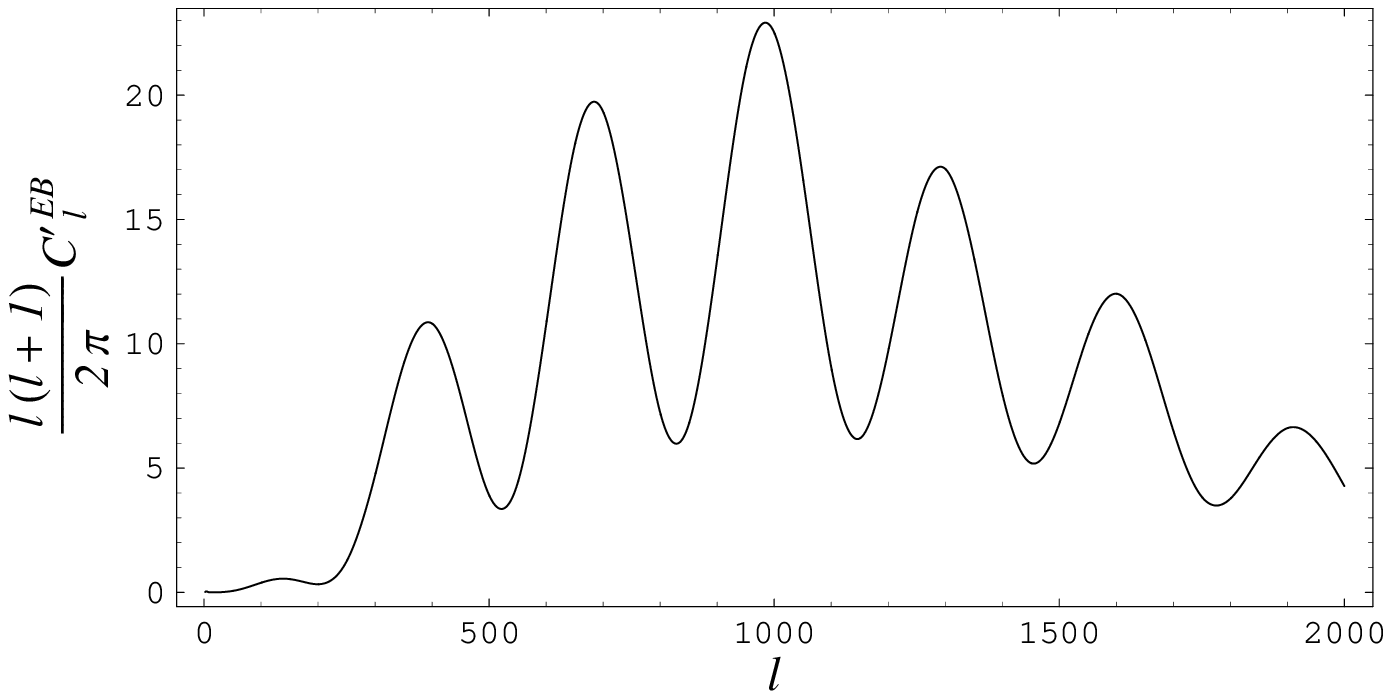}}
\caption[$TB$ and $EB$ spectra under rotation]{CMB $TB$ and $EB$ correlation spectra when polarisation is rotated by an arbitrary angle $\Delta\alpha=\pi/8$.}
\label{f:newTBEB}
\end{center}
\end{figure} 

\begin{figure}[p]
\begin{center}
\subfigure[TE.]{
\includegraphics[width = 0.6\textwidth]{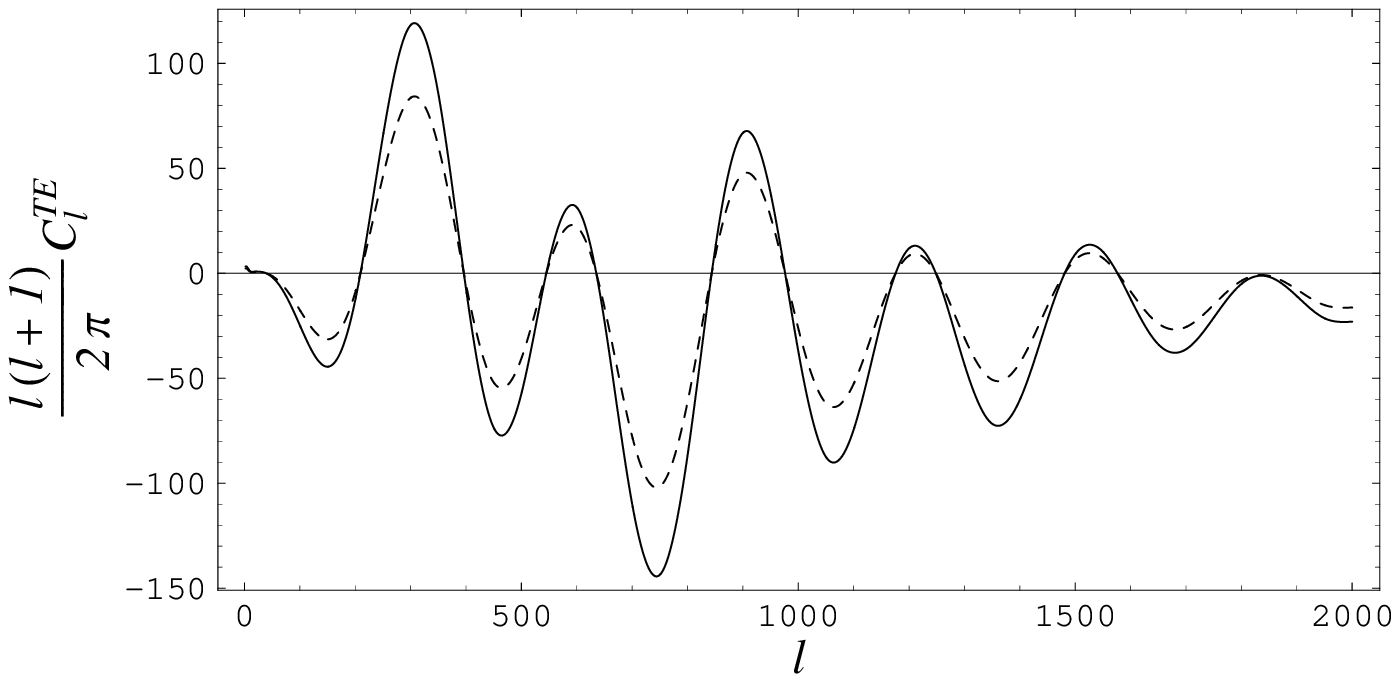}}
\subfigure[EE.]{
\includegraphics[width = 0.6\textwidth]{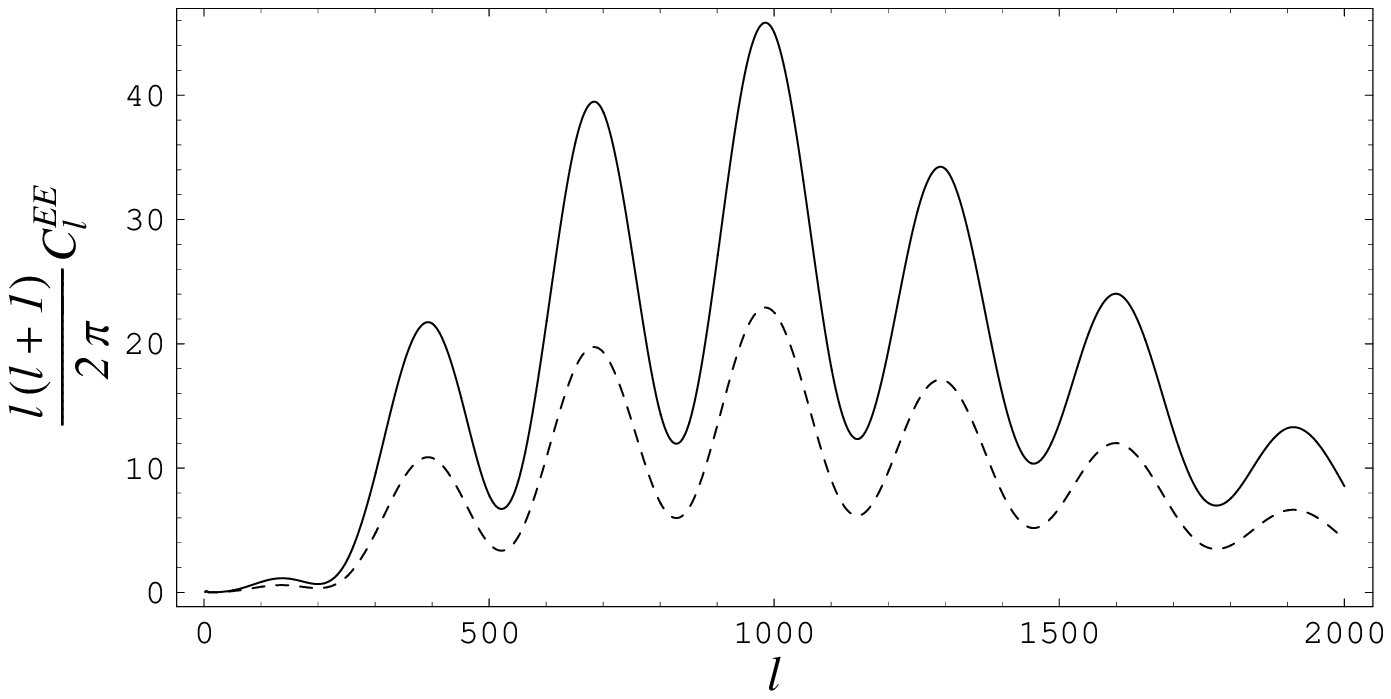}}
\subfigure[BB.]{
\includegraphics[width = 0.6\textwidth]{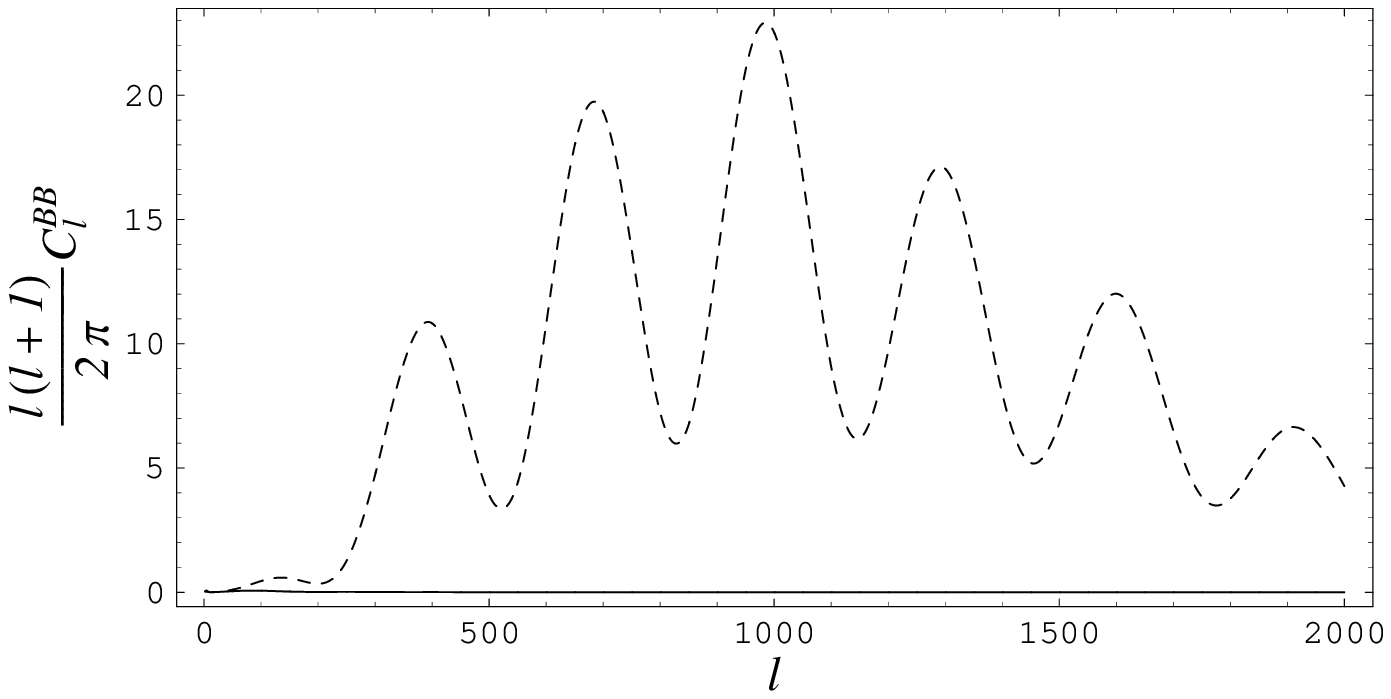}}
\caption[$TE,\,EE,$ and $BB$ spectra under rotation]{CMB $TE,\,EE,$ and $BB$ correlation spectra when polarisation is rotated by an arbitrary angle $\Delta\alpha=\pi/8$ (dashed lines) compared to the spectra in the unrotated case (solid lines).}
\label{f:comparisonspectra}
\end{center}
\end{figure} 

Example correlation spectra can be produced using the Code for Anisotropies in the Microwave Background (CAMB).\footnote{Available at http://camb.info/.} Figure~\ref{f:stdspectra} shows $TT,\,TE,\,EE$ and $BB$ spectra for standard parameters, without any polarisation rotation (recall $TB$ and $EB$ vanish). Figure~\ref{f:newTBEB} shows the $TB$ and $EB$ spectra that would result if a rotation of polarisation occurs; an arbitrary angle $\Delta\alpha=\pi/8$ has been chosen to illustrate the effect. The changes in $TE$, $EE$, and $BB$ spectra under this rotation are shown in Figure~\ref{f:comparisonspectra}. Note that there is some difficulty determining what values of $\Delta\alpha$ are allowed by current data. For example, Xia et al.~\cite{Xia:2009ah} give differing (and seemingly contradictory at $68\%\,C.L.$) values of $\Delta\alpha =-2.62\pm0.87^{\circ}$, $\Delta\alpha =0.59\pm0.42^{\circ}$, and $\Delta\alpha =0.09\pm0.36^{\circ}$ depending on which subset of CMB experiments they analyse. Note that as they assume $\Delta\alpha$ is frequency independent these values cannot be used to constrain the drift parameter $\delta_2$. Recall that $\alpha = \delta_2 t^{\prime}/k$ when expansion of the universe is included. Taking $t^{\prime}=3/5\times 10^{60}$, and choosing one of the WMAP observation frequencies, $40$GHz, shows that the (probably too large) angle $\Delta\alpha =\pi/8$ in fact corresponds to $\delta_2\sim 10^{-93}$, already several orders of magnitude smaller than the bound placed on $\delta_2$ with radio galaxy data in Section~\ref{ss:radiogalaxy}.

\pagebreak[4]
The discussion above has all been concerned with the effect of the rotation of the plane of polarisation on CMB data. Effects such as the suppression of polarisation predicted by the polarisation diffusion equation do not seem to have been discussed much in the literature. In terms of the correlation spectra above, the diffusion will simply reduce the amplitude of each spectra by some fixed, energy dependent, amount.
To properly constrain the drift and diffusion parameters, CMB data from a single frequency band must be compared with the corresponding predicted spectra generated by CAMB and modified for drift and diffusion. Comparing over a range of drift and diffusion parameters will allow the values that best fit current CMB observations to be found. Unfortunately this process requires modifying the Markov Chain Monte Carlo code, CosmoMC, and is outside the scope of this thesis. This work has recently been undertaken by Carlo Contaldi and the results appear in~\cite{Contaldi:2010fh}.

%% file: PolarizationConclusion.tex
\section{Concluding remarks}
\label{s:polarconclusion}

When studying massless particles with polarisation in a discrete spacetime, the assumption of Lorentz invariance turns out to be insufficient to completely constrain the phenomenological model. To make concrete predictions it is necessary to restrict attention to linear polarisation -- fortunately it is precisely linear polarisation that is of interest in astrophysics and cosmology.

As yet there is no underlying microscopic model for polarised photons on a causal set (or even, as discussed in Chapter~\ref{c:massless}, unpolarised photons). When particles in causal set theory are better understood it is hoped that useful models including circular polarisation will be able to be developed and the validity of the results obtained in the current restriction to linear polarisation checked.

For linear polarisation a polarisation diffusion equation was derived in Sections~\ref{s:polaraffinetime} and~\ref{s:polarcosmictime} in terms of affine and cosmic time respectively. The tight constraints on energy diffusion and drift obtained in Chapter~\ref{c:massless} allow the polarisation diffusion equation to be simplified to a diffusion equation in terms of polarisation angle with two free parameters. A distribution of linearly polarised photons is found to experience a suppression in polarisation determined by the diffusion parameter $\delta_1$ and a rotation of polarisation determined by the drift parameter $\delta_2$.

The suppression of polarisation over the travel time of photons from distant sources to us does not seem to have been much considered. Unfortunately it is, for the most part, a difficult effect to constrain since the polarisation fraction at the source must first be well understood.

There are a number of different proposed models that would generate polarisation rotation over the travel time of photons. 
The model proposed here is distinct, however, in the $1/k$ dependence of its rotation and in that it preserves Lorentz invariance. The polarisation rotation is easily constrained by existing data. A natural correlation between the polarisation angle and the axis of radio galaxies allowed a constraint of $\delta_2< 2\times 10^{-37}kgm^2s^{-3}$ to be placed in Section~\ref{s:polarobs}. The photons from the CMB, having had more time to experience the underlying discreteness, impose a tighter constraint on  the rotation and also constrain the suppression. These constraints are discussed in~\cite{Contaldi:2010fh}.

%% file: Conclusion.tex
\chapter{Conclusions and future directions}

Lorentz invariance is not optional in the causal set approach to quantum gravity. It is an essential feature of the theory and proves a very powerful constraint in the development of causal set phenomenology. A surprising amount can be deduced beginning with only the simple assumptions that spacetime is discrete and Lorentz invariant. 

The behaviour of massive particles in a discrete spacetime was shown to be described by a diffusion equation with one free parameter in the continuum limit. The derivation of this equation did not rely on any underlying particle model, but simply on the assumptions of discreteness and Poincar\'e invariance. It could be argued that the derivation does not guarantee a finite diffusion constant and thus a truly observable effect. To rebut this objection it was shown analytically that the continuum diffusion parameter was necessarily finite for the swerves model of particle propagation on a causal set. Numerical simulations of the swerves model and two `intrinsic' models gave results that clearly demonstrated diffusion behaviour, and in the case of the swerves model it was possible to determine the relationship between the diffusion parameter and the underlying model parameters. 

In the causal set approach, spacetime discreteness is expected to be of the order of the Planck scale. The constraints of computer memory limit causal set sizes simulated on a standard computer to around $2^{17}$ elements. We could question the point of even attempting to investigate large scale behaviour in causal sets that correspond to such absurdly small regions of spacetime (even a volume of $1mm^3\times1s$ would contain on the order of $10^{137}$ elements). Yet remarkably, the simulations do agree with the expected continuum behaviour. The results suggest that causal set simulations will have an important role to play in the further development of the theory and its phenomenology.

Lorentz invariance also allowed a useful phenomenological model for massless particles in discrete spacetime to be developed. Massless particles were shown to experience a drift and diffusion in energy in the continuum limit. Crucially, all massless particles still travel at the speed of light in this model. The accurate data available on the cosmic microwave background allowed both the drift and diffusion parameters to be tightly constrained. Whether the constraints are so tight as to exclude all possibility of future observations of the effect remains to be seen. 

The common word `particle' erroneously suggests massless and massive particles are variations of the same fundamental object. The diffusion experienced by massless particles is not simply the massless limit of the massive particle diffusion, suggesting that massive and massless particles will appear in quite different forms in any future complete causal set theory. How exactly will particles fit into the causal set approach? The massive particle models considered here were classical, point particle models and, as yet, we have no candidate models at all for massless particles. To produce any underlying model for massless particles, it seems likely quantum behaviour will have to be considered. 

Placing particles `on' a fixed causal set seems highly unsatisfactory. The causal set should be a dynamic structure, growing in number of elements. On large scales the theory must reproduce the relationship between mass and curvature, suggesting that the presence of particles will influence the growth of the causal set. The need to reproduce quantum effects suggests assigning amplitudes to each possible step in the causal set growth. It would be nice if we could just devise a theory now that satisfies all our requirements, but the problem is simply too big. Progress is made by looking at what can be learned from phenomenology, by developing quantum field theory on causal sets, by looking at the structure of curves spacetime in causal sets, and so on. Steps in the direction of a complete theory will hopefully lead to further phenomenology and, ultimately, falsifiable predictions. 

The power of the Lorentz invariance assumption has its limitations. When considering the polarisation of massless particles, Lorentz invariance proved insufficient to constrain the model. Identifying the assumptions necessary to constrain the model further will have to wait for the next step in understanding particles in causal set theory. In the meantime, restricting to linear polarisation has allowed progress to be made. In a discrete spacetime linearly polarised photons were found to experience rotation in polarisation angle and suppression in polarisation fraction. Rotation of polarisation angle is an effect predicted by a number of other approaches to quantum gravity, but the particular energy dependence of this model is unique. Again, cosmological data allowed the free parameters of the model to be constrained. 

It is important to remember that the results discussed here do not, for the most part, call on specific features of causal set theory. They rely on the assumptions of Lorentz invariance and discreteness and may, hopefully, be generalisable to other quantum gravity theories. A lot has been achieved looking at the large scale behaviour of particles in discrete spacetime, despite the incomplete nature of the theory. Where next can we look for signatures of spacetime discreteness?

The importance of phenomenology in the search for a theory of quantum gravity should not be underestimated. It is crucial in the development of the theories and, more importantly, reminds us to look out at the universe once in a while and remember why it is we seek such a theory to begin with.

%% file: PolarizationLI.tex
\chapter{Explicitly Lorentz invariant polarisation diffusion}
\label{a:polarLI}

The state space for a photon with polarisation is $\mathcal{M}=\mink\times\Lob_0\times\mathbb{P}$, where $\mathbb{P}$ is the polarisation state space. For an explicitly Lorentz invariant formulation of the polarisation diffusion equation, $\mathbb{P}$ is the polarisation state space with coordinates $P_{\mu\nu}$, where $P_{\mu\nu} = k_{\mu}a_{\nu}-a_{\mu}k_{\nu}$ is the complex two-form satisfying the conditions set out in Section~\ref{s:polarstatespace}. If we wish to discuss only linear polarisation, $P_{\mu\nu}$ can be taken to be a \textit{real} two-form satisfying the conditions
\begin{enumerate}
\item $P_{\mu\nu}=-P_{\nu\mu}\,$;
\item $P_{\mu\nu}P^{\mu\nu}=0\,$;
\item $P^{\mu\nu}k_{\mu}=0\,$;
\item $P^{\mu\nu}P_{\mu\sigma}=k^{\nu}k_{\sigma}\,$.
\end{enumerate}
Again, any $P_{\mu\nu}$ satisfying these conditions can be expressed as $P_{\mu\nu} = k_{\mu}a_{\nu}-a_{\mu}k_{\nu}$, where $k_{\mu}a^{\mu}=0$ and $a_{\mu}a^{\mu}=1$. $a^{\mu}$ is now a real four-vector. Condition (1) above provides six constraint equations, condition (2) a further one constraint. Although condition (3) is four equations, it gives only three new independent constraints. Finally condition (4), although 16 equations, gives only one new constraint. The state space $\mathbb{P}$ therefore has only one degree of freedom. 

\section{Invariant vectors}
Before writing down the diffusion equation, $K^{AB}$, $u^A$, and the density of states $n$ must be known. Consider the submanifold $\Lob_0\times\mathbb{P}$. There are two Lorentz invariant vector fields on the submanifold: $v_1^A = \left(k^{\mu},P^{\mu\nu}\right)$ and $v_2^A = \left(0,*P^{\mu\nu}\right)$, where $*P^{\mu\nu}$ is the Hodge dual defined in the usual way as $*P^{\mu\nu}=\frac{-1}{2}\epsilon^{\rho\sigma\mu\nu}P_{\rho\sigma}$.  To see this, first consider the more general vector fields $w_1^A = \left(\alpha_1 k^{\mu},\beta_1 P^{\mu\nu}\right)$ and  $w_2^A = \left(\alpha_2 k^{\mu},\beta_2 *P^{\mu\nu}\right)$. These vectors are clearly Lorentz invariant, as is the submanifold $\Lob_0\times\mathbb{P}$. It remains to show that the vectors are tangent to the submanifold.

The normals to the submanifold are derived from the constraint equations in the usual way:
%%%%%%%%%%%%%%%%%%%%%%%%%%%%%%%%%%%%%%%%%%%%%%%%%%%%%%%%%%%%%%%%%%%%%%%%%%%%%%
\begin{eqnarray}
k^{\mu}k_{\mu} = 0 &\Rightarrow n^1_A &= \left(\frac{\partial\left(k^{\nu}k_{\nu}\right)}{\partial k^{\mu}},\frac{\partial\left(k^{\nu}k_{\nu}\right)}{\partial P^{\rho\sigma}}\right)\nonumber\\
 &&= \left(2k_{\mu},0\right)\,,\\
P^{\mu\nu}P_{\mu\nu} = 0 &\Rightarrow n^2_A &= \left(\frac{\partial\left(P^{\rho\sigma}P_{\rho\sigma}\right)}{\partial k^{\mu}},\frac{\partial\left(P^{\mu\nu}P_{\mu\nu}\right)}{\partial P^{\rho\sigma}}\right)\nonumber\\
&&= \left(0,2P_{\rho\sigma}\right)\,,\\
P^{\mu\alpha}k_{\mu} = 0 &\Rightarrow n^{\alpha}_A &= \left(\frac{\partial\left(P^{\nu\alpha}k_{\nu}\right)}{\partial k^{\mu}},\frac{\partial\left(P^{\nu\alpha}k_{\nu}\right)}{\partial P^{\rho\sigma}}\right)\nonumber\\
&&= \left({P_{\mu}}^{\alpha},{\delta^{\alpha}}_{\sigma}k_{\rho}\right)\,,\\
P^{\mu\nu}P_{\mu\sigma}=k^{\nu}k_{\sigma}&\Rightarrow {n^{\nu}_{\sigma}}_A &= \left(\frac{\partial}{\partial k^{\alpha}}\left(P^{\mu\nu}P_{\mu\sigma}-k^{\nu}k_{\sigma}\right),\frac{\partial}{\partial P^{\alpha\beta}}\left(P^{\mu\nu}P_{\mu\sigma}-k^{\nu}k_{\sigma}\right)\right)\nonumber\\
 &&= \left(-g_{\sigma\alpha}k^{\nu}-\delta^{\nu}_{\alpha}k_{\sigma},\delta^{\nu}_{\beta}P_{\alpha\sigma}+g_{\sigma\beta}{P_{\alpha}}^{\nu}\right).
\end{eqnarray}
%%%%%%%%%%%%%%%%%%%%%%%%%%%%%%%%%%%%%%%%%%%%%%%%%%%%%%%%%%%%%%%%%%%%%%%%%%%%%%
The vector $w_1$ is clearly orthogonal to the first three normals:
%%%%%%%%%%%%%%%%%%%%%%%%%%%%%%%%%%%%%%%%%%%%%%%%%%%%%%%%%%%%%%%%%%%%%%%%%%%%%%
\begin{eqnarray}
n^1_A w_1^A &=& 2\alpha_1 k_{\mu}k^{\mu}\nonumber\\
&=&0,\\
n^2_A w_1^A &=& 2\beta_1 P_{\rho\sigma} P^{\rho\sigma}\nonumber\\
&=& 0,\\
n^{\alpha}_A w_1^A &=& \alpha_1 {P_{\mu}}^{\alpha}k^{\mu} + \beta_1 {\delta^{\alpha}}_{\sigma}k_{\rho}P^{\rho\sigma}\nonumber\\
&=& 0.
\end{eqnarray}
%%%%%%%%%%%%%%%%%%%%%%%%%%%%%%%%%%%%%%%%%%%%%%%%%%%%%%%%%%%%%%%%%%%%%%%%%%%%%%%%
For the last normal
%%%%%%%%%%%%%%%%%%%%%%%%%%%%%%%%%%%%%%%%%%%%%%%%%%%%%%%%%%%%%%%%%%%%%%%%%%%%%%%%
\begin{eqnarray}
{n^{\nu}_{\sigma}}_Aw_1^A &=& \left(-g_{\sigma\alpha}k^{\nu}-\delta^{\nu}_{\alpha}k_{\sigma}\right)\alpha_1 k^{\alpha} + \left(\delta^{\nu}_{\beta}P_{\alpha\sigma}+g_{\sigma\beta}{P_{\alpha}}^{\nu}\right)\beta_1 P^{\alpha\beta}\nonumber\\
&=&-\alpha_1\left(k^{\nu}k_{\sigma}+k^{\nu}k_{\sigma}\right)+\beta_1\left(P_{\alpha\sigma}P^{\alpha\nu}+{P_{\alpha}}^{\nu}{P^{\alpha}}_{\sigma}\right)\nonumber\\
&=& 2\left(-\alpha_1k^{\nu}k_{\sigma} + \beta_1P_{\alpha\sigma}P^{\alpha\nu}\right)\,.
\end{eqnarray}
This equals zero only if $\alpha_1 = \beta_1$. Thus the vector $v_1^A=\left(k^{\mu}, P^{\rho\sigma}\right)$ is normal to the surface.

$w_2$ is also obviously orthogonal to the first three normals if it is first noted that $P_{\mu\nu}*P^{\mu\nu}=0$ and $*P^{\mu\nu}k_{\mu} = 0$ if both $P_{\mu\nu}P^{\mu\nu}=0$ and $P^{\mu\nu}k_{\mu} = 0$. These results can be shown simply if $P_{\mu\nu}$ is expressed as $P_{\mu\nu}=k_{\mu}a_{\nu}-a_{\mu}k_{\nu}$.
%%%%%%%%%%%%%%%%%%%%%%%%%%%%%%%%%%%%%%%%%%%%%%%%%%%%%%%%%%%%%%%%%%%%%%%%%%%%%%%%%%%%%%%%%%%
\begin{eqnarray}
n^1_A w_2^A &=& 2\alpha_2 k_{\mu}k^{\mu}\nonumber\\
&=&0,\\
n^2_A w_2^A &=& 2\beta_2 P_{\rho\sigma} *P^{\rho\sigma}\nonumber\\
&=& 0,\\
n^{\alpha}_A w_2^A &=& \alpha_2 {P_{\mu}}^{\alpha}k^{\mu} + \beta_2 {\delta^{\alpha}}_{\sigma}k_{\rho}*P^{\rho\sigma}\nonumber\\
&=& 0.
\end{eqnarray}
%%%%%%%%%%%%%%%%%%%%%%%%%%%%%%%%%%%%%%%%%%%%%%%%%%%%%%%%%%%%%%%%%%%%%%%%%%%%%%%%%%%%%%%%%%%%%
The remaining normal gives
\begin{eqnarray}
{n^{\nu}_{\sigma}}_A w_2^A &=& \left(-g_{\sigma\alpha}k^{\nu}-\delta^{\nu}_{\alpha}k_{\sigma}\right)\alpha_2 k^{\alpha} + \left(\delta^{\nu}_{\beta}P_{\alpha\sigma}+g_{\sigma\beta}{P_{\alpha}}^{\nu}\right)\beta_2 *P^{\alpha\beta}\nonumber\\
&=&-\alpha_2\left(k^{\nu}k_{\sigma}+k^{\nu}k_{\sigma}\right)+\beta_2\left(P_{\alpha\sigma}*P^{\alpha\nu}+{P_{\alpha}}^{\nu}*{P^{\alpha}}_{\sigma}\right)\nonumber\\
&=& 2\left(-\alpha_2 k^{\nu}k_{\sigma} + \beta_2P_{\alpha\sigma}*P^{\alpha\nu}\right)\,.
\end{eqnarray}
It turns out that $P_{\alpha\sigma}*P^{\alpha\nu}=0$, this is easily checked by choosing a specific frame, say $k^{\mu}=(k,0,0,k)$.
Therefore $w_2$ is only tangent to the surface if $\alpha_2 = 0$. 
Thus $v_2^A=\left(0, *P^{\mu\nu}\right)$ is an invariant vector.

\section{The diffusion equation}

Given the above invariant vectors $v_1$ and $v_2$, the most general Lorentz invariant, symmetric, positive semi-definite matrix $K^{AB}$ on the submanifold $\Lob_0\times\mathbb{P}$ is
\begin{equation}
K^{AB} = av_1^Av_1^B + b\left(v_1^Av_2^B + v_2^Av_1^B\right) + c v_2^Av_2^B\,,
\end{equation}
the general invariant vector is
\begin{equation}
u^A = dv_1^A+ev_2^A\,,
\end{equation}
where $a,\,b,\,c,\,d,\,e$ are constants and $ac-b^2\geq 0$.
The spacetime components of $K^{AB}$ and $u^A$ on the full state space $\mathcal{M}$ are unchanged from the zero spin massless particle case: $K^{\mu A}=0$, $u^{\mu}=k^{\mu}$.

The diffusion equation can thus be written down from Equation~\ref{e:diffequA}:
\begin{eqnarray}
\pd{\rho_{\lambda}}{\lambda}&=&\partial_A\left(K^{AB}n\partial_B\left(\frac{\rho_{\lambda}}{n}\right)-u^A\rho_{\lambda}\right)\nonumber\\
&=& -k^{\mu}\pd{\rho_{\lambda}}{x^{\mu}}\nonumber\\
&& + \pd{}{k^{\mu}}\left[ak^{\mu}k^{\nu}\pd{}{k^{\nu}}\left(\frac{\rho_{\lambda}}{n}\right) + \left(ak^{\mu}P^{\rho\sigma}+bk^{\mu}*P^{\rho\sigma}\right)n\pd{}{P^{\rho\sigma}}\left(\frac{\rho_{\lambda}}{n}\right)-dk^{\mu}\rho_{\lambda}\right]\nonumber\\
&& + \pd{}{P^{\rho\sigma}}\left[\left(ak^{\mu}P^{\rho\sigma}+bk^{\mu}*P^{\rho\sigma}\right)n\pd{}{k^{\mu}}\left(\frac{\rho_{\lambda}}{n}\right)\right.\nonumber\\
&&\left. + \left(aP^{\rho\sigma}P^{\gamma\delta} + b\left(P^{\rho\sigma}*P^{\gamma\delta}+*P^{\rho\sigma}P^{\gamma\delta}\right)+c*P^{\rho\sigma}*P^{\gamma\delta}\right)n\pd{}{P^{\gamma\delta}}\left(\frac{\rho_{\lambda}}{n}\right)\right.\nonumber\\
&& \left.-\left(dP^{\rho\sigma}+e*P^{\rho\sigma}\right)\rho_{\lambda}\vphantom{\pd{}{k^{\mu}}}\right]\,.
\end{eqnarray}

Although it explicitly demonstrates the Lorentz invariance of the process, this equation is hardly user-friendly. To understand the physical consequences of the diffusion, the equation derived in Section~\ref{s:polaraffinetime} using the Bloch sphere as the polarisation state space is far more useful. It is important to confirm however, that these equations are in fact equivalent.  
First consider polar coordinates on $\Lob_0$: $\{k,\theta,\phi\}$. In these coordinates the vector $k^{\mu}$ has components $(k,0,0)$. As mentioned earlier, the polarisation state space $\mathbb{P}$ has only one degree of freedom -- specifically, it can be shown that the constraints on $P^{\mu\nu}$ reduce the polarisation state space to a circle. The density of states, $n$, on this circle is simply a constant and thus the density of states for the full state space $\mathcal{M}$ is $n=k\sin\theta$. The submanifold $\Lob_0\times\mathbb{P}$ is thus the momentum space, a null cone, with a copy of the polarisation space, a circle, at each point. The invariant vector $v_1^A = (k^{\mu},P^{\mu\nu})$ is simply the vector at a point on the cone that points along the cone, away from the tip. The invariant vector $v_2^A=(0,*P^{\mu\nu})$ is the tangent vector to the polarisation circle. Denoting the coordinate on the circle $\psi$, $v_1^A$ and $v_2^A$ become $v_1^A=(k,0,0,0)$ and $v_2^A=(0,0,0,1)$ in coordinates $\{k,\,\theta,\,\phi,\,\psi\}$. Thus, this formalism is exactly the same as that used in Section~\ref{s:polaraffinetime}.

%% file: thesis.bbl
\begin{thebibliography}{10}

\bibitem{Popper:1992}
K.~R. {Popper}.
\newblock {\em {The logic of scientific discovery}}.
\newblock Routledge, London, 1992.

\bibitem{Liberati:2009pf}
Stefano Liberati and Luca Maccione.
\newblock {Lorentz Violation: Motivation and new constraints}.
\newblock 2009, 0906.0681.

\bibitem{Dowker:2003hb}
Fay Dowker, Joe Henson, and Rafael~D. Sorkin.
\newblock {Quantum gravity phenomenology, Lorentz invariance and discreteness}.
\newblock {\em Mod. Phys. Lett.}, A19:1829--1840, 2004, gr-qc/0311055.

\bibitem{Spivak:1979}
Michael Spivak.
\newblock {\em ~A Comprehensive Introduction to Differential Geometry},
  volume~2.
\newblock Publish or Perish Inc, Berkeley CA, 1979.

\bibitem{Whitehead:1948}
Alfred~North Whitehead.
\newblock {\em Essays in Science and Philosophy}.
\newblock Rider and Company, London, 1948.

\bibitem{Stachel:1986}
John Stachel.
\newblock Einstein and the quantum: Fifty years of struggle.
\newblock In R.G. Colodny, editor, {\em From Quarks to Quasars, Philosophical
  Problems of Modern Physics}, page 379. U. Pittsburgh Press, 1986.

\bibitem{Schrodinger:1954}
Erwin Schr{\"o}dinger.
\newblock {\em Nature and the Greeks and Science and Humanism}, page~65.
\newblock Cambridge University Press, Cambridge, 1996.

\bibitem{Abramenko:1958}
B.~Abramenko.
\newblock {On dimensionality and continuity of physical space and time}.
\newblock {\em British Journal for the Philosophy of Science}, IX(34):89--109,
  1958.

\bibitem{Snyder:1946qz}
Hartland~S. Snyder.
\newblock {Quantized space-time}.
\newblock {\em Phys. Rev.}, 71:38--41, 1947.

\bibitem{Bombelli:1987aa}
Luca Bombelli, Joo-Han Lee, David Meyer, and Rafael Sorkin.
\newblock Space-time as a causal set.
\newblock {\em Phys. Rev. Lett}, 59:521, 1987.

\bibitem{Malament:1977}
David~B. Malament.
\newblock The class of continuous timelike curves determines the topology of
  spacetime.
\newblock {\em J. Math. Phys.}, 18:1399--1404, 1977.

\bibitem{Levichev:1987}
A.~V. Levichev.
\newblock Prescribing the conformal geometry of a {L}orentz manifold by means
  of its causal structure.
\newblock {\em Soviet Math. Dokl.}, 35:452--455, 1987.

\bibitem{Hawking:1976fe}
S.~W. Hawking, A.~R. King, and P.~J. McCarthy.
\newblock A new topology for curved space-time which incorporates the causal,
  differential, and conformal structures.
\newblock {\em J. Math. Phys.}, 17:174--181, 1976.

\bibitem{Myrheim:1978}
J.~Myrheim.
\newblock Statistical geometry, 1978.
\newblock {C}ERN preprint TH-2538.

\bibitem{tHooft:1979}
G.~'t~Hooft.
\newblock Quantum gravity: a fundamental problem and some radical ideas.
\newblock In M.~Levy and S.~Deser, editors, {\em Recent Developments in
  Gravitation (Proceedings of the 1978 Cargese Summer Institute)}. Plenum, New
  York, 1979.

\bibitem{Henson:2006kf}
Joe Henson.
\newblock {The causal set approach to quantum gravity}.
\newblock In D.~Oriti, editor, {\em {Approaches to Quantum Gravity: Towards a
  New Understanding of Space, Time and Matter}}, chapter~21. Cambridge
  University Press, Cambridge, 2009, gr-qc/0601121.

\bibitem{Dowker:2006sb}
F.~Dowker.
\newblock Causal sets as discrete spacetime.
\newblock {\em Contemp. Phys.}, 47:1--9, 2006.

\bibitem{Sorkin:1990bh}
Rafael~D. Sorkin.
\newblock First steps with causal sets.
\newblock In R.~Cianci, R.~de~Ritis, M.~Francaviglia, G.~Marmo, C.~Rubano, and
  P.~Scudellaro, editors, {\em Proceedings of the ninth Italian Conference on
  General Relativity and Gravitational Physics, Capri, Italy, September 1990},
  pages 68--90. World Scientific, Singapore, 1991.

\bibitem{Sorkin:1990bj}
Rafael~D. Sorkin.
\newblock Space-time and causal sets.
\newblock In J.~C. D'Olivo, E.~Nahmad-Achar, M.~Rosenbaum, M.~P. Ryan, L.~F.
  Urrutia, and F.~Zertuche, editors, {\em Relativity and Gravitation: Classical
  and Quantum, Proceedings of the SILARG VII Conference, Cocoyoc, Mexico,
  December 1990}, pages 150--173. World Scientific, Singapore, 1991.

\bibitem{Sorkin:2003bx}
Rafael~D. Sorkin.
\newblock Causal sets: Discrete gravity (notes for the {V}aldivia summer
  school).
\newblock In Andr{\'e}s Gomberoff and Don Marolf, editors, {\em Lectures on
  Quantum Gravity, Proceedings of the Valdivia Summer School, Valdivia, Chile,
  January 2002}, pages 305--327. Springer, New York, 2005, gr-qc/0309009.

\bibitem{Bombelli:2006nm}
Luca Bombelli, Joe Henson, and Rafael~D. Sorkin.
\newblock {Discreteness without symmetry breaking: A theorem}.
\newblock {\em Mod. Phys. Lett.}, A24:2579--2587, 2009, gr-qc/0605006.

\bibitem{Brightwell:1991}
Graham Brightwell and Ruth Gregory.
\newblock Structure of random discrete spacetime.
\newblock {\em Phys. Rev. Lett}, 66:260--263, 1991.

\bibitem{Ilie:2005qg}
Raluca Ilie, Gregory~B. Thompson, and David~D. Reid.
\newblock {A numerical study of the correspondence between paths in a causal
  set and geodesics in the continuum}.
\newblock {\em Class. Quant. Grav.}, 23:3275--3286, 2006, gr-qc/0512073.

\bibitem{Sorkin:1997gi}
Rafael~D. Sorkin.
\newblock Forks in the road, on the way to quantum gravity.
\newblock {\em Int. J. Theor. Phys.}, 36:2759--2781, 1997, gr-qc/9706002.

\bibitem{Ahmed:2002mj}
Maqbool Ahmed, Scott Dodelson, Patrick~B. Greene, and Rafael Sorkin.
\newblock Everpresent lambda.
\newblock {\em Phys. Rev.}, D69:103523, 2004, astro-ph/0209274.

\bibitem{Sorkin:2007bd}
Rafael~D. Sorkin.
\newblock {Is the cosmological `constant' a nonlocal quantum residue of
  discreteness of the causal set type?}
\newblock {\em AIP Conf. Proc.}, 957:142--153, 2007, 0710.1675.

\bibitem{Barrow:2006vy}
John~D. Barrow.
\newblock {A strong constraint on ever-present Lambda}.
\newblock {\em Phys. Rev.}, D75:067301, 2007, gr-qc/0612128.

\bibitem{Zuntz:2008zza}
Joseph~A. Zuntz.
\newblock {The cosmic microwave background in a causal set universe}.
\newblock {\em Phys. Rev.}, D77:043002, 2008, 0711.2904.

\bibitem{Philpott:2008vd}
Lydia Philpott, Fay Dowker, and Rafael~D. Sorkin.
\newblock {Energy-momentum diffusion from spacetime discreteness}.
\newblock {\em Phys. Rev.}, D79:124047, 2009, 0810.5591.

\bibitem{Kaloper:2006pj}
Nemanja Kaloper and David Mattingly.
\newblock {Low energy bounds on Poincare violation in causal set theory}.
\newblock {\em Phys. Rev.}, D74:106001, 2006, astro-ph/0607485.

\bibitem{Philpott:2009fz}
Lydia Philpott.
\newblock {Particle simulations in causal set theory}.
\newblock {\em Class. Quantum Grav}, 27(042001), 2010, 0911.5595.

\bibitem{Johnston:2008za}
Steven Johnston.
\newblock {Particle propagators on discrete spacetime}.
\newblock {\em Class. Quant. Grav.}, 25:202001, 2008, 0806.3083.

\bibitem{Johnston:2009fr}
Steven Johnston.
\newblock {Feynman propagator for a free scalar field on a causal set}.
\newblock {\em Phys. Rev. Lett.}, 103:180401, 2009, 0909.0944.

\bibitem{Sorkin:1986}
R.D. Sorkin.
\newblock Stochastic evolution on a manifold of states.
\newblock {\em Ann. Phys.}, 168:119--147, 1986.

\bibitem{Mattingly:2007be}
David Mattingly.
\newblock {Causal sets and conservation laws in tests of Lorentz symmetry}.
\newblock {\em Phys. Rev.}, D77:125021, 2008, 0709.0539.

\bibitem{Cactus}
http://www.cactuscode.org/.

\bibitem{Sheskin:2004}
David~J. Sheskin.
\newblock {\em {Handbook of parametric and nonparametric statistical
  procedures}}.
\newblock Chapman and Hall/CRC, Florida, 2004.

\bibitem{Taylor:1997}
John~Robert Taylor.
\newblock {\em {An introduction to error analysis}}.
\newblock University Science Books, Sausalito, 1997.

\bibitem{Sorkin:2007qi}
Rafael~D. Sorkin.
\newblock {Does locality fail at intermediate length-scales?}
\newblock In D.~Oriti, editor, {\em {Approaches to Quantum Gravity: Towards a
  New Understanding of Space, Time and Matter}}, chapter~3. Cambridge
  University Press, Cambridge, 2009, gr-qc/0703099.

\bibitem{Sorkin:2009bp}
Rafael~D. Sorkin.
\newblock {Light, Links and Causal Sets}.
\newblock {\em J. Phys. Conf. Ser.}, 174:012018, 2009, 0910.0673.

\bibitem{Einstein:1972}
Albert Einstein and Michele Besso.
\newblock {\em Correspondance 1903-1955}.
\newblock Hermann, Paris, 1972.

\bibitem{Dowker:2009}
Fay Dowker, Joe Henson, and Rafael~D. Sorkin.
\newblock The coherence of light from distant sources in a discrete universe.
\newblock In preparation.

\bibitem{Kolb:1990}
E.~W. Kolb and M.~S. Turner.
\newblock {\em The Early Universe}.
\newblock Addison-Wesley, Reading MA, 1990.

\bibitem{Fixsen:1996nj}
D.~J. Fixsen et~al.
\newblock {The Cosmic Microwave Background Spectrum from the Full COBE/FIRAS
  Data Set}.
\newblock {\em Astrophys. J.}, 473:576, 1996, astro-ph/9605054.

\bibitem{Fan:2001ff}
Xiaohui Fan et~al.
\newblock {A Survey of z$>$5.8 Quasars in the Sloan Digital Sky Survey I:
  Discovery of Three New Quasars and the Spatial Density of Luminous Quasars at
  z$\sim$6}.
\newblock {\em Astron. J.}, 122:2833, 2001, astro-ph/0108063.

\bibitem{Bergstrom:2006}
L.~{Bergstr{\"o}m} and A.~{Goobar}.
\newblock {\em {Cosmology and particle astrophysics.}}
\newblock Springer-Praxis, Chichester, 2006.

\bibitem{Hinshaw:2008kr}
G.~Hinshaw et~al.
\newblock {Five-Year Wilkinson Microwave Anisotropy Probe (WMAP 1)
  Observations:Data Processing, Sky Maps, \& Basic Results}.
\newblock {\em Astrophys. J. Suppl.}, 180:225--245, 2009, 0803.0732.

\bibitem{Becker:2001ee}
Robert~H. Becker et~al.
\newblock {Evidence for Reionization at z$\sim$6: Detection of a Gunn- Peterson
  Trough in a z=6.28 Quasar}.
\newblock {\em Astron. J.}, 122:2850, 2001, astro-ph/0108097.

\bibitem{Contaldi:2010fh}
Carlo Contaldi, Fay Dowker, and Lydia Philpott.
\newblock {Polarization Diffusion from Spacetime Uncertainty}.
\newblock 2010, 1001.4545.

\bibitem{Weinberg:1995}
S.~{Weinberg}.
\newblock {\em {The Quantum Theory of Fields}}.
\newblock Cambridge University Press, June 1995.

\bibitem{Born:1980}
M.~Born and E.~Wolf.
\newblock {\em Principles of optics}.
\newblock Pergamon Press, Oxford, 1980.

\bibitem{Kosowsky:1994cy}
Arthur Kosowsky.
\newblock {Cosmic microwave background polarization}.
\newblock {\em Ann. Phys.}, 246:49--85, 1996, astro-ph/9501045.

\bibitem{Dym:1972}
H.~Dym and H.~P. Mc{K}ean.
\newblock {\em {Fourier series and integrals}}.
\newblock Academic Press, New York, 1972.

\bibitem{Klimov:1997}
A.~B. {Klimov} and S.~M. {Chumakov}.
\newblock {Gaussians on the circle and quantum phase}.
\newblock {\em Physics Letters A}, 235:7--14, February 1997.

\bibitem{Luan:2004}
P.-G. {Luan} and Y.-M. {Kao}.
\newblock {Drifting diffusion on a circle as continuous limit of a multiurn
  Ehrenfest model}.
\newblock {\em \pre}, 69(2):022102, February 2004, arXiv:physics/0308023.

\bibitem{Gambini:1998it}
Rodolfo Gambini and Jorge Pullin.
\newblock {Nonstandard optics from quantum spacetime}.
\newblock {\em Phys. Rev.}, D59:124021, 1999, gr-qc/9809038.

\bibitem{Gleiser:2001rm}
Reinaldo~J. Gleiser and Carlos~N. Kozameh.
\newblock {Astrophysical limits on quantum gravity motivated birefringence}.
\newblock {\em Phys. Rev.}, D64:083007, 2001, gr-qc/0102093.

\bibitem{Maccione:2008tq}
Luca Maccione, Stefano Liberati, Annalisa Celotti, John~G. Kirk, and Pietro
  Ubertini.
\newblock {Gamma-ray polarization constraints on Planck scale violations of
  special relativity}.
\newblock {\em Phys. Rev.}, D78:103003, 2008, 0809.0220.

\bibitem{Gubitosi:2009eu}
Giulia Gubitosi, Luca Pagano, Giovanni Amelino-Camelia, Alessandro Melchiorri,
  and Asantha Cooray.
\newblock {A Constraint on Planck-scale Modifications to Electrodynamics with
  CMB polarization data}.
\newblock {\em JCAP}, 0908:021, 2009, 0904.3201.

\bibitem{Carroll:1998zi}
Sean~M. Carroll.
\newblock {Quintessence and the rest of the world}.
\newblock {\em Phys. Rev. Lett.}, 81:3067--3070, 1998, astro-ph/9806099.

\bibitem{Liu:2006uh}
Guo-Chin Liu, Seokcheon Lee, and Kin-Wang Ng.
\newblock {Effect on cosmic microwave background polarization of coupling of
  quintessence to pseudoscalar formed from the electromagnetic field and its
  dual}.
\newblock {\em Phys. Rev. Lett.}, 97:161303, 2006, astro-ph/0606248.

\bibitem{Lue:1998mq}
Arthur Lue, Li-Min Wang, and Marc Kamionkowski.
\newblock {Cosmological signature of new parity-violating interactions}.
\newblock {\em Phys. Rev. Lett.}, 83:1506--1509, 1999, astro-ph/9812088.

\bibitem{Li:2008tma}
Mingzhe Li and Xinmin Zhang.
\newblock {Cosmological CPT violating effect on CMB polarization}.
\newblock {\em Phys. Rev.}, D78:103516, 2008, 0810.0403.

\bibitem{Carroll:1989vb}
Sean~M. Carroll, George~B. Field, and Roman Jackiw.
\newblock {Limits on a Lorentz and Parity Violating Modification of
  Electrodynamics}.
\newblock {\em Phys. Rev.}, D41:1231, 1990.

\bibitem{Feng:2004mq}
Bo~Feng, Hong Li, Ming-zhe Li, and Xin-min Zhang.
\newblock {Gravitational leptogenesis and its signatures in CMB}.
\newblock {\em Phys. Lett.}, B620:27--32, 2005, hep-ph/0406269.

\bibitem{Prasanna:2001xk}
A.~R. Prasanna and S.~Mohanty.
\newblock {Gravitational wave induced rotation of the plane of polarization of
  pulsar signals}.
\newblock {\em Europhys. Lett.}, 57:651--655, 2002, astro-ph/0110606.

\bibitem{Carroll:1997tc}
Sean~M. Carroll and George~B. Field.
\newblock {Is there evidence for cosmic anisotropy in the polarization of
  distant radio sources?}
\newblock {\em Phys. Rev. Lett.}, 79:2394--2397, 1997, astro-ph/9704263.

\bibitem{Cimatti:1994yc}
Andrea Cimatti, Sperello di~Serego~Alighieri, George~B. Field, and Robert A.~E.
  Fosbury.
\newblock {Stellar and scattered light in a radio galaxy at z = 2.63}.
\newblock {\em Astrophys. J.}, 422:562, 1994.

\bibitem{Dunkley:2008mk}
J.~Dunkley et~al.
\newblock {Five-Year Wilkinson Microwave Anisotropy Probe (WMAP) Observations:
  Bayesian Estimation of CMB Polarization Maps}.
\newblock {\em \apj}, 701:1804--1813, August 2009, 0811.4280.

\bibitem{Chiang:2009xsa}
H.~C. Chiang et~al.
\newblock {Measurement of CMB Polarization Power Spectra from Two Years of
  BICEP Data}.
\newblock 2009, 0906.1181.

\bibitem{Brown:2009uy}
M.~L. Brown et~al.
\newblock {Improved measurements of the temperature and polarization of the CMB
  from QUaD}.
\newblock {\em Astrophys. J.}, 705:978--999, 2009, 0906.1003.

\bibitem{Dodelson:2003}
S.~{Dodelson}.
\newblock {\em {Modern cosmology}}.
\newblock Academic Press, Amsterdam, 2003.

\bibitem{Feng:2006dp}
Bo~Feng, Mingzhe Li, Jun-Qing Xia, Xuelei Chen, and Xinmin Zhang.
\newblock {Searching for CPT violation with WMAP and BOOMERANG}.
\newblock {\em Phys. Rev. Lett.}, 96:221302, 2006, astro-ph/0601095.

\bibitem{Xia:2009ah}
Jun-Qing Xia, Hong Li, and Xinmin Zhang.
\newblock {Probing CPT Violation with CMB Polarization Measurements}.
\newblock 2009, 0908.1876.

\end{thebibliography}
